\documentclass[aps,prd,floatfix,superscriptaddress,onecolumn,showpacs,showkeys,amssymb,10pt]{revtex4-1}

\usepackage{amssymb,amsmath,verbatim}
\usepackage{graphicx}
\usepackage{epsfig}
\usepackage{dcolumn}
\usepackage{bm}

\newcommand{\be}{\begin{equation}}
\newcommand{\ee}{\end{equation}\noindent}
\newcommand{\eei}{\end{equation}}
\newcommand{\bea}{\begin{eqnarray}}
\newcommand{\eea}{\end{eqnarray}\noindent}
\newcommand{\eeai}{\end{eqnarray}}

\newcommand{\hf} {\frac{1}{2}}
\newcommand{\nn}{\nonumber\\}

\def\eq#1{(\ref{#1})}

\def\ord#1{{\cal O}(#1)}

\def\t#1{{\tilde#1}}
\def\c#1{{\cal#1}}
\def\b#1{{\bar#1}}
\def\mf#1{{\mathfrak#1}}
\def\pe{\perp}
\def\pa{\parallel}

\def\Tr{{\mathrm{Tr}}}

\def\u#1{{\underline{#1}}}

\def\fd#1#2{\frac{\delta#1}{\delta#2}}
\def\fdd#1#2#3{\frac{\delta^2#1}{\delta#2\delta#3}}
\def\fddd#1#2#3#4{\frac{\delta^3#1}{\delta#2\delta#3\delta#4}}
\def\fdddd#1#2#3#4#5{\frac{\delta^4#1}{\delta#2\delta#3\delta#4\delta#5}}

\begin{document}

\title{Modified renormalization group method applied to the $O(1)$ ghost model with periodic condensate}

\author{Z. P\'eli}
\affiliation{Department of Theoretical Physics, University of Debrecen,
P.O. Box 5, H-4010 Debrecen, Hungary}

\author{S. Nagy}
\affiliation{Department of Theoretical Physics, University of Debrecen,
P.O. Box 5, H-4010 Debrecen, Hungary}

\author{K. Sailer}
\affiliation{Department of Theoretical Physics, University of Debrecen,
P.O. Box 5, H-4010 Debrecen, Hungary}

\date{\today}

\begin{abstract}
In order to discuss the occurrence of a periodic condensate  in the  Euclidean 3-dimensional ghost $O(1)$ model, a modified version of the effective average action (EAA) renormalization group (RG) method is developed, called by us Fourier-Wetterich RG approach.  It is proposed to start with an ansatz for the EAA, that contains terms, in addition to the usual ones, induced by the various Fourier-modes of the periodic condensate and to expand the EAA in functional Taylor-series around the periodic background. The RG flow equations  are derived in the next-to-next-to-leading order of the gradient expansion (GE). No field-dependence of the derivative couplings have been taken into account and $Z_2$ symmetry of  the EAA is preserved. Preliminary numerical results have been obtained under various additional simplifying assumptions. The characteristics of the Wilson-Fisher  fixed point  and the phase structure of the model have been determined numerically in the local potential approximation and in the next-to-leading order  of the GE, when the periodic condensate has been modelled by a single cosine mode in one spatial direction. From the  preliminary results important information is gained on further possibilities to improve  the proposed RG scheme. 
\end{abstract}



\maketitle

\section{Introduction}

There are various models in solid state physics and quantum field theory, in which the ground state exhibits a periodic structure, because the periodic vacuum
provides a deeper minimum of the effective action than the homogeneous one.
Such cases are, for example, the Larkin-Ovchinnikov-Fulde-Ferrell type spatially inhomogeneous  superconducting states \cite{Larkin1964,Fulde1964}, the phase with periodically modulated chiral fields in effective chiral quark models \cite{Kutschera1990,Sadzik2000}, the spin waves in the Heisenberg ferromagnet\cite{Holstein1940}, and the charged massive Schwinger model \cite{Nagy2004} (without pretending for completeness).  Inhomogeneity of the ground state suggests that strongly distance-dependent  interactions should be present in the system.  It is reasonable to expect that for such interactions the gradient terms with sufficiently strong couplings are responsible rather than the ultralocal potential terms in the effective action.
In such models, the expectation value of the kinetic-energy operator is nonvanishing, which is constructed  from the derivatives of the fundamental fields. The nonvanishing expectation value arises generally due to the alternating signs of the gradient terms of various orders (see \cite{Fin1997,Four2000,Basar2009} and Ref. [7] of \cite{Nagy2004}). This so-called kinetic condensation
means that quasiparticles with non-zero momentum appear and condense forming a periodic classical background field, which breaks spatial symmetries of the system spontaneously. As opposed to this, in the case of the Nambu-Goldstone type spontaneous symmetry breaking quasiparticles with vanishing momentum condense into a homogeneous classical background field.

In the present paper we consider a rather simple model in which a periodic vacuum state can occur, the 3-dimensional, Euclidean  ghost scalar $O(1)$ model, where the usual quadratic gradient term has the wrong sign, i.e., the wave function renormalization $Z=-1$ is negative and the positive quartic gradient term with the coupling $Y>0$ ensures the boundedness of the action from below. In modern quantum gravity research much attention has been paid recently to the role of the ghosts in various cosmological scenarios close to the Planck scale and beyond it, e.g., spin-2 states in gravity theories with quadratic curvature invariants \cite{Pais1950}-\cite{Ivanov2016}, the Veneziano ghost dark energy model \cite{Urban2009a}-\cite{Khurshudyan13}, ghost scalar fields in phantom dark matter \cite{Caldwell2002,Brown2008}, and  in `quintom' models for bouncing cosmology \cite{Feng2005,Feng2006}. Among those the kinetic condensation in the simple ghost complex scalar $U(1)$ model  has some relevance in studies of the conformal degree of freedom in gravity and has been discussed in  \cite{Reuter2000,Reuter2013}. There, the simplicity of the model allowed  the authors to find analytically the true ground state and  calculate the scale dependence of the EAA and therefore the scale dependence of the dressed inverse propagator by the means of the saddle-point approximation. The ground state has proven to be a family of plane waves which causes the spontaneous breaking of the $U(1)$ symmetry and also that of the rotational and translational symmetries in spacetime, because of the fixed phase and wave vector of the given ground state.   It was also found, that the renormalization effects are dominated by the instability of the trivial saddle point, rather than by the quantum fluctuations.

In the present paper we develop a  modified version of EAA RG  approach \cite{Wetterich1993} in order to investigate the possibility of spatially periodic vacuum states in 3-dimensional Euclidean ghost $O(1)$ model. Flow equations  are derived  for the kinetic energy and the potential energy pieces of the EAA in  NNLO of GE, i.e., including  the derivative term of the order $\ord{\partial^4}$.
Preliminary numerical results are obtained for the  characteristics of the WF FP and the phase structure of the model  in LPA and in NLO of GE. These seem to provide  guidelines for the potential improvements of the proposed RG approach.
It is considered the rather simple ghost model  of the one-component scalar field $\phi_x$ with the kinetic energy operator $\c{Z}(-\Box)=-Z_k \Box+ Y_k \Box^2$  (with the 3-dimensional Laplace-operator $\Box=\partial_\mu\partial_\mu$) with $Z_k<0$ and $Y_k>0$. The field-dependence of the derivative couplings are neglected for the sake of simplicity in developing the modified RG framework, although later improvements in this line may be as important as in the case of the  anomalous dimension $\eta$ of  ordinary $O(N)$ models in the usual EAA framework \cite{Canet2003}.
The opposite signs of the quadratic and quartic derivative terms ensure that at any value of the gliding scale $k$ there may exist Fourier-modes of the field with a particular momentum $P_k=\sqrt{-Z_k/(2Y_k)} $  for which the kinetic energy density takes the negative minimal value $Z_k P_k^2+Y_kP_k^4=-Z_k^2/(4Y_k)<0$. 
 Then it may happen that for a configuration with periodic condensate of  momentum $P_k$ the value of the Euclidean action gets smaller 
 than that for any homogeneous field configurations. The purpose of the present paper is to develop a modified EAA RG framework that
enables one to study the scale-dependence of the periodic condensate. 
The basic idea is that the dynamical symmetry breaking caused by the appearing of the periodic condensate is mimicked by explicit breaking of translation symmetry of the EAA. Similar approach is widely used in ordinary $O(1)$ scalar models when the $Z_2$ symmetric double-well potential is approximated by its truncated Taylor-expansion around one of its mininima \cite{Tetradis1994,Berges2002}, and  has also been successfully used in the case of the ordinary
$O(N)$ scalar models, when spontaneous breaking of $O(N)$ symmetry has been partially mimicked by the inclusion of explicit symmetry breaking gradient terms into the EAA \cite{Peli2018}. In a phase with a periodic condensate in the vacuum the EAA should have its minimum at the corresponding periodic field configuration. Therefore we shall try to expand the EAA around  the periodic field configuration minimizing it. It seems to be a too involved task to determine the exact form of the field configuration minimizing the EAA. Therefore we shall look for it in a restricted subspace of field configurations, among those characterized by a particular momentum $P_k$ at the gliding scale $k$ and being periodic in the single spatial direction of the unit vector $e_{\mu}$ ($x_\pa=ex=\sum_{\mu=1}^d e_\mu x_\mu=x_1$),
\bea\label{chifouexp}
 \chi_x &=& \sum_{n=1}^{N_m} 2\sigma_{n~k} \cos (nP_k  x_\pa) .
\eea
Here the various Fourier-amplitudes $\sigma_{n~k}$ depend on the gliding scale $k$. Based on the truncation of the periodic background at the increasing number $N_m$ of its Fourier-modes an approximation scheme is put forward which we call Fourier-Wetterich RG approach. The characteristic momentum $P_k$ and the Fourier-amplitudes    $\sigma_{n~k}$ are determined by repeated minimization of the EAA at each values of the scale $k$. 
 The usual technique of GE should then be applied in the following manner: {\em (i)} the scalar field is split into the sum of the homogeneous background component $\Phi$, the periodic background $\chi_x$, and the fluctuating field $\eta_x$, $\phi_x=\phi_{Bx} + \eta_x$, $\phi_{Bx}=\Phi+ \chi_x$; {\em (ii)} the EAA as well as the expression under the trace on the right-hand side of the WE is Taylor-expanded in powers of $\eta_x$ including the terms up to the quadratic ones; and then
{\em (iii)} separate RG flow equations are read off for the potential and the wave function renormalization. We  neglect the field-dependence of the derivative couplings. As a consequence of item {\em (i)} the Fourier-amplitudes $\sigma_{n~k}$ acquire  dependencies  on the homogeneous background $\Phi$.  In order to get a closed set of equations, one assumes that the periodic condensate has a back-reaction on the various terms of the EAA. Therefore, the potential energy piece of the EAA is chosen in the form
\bea
 \int_x \sum_{n=0}^{N_m} U_{n~k}(\phi_x) \cos (nP_k x_\pa)
\eea
where $U_{0~k}(\phi)$ is the usual local potential, but additional potentials $ U_{n~k}(\phi) $ with $n\ge 1$ are included which  are supposed to be induced by the $n$-th Fourier-modes of the periodic background. Similarly, the gradient terms of the EAA
are chosen in the form
\bea
 \int_x \sum_{n=0}^{N_m} \hf \phi_x\lbrack \c{Z}_{n~k}(-\Box_x) \phi_x\rbrack \cos(nP  x_\pa),
\eea
where $\c{Z}_{0~k}(p^2)$ is the usual momentum-dependent 
 wave function renormalization, while $\c{Z}_{n~k}(p^2)$ with $n\ge 1$  are
 the induced momentum-dependent wave function renormalizations.
Then an approximation scheme is outlined in which the fundamental  mode with $n=1$ and
the higher harmonics (with $n>1$) of the condensate can be successively taken into account.
The higher harmonics of the fundamental momentum $P_k$ correspond condensate modes of smaller wavelength and are supposed to be gradually suppressed by the dynamics with decreasing  scale $k$. Therefore, one expects that even the approximation containing only the fundamental mode with momentum $P_k$ may provide physically reliable results for the low-energy behaviour of the model. Therefore, we restrict ourselves 
 to the one-mode approximation with $N_m=1$ when the homogeneous background $\Phi$ (the zero-mode) and  the fundamental periodic mode of the background are taken into account. As a further simplification we neglect the effect of the condensate induced on the derivative terms in our preliminary numerical work. We are completely aware of that our  present work is a first step
towards developing the proposed approach to handle phases with periodic condensates in the EAA framework and consider our numerical results as guidance for later improvements of the Fourier-Wetterich RG approach. The zero-mode approximation (with $N_m=0$), applicable to discuss phases without any periodic condensate represents just the usual EAA RG framework, which has  been successfully used to treat -- among others -- the ordinary $O(1)$ and $O(N)$ models. These are the cases when the minimum of the EAA occurs for some particular (vanishing or non-vanishing) value of the homogeneous background and all amplitudes $\sigma_{n~k}$ and all induced terms, i.e., $U_{n~k}$ and $\c{Z}_{n~k}$ with $n\ge 1$ vanish identically at any scale $k$. In particular, these are the symmetric phase and the phase with spontaneously broken $Z_2$ symmetry in the ordinary 3-dimensional, Euclidean $O(1)$ scalar model. The possibility of occurring a periodic condensate implies the existence of a  phase or phases in which translational symmetry along one of the spatial axis is spontaneously broken.  Such a condensate distinguishes one spatial axis but not any direction along that axis and preserves rotational symmetry along the axis. This should be taken into account in the ansatz for the EAA, especially in its derivative terms, by distinguishing the momenta
transverse $(p_\pe)$ and longitudinal $(p_\pa)$ with respect to the symmetry axis. It makes plausible to integrate in momentum space over a cylindrical region
of linear sizes $\sim k$ instead of a sphere of radius $\sim k$ and requires the matching of the regulator function with the cylindrical symmetry of the system. 

The EAA includes now extra derivative terms and potentials induced by the periodic condensate which  describe a kind of  back-reaction of the periodic condensate on the quantum fluctuating component of the scalar field. They do it in a self-consistent manner, because the induced  terms are functionals of the total field, the sum of the background and the fluctuating piece.   Whenever the Fourier-amplitudes  $\sigma_{n\ge1~ k} (\Phi)$ vanish at some scale $k_{\rm{fr}}$, the corresponding induced terms $V_{n~k}$ and $\c{Z}_{n~k}$ are  assumed to freeze out for $k<k_{\rm{fr}}$.

 The Introduction of the present paper is followed by the elaboration of the Fourier-Wetterich RG approach in Section \ref{FWRG}, then we present our preliminary numerical results for the ghost $O(1)$ model in Section \ref{numres}. The present status of the proposed approximation scheme and it possible improvements are discussed in Section \ref{summary}. Appendix \ref{fderiv} presents  the first few functional derivatives of the rEAA in the one-mode approximation. The consistent treatment of the various Fourier-modes on the left-hand side of the WE is outlined in Appendix \ref{lhswe} The closed form of the full propagator is derived in the one-mode approximation in Appendix \ref{a:fullprop}. Appendix \ref{traces} outlines the technique of evaluating the traces on the right-hand side of the WE  and summarizes its results. The  extrema of the rEAA are discussed in Appendix \ref{minimi} and finally the explicit forms of the flow equations obtained in LPA and in NLO of the GE are given in Appendices \ref{a:lpaflow} and \ref{a:nloflow}, respectively.

\section{Fourier-Wetterich renormalization group approach}\label{FWRG}

\subsection{Structure of the RG equation}\label{eearg}

Applying the EAA RG approach to the  one-component scalar field $\phi_x$   one splits the EAA ${\b{\Gamma}}_k[\phi]=\Gamma_k[\phi] +\Delta \Gamma_k[\phi]$ into the reduced EAA (rEAA) $\Gamma_k[\phi]$ and the regulator piece
\bea
\Delta \Gamma_k[\phi]&=& \frac{1}{2}\int_x \phi_x \c{R}_{k~x,y} \phi_y,
\eea
where 
\bea
   \c{R}_{k~x,y} &=&  R_k(-\Delta_\pe, -\partial_\pa^2)\delta(x-y)
\eea
with the choice of a field-independent infrared (IR) cutoff function $R_k(u_\pe, u_\pa)$. Here and below  the differential operators act always on the spatial coordinates $x_\mu~(\mu=1,2,3)$, with the parallel coordinate $x_\pa=x_1$ and the transverse vector $x_\pe=(x_2,x_3)$, $k$ denotes the running cutoff, $\partial_\pa=\partial_{x_\pa}$, and
$\Delta_\pe=\sum_{\mu=2}^3\partial_{x_\mu}^2$ is the Laplacian in the 2-dimensional plane perpendicular to the symmetry axis ($x_1$-axis).
The WE for the rEAA ${\Gamma}_k$ is given as
\bea\label{we} 
\dot{\Gamma}_k &=& \hf \Tr \biggl( \lbrack\Gamma^{(2)}_k + \c{R}_k  \rbrack^{-1}{\dot {\c{R}}}_k \biggr),
\eea  
where  the dot over the quantities indicates the scale-derivative $k\partial_k$, $\Gamma^{(2)}_k $ is a shorthand for the second functional derivative matrix
$\Gamma_{k,x,y}^{(2)}= \fdd{\Gamma_k[\phi]}{\phi_x }{\phi_y}$. The trace is taken over a complete set of field configurations.
Now we  make an ansatz for the rEAA in the NNLO of the GE. Application of the usual GE techniques involves the split of the field $\phi_x=\phi_{Bx} + \eta_x$ into  the  background piece $\phi_{Bx}$ and the inhomogeneous fluctuating field $\eta_x$. The background field consists of the homogeneous background $\Phi$ and the periodic background $\chi_x$ representing the periodic condensate (see Eq. \eq{chifouexp}), $\phi_{Bx}=\Phi+\chi_x$. 
 For the rEAA we make the general NNLO ansatz
\bea\label{eaaon}
  \Gamma_k[\phi] &=& \hf \int_{x,y}\sum_{n=0}^{N_m} \phi_{ x} D^{-1}_{n~ x, y} \phi_{ y}  \cos (nP_k x_1)
 + \int_x\sum_{n=0}^{N_m} U_{n~k}(\phi_x) \cos (nP_k x_1)\nn
\eea
where  for each of  the $n$-th  modes the kernels
\bea\label{invpro}
  D^{-1}_{n~ x, y}(-\Delta_\pe,-\partial^2_\pa) &=& 
\c{Z}_{n~k}(-\Delta_\pe,-\partial_\pa^2) \delta_{x,y}
\eea
have been introduced with the momentum-dependent wave function renormalizations
$\c{Z}_{n~k}(-\Delta_\pe,-\partial_\pa^2)$. The Fourier-amplitudes $\sigma_{n~k}(\Phi)$ and the momentum $P_k$ are those
  minimizing the EAA at any given scale $k$ for a given homogeneous background $\Phi$.

The one-mode approximation corresponds to the truncation $N_m=1$, when Eq. \eq{chifouexp} reduces to
\bea\label{chi1mode}
  \chi_x&=& 2\sigma_k(\Phi) \cos( P_k x_1)
\eea
 and the notation can be reduced to
$\c{Z}_{0~k}=\c{Z}_k$, $\c{Z}_{1~k}=\c{E}_k$, $U_{0~k}=U_k$, $U_{1~k}=V_k$, and $\sigma_{1~k}=\sigma_k$. In the general formulas for the flow equations we keep the terms containing $\c{E}_k$, but our preliminary numerical results are obtained by setting $\c{E}_k\equiv 0$ due to the complexity of the set of flow equations otherwise.   It is an additional approximation that
 the potentials and the amplitude of the periodic condensate are chosen in the form of truncated Taylor expansions around the vanishing field configuration $\phi=0$,
\bea\label{potsitrun}
 && U_{k}(\phi) = \sum_{n=2}^M \frac{g_{n}(k)}{n!}\phi^n,~~~~
 V_k(\phi)=\sum_{n=3}^{M-1} \frac{v_{n}(k)}{n!}\phi^n,~~~~
 \sigma(\phi) = \sum_{n=0}^{M-2}  \frac{s_{n}(k)}{n!}\phi^n,
 \nn
\eea
respectively. Below we show that
 {\em (i)}
 the lowest-order derivatives of the potentials appearing in the 
propagator are $U_k''(\Phi)$ and $V_k'''(\Phi)$ (the prime denoted differentiation with respect to the variable $\Phi$), dictating the choice of the first non-vanishing terms of the Taylor series; 
{\em (ii)}  for  following the flow of the ordinary potential $U_k$ with the accuracy up to the term  $\ord{\phi^M}$  the induced potential and the amplitude $\sigma_k(\phi)$ should be given with the accuracy up to the term $\ord{\phi^{M-1}}$
and  $\ord{\phi^{M-2}}$, respectively.
We shall use the truncation $M=4$ throughout this paper.
In the one-mode approximation we shall choose the momentum-dependence of the  wave function renormalization as
\bea\label{Zansatz}
 \c{Z}_k(-\Delta_\pe,-\partial_\pa^2) &=&
 -Z_{\pe ~k} \Delta_\pe - Z_{\pa ~k}\partial_\pa^2
 + Y_{\pe~k}   \Delta_\pe^2 + 2Y_{X~k} \Delta_\pe\partial_\pa^2
 + Y_{\pa~k}\partial_\pa^4
 \nn
\eea 
and make a similar assumption for the momentum-dependence of $\c{E}_k$.
The ansatz \eq{Zansatz} takes into account the axial symmetry induced by the periodic condensate, but  respects the reflection symmetry $x_1 \leftrightarrow -x_1$.
At the UV scale we choose the initial conditions with $Z_{\pe \Lambda}=Z_{\pa\Lambda}=-1$, $Y_{\pe~\Lambda}=Y_{\pa~\Lambda}=Y_{X~\Lambda}>0$ (that ensures boundedness of the bare Euclidean action from below) and $\c{E}_\Lambda=0$.

Truncating the  functional Taylor-expansion at the quadratic term of the order $\ord{\eta^2}$,  the left-hand side of the WE  takes the form 
\bea\label{dotga} 
{\dot{\Gamma}}_k[\phi_B+\eta] &=&{\dot{\Gamma}}_k[\phi_B]
+ \int_x {\dot F}_{k~x} \eta_x
+\hf \int_{x,y} \eta_x {\dot A}_{k~x,y} \eta_y,
\eea
while the expansion of the second functional derivative of the EAA can be written as
\bea\label{ga2exp}
\Gamma_{k~x,y}^{(2)}[\phi_B+\eta] &=& A_{k~x,y} + (\eta B)_{k~x,y}+ \hf (\eta C\eta)_{k~ x,y},
\eea
where
\bea
\label{F}
 F_{k~x}&=& \fd{\Gamma_k}{\phi_x} \Biggr|_{\phi=\phi_B}\\
\label{A}
  A_{k~x,y}&=& \Gamma_{k,x,y}^{(2)}[\phi_B],\\
\label{B}
 (\eta B)_{k~x,y}&=& \int_z  \eta_z \fddd{\Gamma_k}{\phi_x}{\phi_y}{\phi_z}\biggr|_{\phi=\phi_B},\\
\label{C}
(\eta C\eta)_{k~ x,y}&=&  \int_{z,u} \eta_z \fdddd{\Gamma_k}{\phi_x}{\phi_y}{\phi_z}{\phi_u}\biggr|_{\phi=\phi_B}  \eta_u.
\eea
The terms linear in $\eta$ on the left-hand side of the WE should vanish, i.e., $F_k=0$, because the amplitude $\sigma_k(\Phi)$ and the characteristic  momentum $P_k$ of the periodic background $\chi_x$ are determined via minimizing the rEAA $\Gamma_k[\phi_B]$ for given scale $k$ and homogeneous background $\Phi$.
The explicit  expressions of the first few functional derivatives of the rEAA are derived in Appendix \ref{fderiv} in the one-mode approximation.
The left-hand side of the WE expanded in powers of the fluctuating field $\eta_x$ at the background configuration $\phi_{Bx}$ is evaluated in Appendix \ref{lhswe} keeping the terms up to the ones quadratic in the fluctuating field $\eta_x$.
 The latter is a technical device to separate the flow of the potentials and that of the gradient terms in the EAA. It is assumed that $\eta_x$ does not contain Fourier-modes $\eta_Q=\int_x e^{iQx} \eta_x$ with zero momentum and with the longitudinal momenta $P$ and its upper harmonics, implying that
the terms linear in $\eta_Q$  vanish exactly on both  sides of the WE.   Below we shall see that the non-vanishing of  the induced derivative piece with $\c{E}_k(-\Delta_\pe, -\partial_\pa^2)$ would violate $Z_2$ symmetry of the rEAA. Therefore, we assume throughout this paper that  $\c{E}_k(-\Delta_\pe, -\partial_\pa^2)=0$.

The explicit expressions for the first few terms of the Taylor-expansion 
of the left-hand side of the WE are derived in Appendix \ref{lhswe} in the one-mode approximation. The lengthy but straightforward analytic manipulations yield for the left-hand side of the WE
\bea\label{gammadot}
 {\dot \Gamma}_k[\phi_B+\eta] &=& {\dot \Gamma}_k[\phi_B]
+ \gamma_k^{(1)} (\eta_{P_ke}+\eta_{-P_ke})
+   \int_Q \gamma_k^{(2)0}(Q_\pe^2,Q_\pa^2)\eta_Q\eta_{-Q}
+ \sum_{\tau=\pm}\int_{Q,Q'} \gamma_k^{(2)1}(Q_\pe^2,Q_\pa^2) \delta_{Q+Q'+\tau P_ke,0}\eta_Q\eta_{Q'}
\nn
\eea
where the rEAA at the background configuration $\phi_{Bx}$ is given as
\bea\label{gamfib}
 \Gamma_k[\phi_B]&=&V\biggl\lbrack 
\sigma_k^2 \c{Z}_k(0,P_k^2) + \hf \Phi \sigma \c{E}_k(0,P_k^2)
+ U_k(\Phi) +\sigma_k V_k'(\Phi)+ \sigma_k^2 U_k''(\Phi) + \hf \sigma_k^3 V_k'''(\Phi)
+ \frac{1}{4}\sigma_k^4 U_k'''' (\Phi)\biggr\rbrack,
\eea
and the integral kernels are
\bea\label{gam1}
\gamma_k^{(1)}&=& 
\sigma_k{\dot{ \c{Z}}}_k(0,P_k^2)
+ \frac{1}{4} \Phi {\dot{\c{E}}}_k(0,P_k^2)
+ \hf{\dot V}_k'(\Phi)
+ \sigma_k {\dot U}_k''(\Phi)
+\frac{3}{4} \sigma_k^2 {\dot V}_k'''(\Phi)
+\hf\sigma_k^3 {\dot U}_k''''(\Phi),
\eea
\bea\label{gam20}
\gamma_k^{(2)0}(Q_\pe^2,Q_\pa^2)&=&
\hf {\dot{\c{Z}}}_k(Q_\pe^2,Q_\pa^2)+ \hf {\dot U}_k''(\Phi)
+\hf \sigma {\dot V}_k'''(\Phi) +\hf \sigma^2 {\dot U}_k''''(\Phi),
\nn
\eea
and
\bea\label{gam21}
\gamma_k^{(2)1}(Q_\pe^2,Q_\pa^2)&=&
\frac{1}{4} {\dot{\c{E}}}_k(Q_\pe^2,Q_\pa^2)+
 \frac{1}{4} {\dot V}_k''(\Phi)
+\hf \sigma_k {\dot U}_k'''(\Phi)
+\frac{3}{8} \sigma_k^2 {\dot V}_k'''(\Phi).
\nn
\eea
The scale-derivative $ {\dot \Gamma}_k[\phi_B] $ of the rEAA  at the given configuration $\phi_B$ should be evaluated by taking the scale-derivatives of the couplings in the expression  \eq{gamfib} at  given $\sigma_k(\Phi)$ and momentum $P_k$ since the latter are parameters in $\phi_B$.

 It is worthwhile mentioning how $Z_2$ symmetry of the rEAA can be realized. The ordinary $Z_2$ symmetric  potential is an even function, $U_k(\Phi)=U_k(-\Phi)$. Then the symmetry of the rEAA requires $\sigma_k(\Phi)$ to be even as well (see the ordinary gradient piece in the right-hand side of Eq. \eq{gamfib}.) The terms containing the induced potential in Eq. \eq{gamfib} are symmetric if and only if $V_k(\Phi)$ is an odd function, $V_k(\Phi)=-V_k(-\Phi)$.
 Furthermore the induced gradient term would only be symmetric if $\c{E}_k(0,P_k^2)$ would change its sign under the transformation $\Phi\to -\Phi$, but for background-independent induced derivative piece this implies   $\c{E}_k(0,P_k^2)=0$ as the only possibility.
Therefore, we shall assume that the induced gradient piece in the one-mode approximation  vanishes, although formally we shall write down how the RG flow equation could be derived for it.

The trace on the right-hand side of the WE \eq{we} can  be split in terms of increasing order of $\eta_x$,
\bea\label{neuexp}
 \Tr \biggl( \lbrack\Gamma^{(2)} + \c{R}_k  \rbrack^{-1}{\dot {\c{R}}}_k \biggr)
&=& T_0 + T_1 +T_{2B}+T_{2C},
\eea
with 
\bea\label{Ts0}
 T_0 &=&  \Tr[\c{G}\cdot {\dot {\c{R}}}_k],\\
\label{Ts1}
 T_1 &=& -\Tr[\c{G} \cdot (\eta B)\cdot \c{G}\cdot {\dot {\c{R}}}_k],\\
\label{Ts2B}
T_{2B} &=& \Tr[\c{G} \cdot (\eta B)\cdot \c{G} \cdot (\eta B)\cdot \c{G} \cdot
{\dot{\c{ R}}}_k],\\
\label{Ts2C}
T_{2C} &=& -\hf\Tr[\c{G} \cdot (\eta C \eta)\cdot \c{G} \cdot {\dot{\c{ R}}}_k].
\eea
Here $\c{G}$ stands for the IR regulated full propagator discussed below in Section \ref{fullpropag}. 
The explicit expressions for the matrices $B$ and $C$ are obtained by taking the Fourier-transforms of
the third and fourth functional derivatives given in Eqs. \eq{fdgam3} and \eq{fdgam4} at the field configuration $\phi_B$. In the flow equations for the derivatives terms we need the values of the matrices $B$ and $C$ only at $\Phi=0$, since we restricted ourselves to field-independent derivative couplings,
\bea\label{bpqr}
&&  B_{p,q,r}|_{\Phi=0}= b_0 \delta_{p+q+r,0} + b_1 \sum_{\tau=\pm 1} \delta_{p+q+r+\tau Pe,0},\\
\label{cpqrs}
&& C_{p,q,r,s}|_{\Phi=0}= c_0 \delta_{p+q+r+s,0} + c_1 \sum_{\tau=\pm 1} \delta_{p+q+r+s+\tau Pe,0},
\eea
with 
\bea\label{bcdef}
  && b_0=U_k'''(0) + \sigma_k(0) V_k''''(0),~~~~b_1=\hf V_k'''(0) +\sigma_k(0) U_k''''(0),\nn
&&~~~~~~~~~~~~~~~~~~~c_0=U_k''''(0), ~~~~c_1=\hf V_k''''(0)
\eea
which reduce to
\bea\label{bcsym}
 && b_0=0,~~~~b_1=\hf v_3 + \sigma(0) g_4,
~~~~c_0=g_4,~~~~c_1=0
\eea
when the RG flow keeps $Z_2$ symmetry of the rEAA (see the discussion below Eq. 
\eq{gam21}).
The explicit expressions of the traces are derived in Appendix \ref{traces}.

As discussed below in  \ref{regul}, the regulator is chosen in accordance with the cylindrical symmetry of the system in such a manner that the RG transformation integrates out for each infinitesimal change of the scale $k$ to $k-d k$ essentially the modes of $\eta_Q$ with
momenta $Q$ in the cylindrical momentum shell of thickness $dk$ at $Q_\pe =|\vec {Q}_\pe|= k/\sqrt{2} $ and $Q_\pa=\pm k/\sqrt{2}$.
Then the traces involve the loop-integrals  of the type
\bea\label{momintegral}
  \int_p f(p_\pe^2, p_\pa) &=&  \alpha_3\int_{-k/\sqrt{2}}^{k/\sqrt{2}} dp_\pa \int_0^{k/\sqrt{2}} dp_\pe p_\pe  f(p_\pe^2, p_\pa)
\eea
in cylindrical momentum coordinates with $\alpha_3=(2\pi)(2\pi)^{-3}=(4\pi^2)^{-1}$.

\subsection{Full propagator}\label{fullpropag}

As shown in Appendix \ref{fderiv} the full propagator $\c{G}_{p,q}$ is given via the relation
\bea\label{fullpro1}
  \c{G}^{-1}_{p,q} &=& \Gamma_{k~p,q}^{(2)}|_{\phi_B}
= G^{-1}(p_\pe^2, p_\pa^2,\Phi) \delta_{p+q,0} +\sum_{n=1}^{N_m} \mf{V}_{n~p,q}
\eea
where the reduced inverse propagator is given by
\bea\label{rpro}
  G^{-1}(p_\pe^2, p_\pa^2,\Phi) &=& \c{Z}_k(p_\pe^2,p_\pa^2)
 + U''_k(\Phi) + \sigma V'''(\Phi) + \sigma^2 U''''(\Phi)
\eea
and $\mf{V}_{n}$ represent interaction vertexes induced by the periodic condensate. In the one-mode approximation there is only a single induced vertex, $\mf{V}_1=\mf{V}$,
\bea\label{indverm}
 \mf{V}_{p,q}&=& \mf{V}(p_\pe^2,p_\pa^2,q_\pe^2,q_\pa^2,\Phi)\sum_{\tau=\pm 1}
\delta_{p+q+\tau P e,0}
\eea
with
\bea\label{indver}
\mf{V}(p_\pe^2,p_\pa^2,q_\pe^2,q_\pa^2,\Phi)&=&
\hf \biggl( \hf \c{E}_k(p_\pe^2, p_\pa^2) +  \hf \c{E}_k(q_\pe^2, q_\pa^2)
+ V_k''(\Phi) + 2\sigma U_k'''(\Phi)+ \frac{3}{2}\sigma^2 V_k''''(\Phi)
\biggr)
\eea
 Fortunately, in the one-mode approximation the full propagator can be obtained by inversion of $\c{G}^{-1}_{p,q}$ in a closed form by resumming the Neumann-series (see Appendix \ref{a:fullprop}),
\bea\label{fullpro2}
\c{G}_{p,q}
&=&
\c{G}(p_\pe^2,p_\pa,\Phi)\biggl\lbrack 
  \delta_{p+q,0} - G(q_\pe^2,q_\pa^2,\Phi)  {\mf{V}}(p_\pe^2,p_\pa^2,q_\pe^2,q_\pa^2,\Phi)\sum_{\tau=\pm}\delta_{p+q+\tau Pe,0} \biggr\rbrack
\eea
with 
\bea\label{resumg}
\c{G}(p_\pe^2,p_\pa,\Phi)&=& \biggl( 1- G(p_\pe^2,p_\pa^2,\Phi) \sum_{\tau=\pm}
G(p_\pe^2,(p_\pa+\tau P)^2,\Phi)
\times \mf{V}^2(p_\pe^2,p_\pa^2,p_\pe^2,(p_\pa+\tau P)^2,\Phi) \biggr)^{-1} G(p_\pe^2,p_\pa^2,\Phi) .
\eea
The right-hand side of the WE is expressed in terms of the  IR regulated full propagator
$\c{G}_{{\rm{reg}} ~p,q} = ( \Gamma_k^{(2)}[\phi_B])+\c{R} )^{-1}_{p,q}$. The comparison with Eq. \eq{fullpro1} shows that one has to regulate the reduced propagator
$G(p_\pe^2, p_\pa^2,\Phi)$ by adding the corresponding regulator function to its inverse, i.e.,  perform the replacement of  $G(p_\pe^2,p_\pa^2,\Phi)$ given in Eq. \eq{rpro} via
\bea\label{redpro}
G_{\rm{reg}}(p^2_\pe,p_\pa^2,\Phi) &=& 
\biggl(\c{Z}(p_\pe^2,p_\pa^2) +  U''(\Phi) + \sigma^2 U''''(\Phi)
+ \sigma V'''(\Phi)+ R_k(p_\pe^2,p_\pa^2) \biggr)^{-1}.
\eea
For the sake of simplicity, below we shall not indicate the regulation of the propagators in the notation explicitly, so that $\c{G}$ and $G$ shall stand for the regulated expressions.

The full propagator given  by Eqs. \eq{fullpro2} and \eq{resumg} shows the form of a kind of resummed Dyson series. The non-vanishing amplitude of the periodic condensate and the  potential $V_k(\Phi)$ induced by the periodic condensate generate  the induced  vertex $\mf{V}(p_\pe^2, p_\pa^2, p_\pe^2, (p_\pa+\tau P_k)^2, \Phi)$ which describes the interaction of the propagating virtual scalar particle with the periodic condensate. Via this induced interaction the particle acquires the elementary shift $\tau P$ of its longitudinal momentum. Repeated induced interactions may produce longitudinal momentum shifts being  any integer times the elementary  shift. In the one-mode approximation, however, we keep in the propagator only contributions when in total either no shift or an elementary shift is produced in the transition from the state with momentum $p$ to the state with momentum $q$ (see the Dirac-deltas
in the expression of the matrix of the full propagator in Eq. \eq{fullpro2}).
Furthermore, the sub-processes contributing to the transition amplitude $\c{G}(p_\pe^2, p_\pa,\Phi)$ given by Eq. \eq{resumg} are those without a net longitudinal momentum shift.  In the one-mode approximation these are the following sub-processes  
 {\em (i)}  the particle propagates without suffering any longitudinal momentum shift, the transition amplitude of such a propagation is given by the reduced propagator $ G(p_\pe^2,p_\pa^2,\Phi)$;
{\em (ii)} the particle suffers two subsequent interactions with the periodic condensate described by the induced vertex $\mf{V}$ one producing  the longitudinal momentum shift $\pm P_k$
and another producing the shift $\mp P_k$, the corresponding transition amplitude is
\bea
&&\sum_{\tau=\pm 1} G(p_\pa^2)\mf{V}(p_\pa^2,(p_\pa+\tau P_k)^2 )G((p_\pa+\tau P_k)^2 )
\mf{V}((p_\pa+\tau P_k)^2,p_\pa^2 ) G(p_\pa^2).
\eea
The sub-processes of propagation of type {\em (ii)} repeated by several times,
having the amplitude (for $n\ge 2$)
\bea
&& \biggl( \sum_{\tau=\pm 1} G(p_\pa^2)\mf{V}(p_\pa^2,(p_\pa+\tau P_k)^2 )G((p_\pa+\tau P_k)^2 
   \mf{V}((p_\pa+\tau P_k)^2 ) G(p_\pa^2) \biggr)^n
\eea
give additional contributions to the transition amplitude $\c{G}(p_\pe^2,p_\pa)$.
(Here in the notation we suppressed the dependences of the propagators and the vertices  on the transverse momenta and the
homogeneous background.) The longitudinal momentum shifts arise due to the induced interaction of the quantum fluctuating piece of the scalar field with the periodic condensate. It is worthwhile mentioning that the summation over $\tau=\pm 1$ in the building blocks of the Dyson series ensures that the sum
\eq{resumg} is an even function of $p_\pa$. Consequently the integrands of the loop integrals are even functions of the longitudinal loop momentum on the right-hand side of the WE.

For a $Z_2$ symmetric ordinary potential the reduced propagator
$ G(p_\pe^2,p_\pa^2,\Phi)$ and the full propagator $\c{G} (p_\pe^2,p_\pa,\Phi)$
are invariant under the  $Z_2$ transformation $\Phi \to -\Phi$, while the induced vertex $\mf{V}(p_\pe^2,p_\pa^2, q_\pe^2, q_\pa^2,\Phi)$ changes sign. In the case when $\c{E}_k(p_\pe^2, p_\pa^2)=0$, the vertex becomes independent of the momenta, $\mf{V}(\Phi)$.

\subsection{Litim's optimized regulator for cylindrical geometry}\label{regul}

A generalization of Litim's optimized regulator is used, which is reconciled with the cylindrical symmetry of the periodic condensate. Litim's regulator \cite{Litim2001A,Litim2001B,Litim2002} acts only below the gliding momentum scale $k$ by compensating the momentum-dependence of the inverse propagator. For ordinary fields (i.e., those with $Z_k>0$) it should satisfy the criterion of non-negativity.
 In the ordinary Euclidean scalar $O(1)$  model with spherical symmetry Litim's regulator is given as
\bea\label{litimr}
  R_k(p^2)&=& \biggl( Z_k (k^2-p^2) + Y_k(k^4-(p^2)^2)\biggr) \Theta(k^2-p^2)
\eea
after a proper generalization via the inclusion of a quartic term \cite{Litim2011}. The Heaviside function $\Theta(k^2-p^2)$ ensures that no cutoff occurs for large momenta outside of the momentum sphere of radius $k$. The  term $k^2$ is the maximum value of $p^2$ for momenta restricted to the momentum sphere of radius $k$, so that  the expressions
multiplying the derivative couplings are all non-negative. Now we generalize
the cutoff function reconciling it with the cylindrical geometry and non-negativity (for positive $Z_k$ and $Y_k$), as follows.
 Let the inequalities $0\le p_\pe^2\le k^2/2$ and $-k/\sqrt{2}\le  p_\pa\le k/\sqrt{2}$ define the cylindrical region of momentum space over which the loop-integrals are taken (see Eq. \eq{momintegral}). Outside of this region there should be no cutoff, so that the
Heaviside function $ \Theta(k^2-p^2)$ of the spherical case should be replaced by $\Theta (\hf k^2-p_\pe^2)\Theta(\hf k^2-p_\pa^2)$, and the regulator chosen as
\bea\label{litreg}
R_k(p_\pe^2,p_\pa^2) &=& 
\biggl\lbrack Z_{\pe ~k}\biggl(\hf k^2-p_\pe^2\biggr)+ Z_{\pa~k}\biggl(\hf k^2-p_\pa^2\biggr) 
+ Y_{\pe~k} \biggl(\frac{1}{4}k^4 -p_\pe^4\biggr)+ 2Y_{X~k}\biggl(\frac{1}{4}k^4-p_\pe^2p_\pa^2 \biggr) 
+ Y_{\pa~k} \biggl( \frac{1}{4}k^4 -p_\pa^4 \biggr)\biggr\rbrack 
\nn
&& \times \Theta(\hf k^2-p_\pe^2)\Theta(\hf k^2-p_\pa^2).
\eea
Here the various terms are chosen in accordance with choice \eq{Zansatz} of the momentum-dependent wave function renormalization, so that the regulated reduced inverse propagator becomes independent of the momentum $p_\mu$.
 The choice of the boundaries of the cylindrical region results in the constant
 factors ahead of $k^2$ and $k^4$. These factors 
 ensure that the expression 
 $\lbrack Z_k(k^2-p^2)+Y_k(k^4-p^4)\rbrack$ of the spherical case is recovered for $Z_{\pe k}=Z_{\pa k}=Z_k$ and $Y_{\pe k}=Y_{X k}=Y_{\pa k}=Y_k$.

Evaluating the loop-integrals on the right-hand side of the WE, one also needs 
the reduced propagator with shifted longitudinal momenta $p_\pa\pm P_k$.
 In the case of the one-mode approximation we show how  the  Litim-type regulator $R_k^{[\pm ]}\bigl(p_\pe^2,(p_\pa\pm P_k)^2\bigr)$ for shifted momenta $p_\pa\pm P_k$ can be  chosen satisfying the conditions the regulator function should fulfil. In the case with cylindrical symmetry and with the boundaries of the momentum cylinder of scale $k$ described above, the maximal  values of $p_\pe^2$ and $p_\pa^2$ are identical and equal to $\hf k^2$, so that the regulator given in Eq. \eq{litreg} for unshifted longitudinal momentum satisfies the above mentioned criteria. Now for shifted longitudinal momentum the maximal value of $(p_\pa\pm P_k)^2$ is$\biggl( \frac{k}{\sqrt{2}}+P_k\biggr)^2$ in the momentum cylinder considered. Therefore we choose the regulator for shifted longitudinal momenta $p_\pa \pm P_k$ as
\bea\label{regsh}
R_k^{[\pm    ]}\bigl(p_\pe^2,(p_\pa\pm P_k)^2\bigr) &=&
\biggl\{ Z_{\pe~ k}\biggl(\hf k^2-p_\pe^2\biggr)
+ Z_{\pa~k}\biggl\lbrack \biggl( \frac{k}{\sqrt{2}}+P_k\biggr)^2 -(p_\pa\pm P_k)^2\biggr\rbrack 
\nn
&&
~~~+Y_{\pe~k}  \biggl(\frac{1}{4}k^4 -p_\pe^4\biggr) 
+2Y_{X~k}\biggl\lbrack  \frac{k^2}{2} \biggl(\frac{k}{\sqrt{2}}+P_k\biggr)^2-p_\pe^2(p_\pa\pm P_k)^2 \biggr\rbrack 
\nn
&&
~~~+Y_{\pa~k} \biggl\lbrack\biggl( \frac{k}{\sqrt{2}}+P_k\biggr)^4
-(p_\pa\pm P_k)^4 \biggr\rbrack \biggr\}
\Theta(\hf k^2-p_\pe^2)\Theta(\hf k^2-p_\pa^2).
\eea   
Then for positive derivative couplings the non-negativity criterion is automatically satisfied. Vanishing outside of the momentum cylinder is ensured by the Heaviside functions again, and vanishing in the limit $k\to 0$ is also fulfilled independently of the value of $P_k$, because all momenta $p_\pe$ and $p_\pa$ go to zero and $\biggl( \frac{k}{\sqrt{2}}+P_k\biggr)^{2l}-(p_\pa\pm P_k)^{2l} \to P_k^{2l}-P_k^{2l}=0$ for $l=1,2$.
Then one finds the following regulated expressions of the reduced  propagators,
\bea\label{greg}
G_{\rm{reg}}(p_\pe^2,p_\pa^2) &=& \biggl\lbrack (Z_{\pe~k} + Z_{\pa~k}) \hf k^2
+ (Y_{\pe~k}+2Y_{X~k}+ Y_{\pa~k})\frac{1}{4}k^4
\nn
&&
~~~+ U_k''(\Phi) +\sigma_k(\Phi) V_k'''(\Phi) +\hf \sigma_k^2(\Phi) U_k''''(\Phi)\biggr\rbrack^{-1},
\\
\label{gregsh}
G_{\rm{reg}}\bigl(p_\pe^2, (p_\pa\pm P_k)^2\bigr) &=& \biggl\lbrack
Z_{\pe~k} \hf k^2 + Z_{\pa~k}\biggl(  \frac{k}{\sqrt{2}}+P_k\biggr)^2
+ Y_{\pe~k} \frac{1}{4}k^4
+ 2Y_{X~k} \frac{k^2}{2}\biggl(  \frac{k}{\sqrt{2}}+P_k\biggr)^2
+ Y_{\pa~k} \biggl(  \frac{k}{\sqrt{2}}+P_k\biggr)^4 
\nn
&&
~~~+ U_k''(\Phi) +\sigma_k(\Phi) V_k'''(\Phi) +\hf \sigma_k^2(\Phi) U_k''''(\Phi)\biggr\rbrack^{-1}.
\eea
The propagators $G$ with  longitudinal momenta $p_\pa$ and $p_\pa\pm P_k$ shifted additionally by the momentum $Q_\pa$ of $\eta_Q=\eta_{Q_\pa,Q_\pe}$ in the traces \eq{Ts2B}, \eq{Ts2C} are regulated also by $R_k(p_\pe^2, p_\pa^2)$ and $R_k^{[\pm]}\bigl(p_\pe^2, (p_\pa\pm P_k)^2\bigr)$, respectively.

It is the proper place to make one more  remark in connection with the use of the regulator function. Namely, in the ghost scalar models we have $Z_k<0$, but even then we shall use the regulators in the above given forms not caring about the loss of its nonnegativity property. This is the way it has been successfully  used in similar situations, like in Quantum Einstein Gravity \cite{Fischer2006,Litim2012}.

The traces on the right-hand side of the WE contain the scale derivative of the
cutoff function ${\dot R}_k(p_\pe^2,p_\pa^2)$ given by
\bea\label{dotr}
{\dot R}_k(p_\pe^2,p_\pa^2)&=& (Z_{\pe~k}+Z_{\pa ~k})k^2 + (Y_{\pe~k}+2Y_{X~k}
+ Y_{\pa~k}) k^4 + \rho_k(p_\pe^2,p_\pa^2),\nn
\eea
where the piece
\bea\label{rhok}
  \rho_k(p_\pe^2,p_\pa^2)&=& {\dot Z}_{\pe~k}\biggl( \hf k^2- p_\pe^2\biggr)
 + {\dot Z}_{\pa~k}\biggl( \hf k^2- p_\pa^2\biggr)
+ {\dot Y}_{\pe~k} \biggl( \frac{1}{4} k^4-p_\pe^4\biggr)
+ 2{\dot Y}_{X~k} \biggl(  \frac{1}{4} k^4-p_\pe^2 p_\pa^2\biggr)
+ {\dot Y}_{\pa~k}  \biggl( \frac{1}{4} k^4-p_\pa^4\biggr)\nn
\eea
depends on the scale derivatives of the derivative couplings. In the LPA the latter vanishes and in higher orders of the gradient expansion it is generally negligible for ordinary scalar fields. It is, however, far from clear whether the piece $\rho_k(p_\pe^2,p_\pa^2)$ does or does not affect the phase structure and the values of the critical exponents around  the fixed points of the ghost model. 
This should be decided by numerical studies.

We performed our preliminary numerical study using, on the one hand, Litim's regulator, which removes the dependence of the reduced propagators on the loop momentum. There remains yet the dependence of the full propagator \eq{resumg} coming into play via the induced vertex. On the other hand, using an ansatz for the EAA which does not violate $Z_2$ symmetry explicitly (Taylor-expansions in $\Phi$ at $\Phi=0$ and the choice of the vanishing induced wave function renormalization
$\c{E}_k(p_\pe^2, p_\pa^2)=0$), the induced vertex $\mf{V}(\Phi)$ became independent of the loop-momenta. Therefore, any approximation scheme, keeping  $Z_2$ symmetry of the EAA, has on the one hand the drawback of loosing an important piece of dynamics, a large part of the effects caused by the longitudinal momentum shifts falls off. Only the presence of the shifted regulated reduced propagator
\eq{gregsh} in the expression of the full propagator  \eq{resumg} carries information on the shift of the longitudinal momentum. The role of the presence of the periodic condensate occurs basically via the dependence of the reduced propagator  and that of the induced vertex on the amplitude $\sigma_k(\Phi)$. The trace  \eq{Ts0} in the flow equation for the ordinary and the induced potentials may be
 influenced by the presence of the condensate in both ways. The trace in the flow equation for the momentum-dependent wave function renormalization should, however, be taken at the point $\Phi=0$, because we use field-independent wave function renormalizations at Taylor-expansions at $\Phi=0$. Then the induced vertex vanishes, $\mf{V}(\Phi=0)=0$, the full propagator reduces to the reduced one, so that the presence of the periodic condensate may influence the trace \eq{Ts2B} only via the factor $b_1(\Phi=0)=v_3/2$, i.e., via the induced potential rather indirectly.

\subsection{Parameters of the periodic condensate}\label{params}

Let us turn now to the determination of the parameters  $P_k$ and $\sigma_k(\Phi)$ of the periodic condensate \eq{chi1mode} in the one-mode approximation.
The rEAA $ \Gamma_k[\phi_B]$ given by Eq. \eq{gamfib} at any given scale $k$ and given homogeneous background $\Phi$ should be considered as the function of and minimized with respect to
the parameters $P$ and $\sigma$. The necessary conditions of the extrema are
\bea\label{pargaparp}
 \partial \Gamma_k[\phi_B] /\partial P^2&=&0,\\
\label{pargaparsin}
\partial \Gamma_k/\partial \sigma&=&0.
\eea
 Since we have assumed that the induced gradient term containing \\ $\c{E}_k(-\Delta_\pe, -\partial_\pa^2)$ vanishes due to avoiding explicit violation of $Z_2$ symmetry, 
 the  condition \eq{pargaparp} reduces to 
\bea
\partial \c{Z}_k(0,P^2) /\partial {P^2}&=&0
\eea
with the solution for the square of the characteristic momentum
\bea\label{charmom}
  P_{k}^2&=& - \frac{Z_{\pa k}}{ 2Y_{\pa k}}
\eea
which is positive so far $Z_{\pa k}<0$ and $Y_{\pa k}>0$, i.e., one treats a ghost scalar field, and independent of the homogeneous background $\Phi$. One can see that in LPA the characteristic momentum remains scale-independent, while in NLO and NNLO  of the GE it varies with the scale $k$. The existence of the characteristic momentum \eq{charmom}  
 results in the negative contribution
\bea
  V\sigma^2_k(\Phi) \c{Z}_k(0,P_k^2)&=& -V\sigma_k^2(\Phi)\frac{ Z_{\pa k}^2}{4Y_{\pa k} }<0  
\eea
of the   derivative term to the rEAA.

The condition \eq{pargaparsin} yields the cubic equation 
\bea\label{cubiceqsig}
 0&=& V_k'(\Phi) +2\sigma_k(\Phi) \biggl( U_k''(\Phi) - \frac{ Z_{\pa k}^2}{4Y_{\pa k} }
\biggr) +\frac{3}{2} \sigma_k^2(\Phi) V_k'''(\Phi) + \sigma_k^3 (\Phi)U_k''''(\Phi)\nn
\eea 
for the Fourier amplitude $\sigma$. Using the truncation $M=4$ (see the Taylor-expansions in Eq. \eq{potsitrun}) Eq. \eq{cubiceqsig} determines the estimate 
\bea\label{sigest}
\sigma_k(\Phi)&=&\sigma_{0~k}+\hf \sigma_{2~k}\Phi^2
\eea
 as a polynomial of the order $M-2=2$.
In terms of the new variable $s=\sigma+ [V'''/(2U'''')]$ the cubic equation \eq{cubiceqsig}   exhibits either one real root $s^{(0)}$ or three  real roots  $s^{(-)}<s^{(0)}<s^{(+)}  $ (see the discussion in Appendix \ref{minimi}).

In our numerical studies we determined the roots of Eq. \eq{cubiceqsig} by means of Cardano's formulas, so that we were able to follow how the roots move on the complex plane when the scale $k$ is decreased. The estimate of $\sigma(\Phi)$ via a second-order polynomial has then been used when non-trivial real roots occurred. In such cases the value of the rEAA at $\Phi=0$ has been evaluated for both minima and numerically decided which minimum gives the true minimum of the rEAA.
  It is shown in Appendix  \ref{minimi} that at the UV scale when the induced potential vanishes, $V_\Lambda(\Phi)=0$, the necessary and sufficient
condition of existing two non-trivial real roots is that it should hold the inequality
\bea\label{rootineq}
 U_k''(\Phi) - \frac{ Z_{\pa k}^2}{4Y_{\pa k} } &<&0.
\eea
Then the values of the rEAA at both local minimum places are equal and deeper than its value for a homogeneous background. Decreasing the scale $k$ one of the minima gets deeper. 
 During the RG flow it may happen that in the UV region there are 3 real roots, but at some finite scale $k_s$ 2 of them move off from the real axis and a single real root remains present for scales $k<k_s$.

\subsection{Strategy of reading off the RG flow equations}\label{strat}

Here we discuss the general strategy by which one can derive from the WE  \eq{we} the separate RG flow equations  for the potentials and the momentum-dependent wave function renormalizations induced by the various Fourier modes of the periodic condensate given by \eq{chifouexp}, when  the EAA is looked for in the form of Eq. \eq{eaaon}. Both sides of the WE should be Taylor expanded  at the background configuration $\phi_{Bx}$ keeping the terms up to the quadratic ones in the fluctuating piece $\eta_x$ of the scalar field.  The  first-order terms vanish as discussed above, while the comparison of the terms of zeroth order and the quadratic ones in $\eta$ on both sides of the WE provide two   RG equations, one for the various potentials and another for the various  momentum-dependent wave function renormalizations:
\bea
\label{potflow}
&&{\dot \Gamma}_k [\phi_B] = \hf T_0 [ \c{Z}_k,\c{Z}_{1k},\ldots;
U_k(\Phi), U_{1k}(\Phi),\ldots ),\\
\label{wfrflow}
&&\sum_{n=0}^{N_M}\sum_{\tau=\pm 1} \int_{Q,Q'} \gamma_k^{(2)n}(Q_\pe^2, Q_\pa^2) \delta_{Q+Q'+n\tau Pe, 0}\eta_Q\eta_{Q'} =
\hf T_2  [ \c{Z}_k,\c{Z}_{1k},\ldots;\eta; U_k(\Phi), U_{1k}(\Phi),\ldots ).
\eea
Here the right-hand sides are functionals of the wave function renormalizations $\c{Z}_{nk}$ and functions of the potentials $U_{nk}(\Phi)$ and their $\Phi$-derivatives. (The latter are suppressed in the notation).

There may occur that no periodic condensate
is preferred by the system, i.e., the amplitudes $\sigma_n(\Phi)$ minimizing the rEAA vanish at some finite scale $k_{\rm{fr}}$. Then one keeps frozen  all the induced terms of the EAA (and formally the momentum $P_k$) at their values at the scale  $k_{\rm{fr}}$. The only flow equations one needs for scales $k<k_{\rm{fr}}$ are those for the ordinary potential and the ordinary momentum-dependent wave function renormalization, a case similar to the application of the usual EAA RG approach.  For $k\ge k_{\rm{fr}}$, however, one has to identify the separate equations for the ordinary local potential $U_k$
and the induced potentials $U_{nk}$ with $n\ge 1$ as well as for the ordinary wave function renormalization $\c{Z}_k(p_\pe^2,p_\pa^2)$ and the induced ones, $\c{Z}_{nk}(p_\pe^2,p_\pa^2)$ with $n\ge 1$. The separate equations for the various potentials can be read off from Eq. \eq{potflow} by the following algorithm.
 One sets formally $\sigma_n(\Phi)$  to zero for all $n\ge 1$ and
reads off the evolution equation for the ordinary potential $U_{0k}(\Phi)=U_k(\Phi)$. The expression on the right-hand side of  Eq. \eq{potflow} may depend on all of the potentials and their derivatives and on all momentum-dependent wave function renormalizations.
Having determined the flow equation for $U_{0k}(\Phi)$  one
sets formally   $\sigma_n(k)$  to zero for all $n\ge 2$ in Eq.  \eq{potflow}
 and interprets the flow equation obtained as that for the induced potential $U_{1k}(\Phi)$.
 Having determined the flow equations for $U_{nk}(\Phi)$ with $n=0,1$ 
 one
sets formally   $\sigma_n(k)$  to zero for all $n\ge 3$ in Eq.  \eq{potflow} and interprets the flow equation obtained as that for the induced potential $U_{2k}(\Phi)$,  and so on.

The separate flow equations for the various momentum-dependent wave function renormalizations are obtained by using the fact that only particular bilinear expressions of $\eta_Q$ are present in our ansatz for the EAA and its scale derivative. These are of the form
$\sum_{\tau=\pm 1}\int_{Q,Q'} \gamma_k^{(2)n}(Q_\pe^2, Q_\pa^2) \delta_{Q+Q'+n\tau Pe, 0}\eta_Q\eta_{Q'}$
containing shifts between the momenta $Q$ and $Q'$ equal to integer $\pm n$ times  the elementary shift $Pe$ as a result of the 
induced interaction of the fluctuating field with the various modes of the periodic condensate. Therefore we shall keep only these kind of terms on the right-hand side of the flow equation \eq{wfrflow}. Then the separate flow equations for the  wave function renormalizations $\c{Z}_{nk}(p_\pe^2, p_\pa^2) $ with the various orders $n$ can be read off by comparing the terms of the form $\sum_{\tau=\pm 1}\int_{Q,Q'} f_k^{n}(Q_\pe^2, Q_\pa^2) \delta_{Q+Q'+n\tau Pe, 0}\eta_Q\eta_{Q'}$ on both sides of Eq. \eq{wfrflow}.  Here it should however be mentioned that there occur  on the right-hand side of Eq. \eq{wfrflow} bilinear terms with kernels depending on $Q_\pa$ (not only those depending on $Q_\pa^2$) because of shifted arguments  $p-Q+\tau nPe$. The presence of such bilinear expressions would violate the reflection symmetry with respect to the plane $x_\pa=0$. Therefore no such bilinear expressions should be included into the ansatz for the EAA,  and the contributions of similar expressions  should cancel on the right-hand side of the flow equations (formally due to the sums over $\tau=\pm 1$).

Finally the flow equations for the running  couplings of the potentials can be obtained by expansion in powers of $\Phi$ on each sides of the flow equations for the potentials. As discussed above, the minimization of the EAA  at a given scale $k$ with respect to the periodic background configuration yields $\Phi$-dependent Fourier-amplitudes $\sigma_{n~k}(\Phi)$. These field-dependences should be taken into account when the Taylor-expansion in powers of $\Phi$  is performed. 
For  our preliminary numerical  work we perform the Taylor-expansion at $\Phi=0$, although it is well-known that the expansion at the scale-dependent minimum of the ordinary potential provides better numerical results for the ordinary $O(1)$ model.
The flow equations for the running derivative couplings are obtained by Taylor expanding the integrands in powers of $Q_{\pa}=Q_1$ and  $Q_{\pe\mu}$ $(\mu=2,3)$  on the right-hand sides of the various flow equations for the  momentum-dependent wave function renormalizations. 

The solution of the RG evolution equations is performed by the algorithm representing a cycle for the various values of the gliding scale $k$ decreased stepwise
from the UV cutoff value  $\Lambda$ by steps $\Delta k$ to zero, keeping $\delta k=\Delta k/k$ constant. Each step of the cycle consists of the following manipulations:
 \begin{enumerate} 

\item The necessary conditions of the minimum of the rEAA \\ $\Gamma_k[\phi_B]=\Gamma_B( \Phi;\{\sigma_{n~k}\},P_k^2)$ for given homogeneous background  $\Phi$,
\bea\label{mincon}
 0&=& \partial_{P^2}\Gamma_B( \Phi;\{\sigma_{n~k}\},P^2)= 
\partial_{\sigma_{n~k}}\Gamma_B( \Phi;\{\sigma_{n~k}\},P^2),~~n=1,2,\ldots, N_m
\eea
are solved for the characteristic momentum $P_k$ and the \\ Fourier-amplitudes 
$\sigma_{n~k}(\Phi)$ of the periodic condensate. The latter are approximated by polynomials of $\Phi$ of the degree $M-2$. The flow equations for the various couplings of the potentials are obtained by Taylor-expanding  $\Gamma_B( \Phi;\{\sigma_{n~k}(\Phi)\},P_k^2)$ in powers of $\Phi$ at $\Phi=0$. On the right-hand sides of  the flow equations for the various derivative couplings one has to set $\Phi=0$, because  field-independent derivative couplings are used. 

\item Making use of the RG equations for the various couplings $c_i(k)=\{ g_1(k), g_2(k),\ldots, v_1(k),v_2(k),\ldots, Z_{\pe ~k}, Z_{\pa~k},$ $ Y_{\pe~k}, Y_{X~k}, Y_{\pa~k},\ldots \}$ one determines their scale derivatives ${\dot c}_i(k)=k\partial_k c_i$.

\item One determines all couplings at the lower scale $k-\Delta k$ via the relations $c_i(k-\Delta k) =c_i(k) - {\dot c}_i(k)\delta k$.
 
\end{enumerate}
Then one repeats the steps of the cycle for scale $k-\Delta k$ and so on, successively decreasing the scale $k$.

\section{Preliminary numerical studies}\label{numres}

\subsection{On our numerical approach}\label{onnum}
Preliminary numerical data have been obtained on the position and characteristics of the WF FP and the phase diagram of the Euclidean 3-dimensional  ghost $O(1)$ model various approximations of the Fourier-Wetterich RG framework:
in the LPA in the zero-mode approximation (LPA0), in the LPA in the one-mode approximation (LPA1), and in the NLO of the GE in the one-mode approximation (NLO1). The zero-mode approximation in NLO should provide the same result as the
zero-mode approximation in LPA because we restricted our approach to
the usage of field-independent wave function renormalization and it is well-known that in the usual functional RG approach run of the wave function renormalization only occurs in the $Z_2$ symmetric $O(1)$ model when the field-dependence of the wave function renormalization is taken into account at least at the order $\ord{\Phi^2}$. In our Fourier-Wetterich RG approach in the one-mode approximation, however, the derivative couplings $Z_{A~k}$ $(A=\pe,\pa)$ may have flow even if the field-dependence of the wave function renormalization is not taken with.
This is ensured by the multiplier $b_1^2$ in front of the loop-integral on the right-hand side of the flow Eq. \eq{wfrAflnlo}. Moreover, the factor $b_1^2$ may not vanish even if the amplitude of the periodic condensate vanishes but as a remainder of its previous evolution  the coupling $\t{v}_3$ of the induced potential is frozen at a non-vanishing value.

The phase diagram can be mapped out by generating RG trajectories in the space $\{ \t{c}_j \}$ of the dimensionless couplings for various initial conditions.
For the latter we have chosen generally     $Z_{\pa~\Lambda}=Z_{\pe~\Lambda}=1$ and $\t{Y}_\pa=\t{Y}_X=\t{Y}_\pe=\t{Y}\in (0,2]$ and $\t{u}_1(\Lambda)=\t{u}_3(\Lambda)=\t{v}_2(\Lambda)=\t{v}_4(\Lambda)=0$ 
 corresponding to a $Z_2$ symmetric model at the  UV scale $\Lambda=1$.
 The induced wave function renormalization $\c{E}_k(Q_\pe^2,Q_\pa^2)$ was neglected completely. The RG trajectories have been generated for fixed $\t{g}_4(\Lambda)=0.01$ but for various values of $\t{Y}_\Lambda$ and $\t{g}_2(\Lambda)$ and keeping  the ratio $\Delta k/k=10^{-3}$ fixed.
 The  flow equations for the various dimensionless couplings $\t{c}_j$ have the forms
\bea\label{setflow}
   {\dot {\t{c}}}_j &=& \beta_{\t{c}_j} (\t{c}_1,\t{c}_2,\ldots)
\eea
which have been obtained by computer algebraic routines from the flow equations
 given in Appendix \ref{a:lpaflow} for the LPA and in  Appendix \ref{a:nloflow}
for the NLO of the GE. The approximation LPA0 corresponds to the usual EAA RG approach in the LPA, the running couplings are $ \{\t{c}_j\}=( \t{u}_2, \t{u}_3,\t{u}_4 )$ and the set \eq{setflow} of the flow equations are derived from Eq.
\eq{ulpadiml} by Taylor-expanding its both sides in powers of $\Phi$ at $\Phi=0$. In the approximation LPA1   the running couplings are $ \{\t{c}_j\}=( \t{u}_2, \t{u}_3,\t{u}_4, \t{v}_2, \t{v}_3,\t{v}_4 )$ and the set of flow equations found in LPA0 has to be completed by those obtained from Eq. \eq{potlpadiml}. In the approximation NLO1 the running couplings are  $ \{\t{c}_j\}=( \t{u}_2, \t{u}_3,\t{u}_4, \t{v}_2, \t{v}_3,\t{v}_4, Z_{\pa~k}, Z_{\pe~k} )$ and the set  \eq{setflow}
of the flow equations are those obtained from Eqs. \eq{ulpadiml} and \eq{potflnlo} by Taylor-expansion in $\Phi$ at $\Phi=0$ and the ones given by Eq. \eq{wfrAflnlo}.

In the  approximation LPA0 no periodic condensate is taken into account, the amplitude $\sigma_k(\Phi)$ and the induced potential $V_k(\Phi)$ are set to zero.
 Nevertheless, one can estimate the scaling of $\sigma_k(\Phi)$ in the deep IR regime, by taking the IR scaling laws of the couplings and inserting those
 into  the inequality \eq{ineq} and the roots of Eq. \eq{redcubiceq} as discussed  in Appendix \ref{minimi} for the case with $V_k(\Phi)=0$.
 In the approximations LPA1 and NLO1 the set  \eq{setflow} of the flow equations should be solved together with the determination of a polynomial estimate of the amplitude $\sigma_k(\Phi)$ of the periodic condensate for any values of the gliding scale $k$, as discussed at the end of Section \ref{strat}. Typically, the relative change $\delta k=\Delta k/k$ of the scale has been kept constant during the flow and  was chosen  $\delta k\le 10^{-3}-10^{-5}$  in order to obtain stable flows in the IR regime.

There were made several test runs in the various approximations in order to check that $Z_2$ symmetry is kept during the RG flow even in the one-mode approximation and it was found that $\t{u}_1(k)$, $\t{u}_3(k)$, $\t{v}_2(k)$, and $\t{v}_4(k)$ keep their vanishing values,  all within the numerical accuracy of $<10^{-10} $. This means that the proposed Fourier-Wetterich RG framework  satisfies the general expectation that it does not violate $Z_2$ symmetry. Therefore the detailed numerical analysis of the WF FP and that of the phase structure were performed in the reduced parameter spaces, spanned by the couplings not identically vanishing due to $Z_2$ symmetry.

For the numerical determination of the WF FP two methods were combined as a rule both of which have been proven very effective in the case of the determination of the WF FP of the ordinary $O(N)$ models \cite{Litim2007,Nagy2012}. The fine-tuning method
determines the position and characteristics of the WF FP via the evaluation
of the flow of the couplings along almost critical trajectories and identifying
the WF crossover scaling region in which the couplings keep almost constant values. In the root-finder method one solves the FP equations $\beta_j(\t{c}_1,\t{c}_2,\ldots )=0$ with estimated initial values by means of computer algebraic algorithm. A detailed description of the application of both methods was given in \cite{Peli2018}.  The enlargement of the number of dimensions of the parameter space
 in the subsequent  approximations LPA0 $\longrightarrow$ LPA1 $\longrightarrow$ NLO1 requires some additional refinements in the application of the fine-tuning and root-finder methods. Let us turn now to these peculiarities of the determination of the WF FP.

\begin{figure}[htb]
\centerline{
\psfig{file=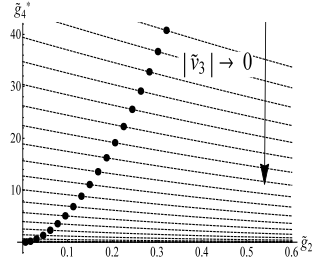,height=3.52cm,width=4.04cm,angle=0}}
\caption{\label{fig:y16_gfp} Determination of the Gaussian FP for $Z=-1$, $\t{Y}=1.6$. The continuous lines  are the solution functions $\t{g}_4^{[1]}(\t{g}_2,\t{v}_3)$ of Eq. \eq{betag4}. The  dots illustrate the  solutions of Eqs. \eq{betag2} and \eq{betag4} for various values of $\t{v}_3$, belonging to the Gaussian branch $\c{F}^{[G]}$. The dot for $\t{v}_3= 0$ corresponds to the Gaussian FP in the approximation LPA1. The solutions depend only on $|\t{v}_3|$ and there are real roots only in the interval  $\t{v}_3 \in \lbrack -20,20 \rbrack$.  The solutions are plotted from $\t{v}_3 = 20$ to $\t{v}_3=0$ with the step size $\Delta \t{v}_3=1$.}
\end{figure}

\begin{figure}[htb]
\centerline{
\psfig{file=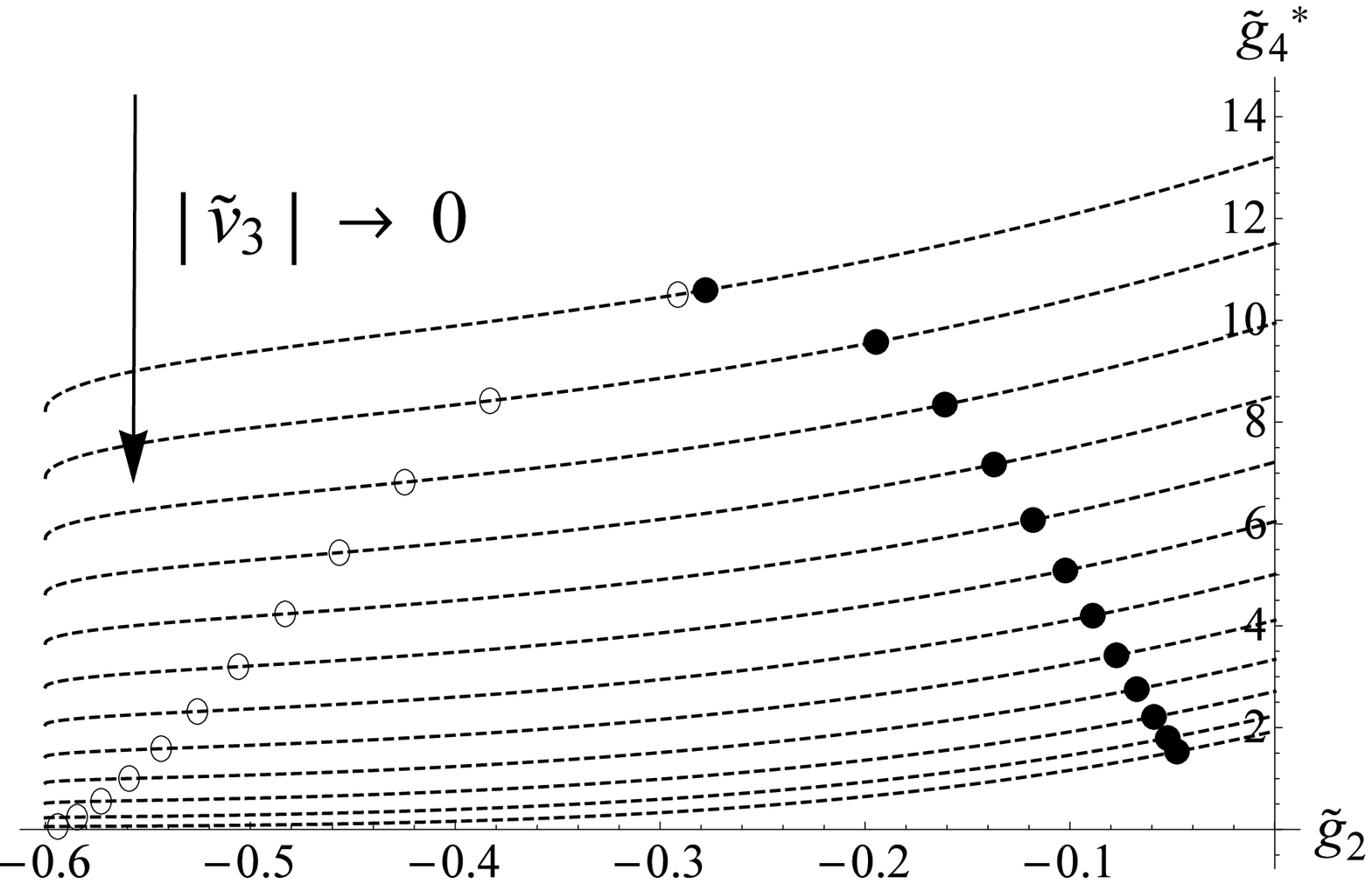,height=3.52cm,width=4.04cm,angle=0}}
\caption{\label{fig:y16_wffp} Determination of the WF FP for $Z=-1$ and $\t{Y}=1.6$. The continuous lines are the solution functions $\t{g}_4^{[2]}(\t{g}_2,\t{v}_3)$ of Eq. \eq{betag4}. There are two branches, the WF branch  $\c{F}^{[WF]}$  (dots) and the IR branch  $\c{F}^{[IR]}$ (empty circles)
of the real solutions of Eqs. \eq{betag2} and \eq{betag4} for $0\le |\t{v}_3|\le 6.002$ and ending up at the same point for  $ |\t{v}_3|=6.002$. The solutions only depend on $|\t{v}_3|$.   The solutions are plotted from $\t{v}_3 = 6$ to $\t{v}_3=0$, with the step size $\Delta \t{v}_3=0.25$.}
\end{figure}

In the approximation LPA0 with the polynomial truncation $M=4$ the FP equations 
$\beta_{\t{g}_2}=\beta_{\t{g}_4}=0$ can be solved and one finds in addition to the Gaussian FP at $\t{g}_2=\t{g}_4=0$ the WF FP at
\bea\label{wffplpa0}
 && \t{g}_2^{*} = -\frac{1}{13} (Z+\t{Y}),~~~~
  \t{g}_4^{*}= \frac{4\pi^2 \sqrt{2}}{3}\frac{14^3}{13^3}\frac{  (Z+\t{Y})^3 }{ Z+2\t{Y}} 
\eea
In the approximation LPA1 there is now way to determine the position of the WF FP analytically. The FP equations have the structure
\bea\label{betag2}
    0&=&\beta_{\t{g}_2} (\t{g}_2, \t{g}_4, \t{v}_3; \t{\sigma}_0=0, \t{\sigma}_2=0),\\
\label{betag4}
 0&=&\beta_{\t{g}_4} (\t{g}_2, \t{g}_4, \t{v}_3; \t{\sigma}_0=0, \t{\sigma}_2=0)
,
\\
\label{betav3}
 0&=&\beta_{\t{v}_3} (\t{g}_2, \t{g}_4, \t{v}_3; \t{\sigma}_0, \t{\sigma}_2),
\eea
where 
\bea\label{sig0}
\t{\sigma}_0=\t{\sigma}_0(\t{g}_2, \t{g}_4,\t{v}_3)
\eea
 and $\t{\sigma}_2=\t{\sigma}_2(\t{g}_2, \t{g}_4, \t{v}_3;\t{\sigma}_0) $ are the coefficients in the polynomial estimate \eq{sigest} of the amplitude $\t{\sigma}_k(\Phi)$ of the periodic condensate. The coupled set of Eqs. \eq{betag2}-\eq{sig0} can only be solved numerically. Our procedure can be outlined as follows. First we consider $\t{v}_3$ as free parameter. Eq.  \eq{betag4} can be solved analytically, it has two roots  $\t{g}_4^{[A]}(\t{g}_2,\t{v}_3)$  $A=1,2$ both being functions of the couplings  $\t{g}_2$ and $\t{v}_3$. These roots define two  one-parameter families $\c{F}^{[A]}$  $(A=1,2)$ of solutions for Eqs. \eq{betag2} and \eq{betag4}. Each family $\c{F}^A$ has various branches of solutions $\c{F}^{[A\alpha]}$ because Eq. \eq{betag2} has several solutions. Then we identify   the  Gaussian $\c{F}^{[G]}$ and the WF $\c{F}^{[WF]}$ branches of the solutions which contain the Gaussian FP and the WF FP, respectively, in the case $\t{v}_3=0$, i.e.,  in the approximation LPA0.  As to the next, we look in the  branches  $\c{F}^{[G]}$ and  $\c{F}^{[WF]}$  for a solution belonging to a particular value $\t{v}_3^*$  for which also Eq. \eq{betav3} is satisfied with high accuracy. In this manner we find the positions of the Gaussian  and the WFs in the approximation LPA1. Our procedure is illustrated for the determination of the Gaussian FP and tht of the WF FP in Fig. \ref{fig:y16_gfp}
and Fig. \ref{fig:y16_wffp}, respectively. 
Taking into account the actual values of the parameters $\t{\sigma}_0$ and $\t{\sigma}_2$ in the polynomial estimate \eq{sigest} of the periodic condensate's amplitude is  another important task in the procedure for the determination of the positions of the FPs in the one-mode approximation. This has been done for both of the branches $\c{F}^{[B]}$ with $B={\rm{G}},~{\rm{WF}}$
by the following algorithm:
 {\em (i)}
 Ones goes through various values $\t{v}_3=\t{v}_{30}$, inserts those  and the corresponding solutions $ \t{g}_n^{[B]}(\t{v}_{30})$ for $ n=2,4$ from  the branch $\c{F}^{[B]}$ into Eq. \eq{sig0}, takes  $\t{\sigma}_0^{[B]}(\t{g}_2^{[B]}(\t{v}_{30}),\t{g}_4^{[B]}(\t{v}_{30}),\t{v}_{30})\equiv\t{\sigma}_0^{[B]}(\t{v}_{30})$ and calculates $\t{\sigma}_2^{[B]}(\t{v}_{30}))$ similarly.
{\em (ii)}
 These values are then inserted
into Eq. \eq{betav3} and the obtained equation
\bea\label{betav3est}
  0&=&\beta_{\t{v}_3} \bigl(\t{g}_2^{[B]}(\t{v}_{30}), \t{g}_4^{[B]}(\t{v}_{30}), \t{v}_3; \t{\sigma}_0^{[B]}(\t{v}_{30}), \t{\sigma}_2^{[B]}( \t{v}_{30})\bigr)
\eea
 is solved for $\t{v}_3$ with a root-finding algorithm, in order to get the root
$ \t{v}_3^{[B]}(\t{v}_{30})$, belonging to the initial choice $\t{v}_{30}$. 
{\em (iii)} Now one looks for the case when the discrepancy $|\t{v}_{30}- \t{v}_3^{[B]}(\t{v}_{30})| $ is minimal, and identify the corresponding value  $\t{v}_3^{[B]}(\t{v}_{30})\approx \t{v}_{30}$ 
as the best estimate of the fixed-point value  $ \t{v}_3^{[B]*}$. Then also the FP values of the other couplings $\t{g}_n^{[B]*}= \t{g}_n^{[B]}(\t{v}_{3}^{[B]*}) $ with $n=2,4$ and $\t{\sigma}_n^{[B]*}=\t{\sigma}_n^{[B]}(\t{v}_{3}^{[B]*}) $ for $n=0,2$ are calculated.
{\em (iv)} Finally, the FP values are inserted into the   right-hand sides of Eqs. \eq{betag2}-\eq{betav3} in order to test that the estimate of the position of the FP is good enough to make the beta-functions to vanish.
In each cases studied numerically we have found the Gaussian FP and the WF FP  with the
 accuracies $|\beta_{c_i}|\le 10^{-9}$ and  $|\beta_{c_i}|\le 10^{-15}$, respectively,  for all couplings $c_i=\t{g}_2$, $\t{g}_4$,  $\t{v}_3$.

 It can be seen in Fig. \ref{fig:y16_wffp} that there is another branch $\c{F}^{[IR]}$ of the real solutions of Eqs. \eq{betag2} and \eq{betag4}. The point of the branch $\c{F}^{[IR]}$ corresponding to  $\t{v}_3=0$  is positioned at $\t{g}_2=\t{g}_2^{[IR]}= -Z-\t{Y}$, $\t{g}_4^{[IR] }=0$. In the ordinary $O(1)$ model this is an infrared (IR) singularity point:  the flow equations along nearly critical trajectories, running in the symmetry broken phase, develop a singularity at it and their solutions end at it for some small but finite $k^{[IR]}$ values, where $k^{[IR]}\to 0$ when the trajectory approaches the separatrix. Increasing the value of $\t{v}_3$ with the step size $\Delta \t{v}_3=0.1$ from zero to $\t{v}_{3\rm{max}}$, we have not found vanishing of $\beta_{\t{v}_3}$. This indicates that the points of  branch  $\c{F}^{[IR]}$ are points of singularities belonging to the various values of $\t{v}_3$, but there exists no IR FP 
either in the ghost $O(1)$ model.

It should be pointed out that the rather involved search for the WF FP can be simplified in any interval of $\t{Y}$ in which the FP values of the couplings
vary continuously with $\t{Y}$.  Namely, one goes through the whole procedure
 say at some particular value of $\t{Y}$,  determines the FP values of the couplings, and  then decreases (or increases) $\t{Y}$ in small steps, using the  FP values found in the previous step as educated guess of the roots for the root-finding algorithm applied for solving the set of Eqs. \eq{betag2}-\eq{sig0}.

In the approximation NLO1  it was found by the fine-tuning method that the beta-functions for the wave function renormalizations $Z_{A~k}$ with $(A=\pe, \pa)$ exhibit no WF crossover regions. In accordance with that, we could not find solutions  of the set of the FP equations \eq{betag2}-\eq{sig0} and $\beta_{Z_A}=0$ by means
 of the appropriate modification of the above described root-finder algorithm.
Therefore,   the root-finder method  reduces essentially to the one applied in the approximation LPA1 while the wave function renormalizations  $Z_{A~k}$ with $(A=\pe, \pa)$ should be set to given values (see  the discussion in the next section).

\subsection{Characteristics of the FPs}\label{charwffp}

The Gaussian FP has been found at $\t{g}_2^{[G]}=\t{g}_4^{[G]}=0$ in the approximation LPA0 and at  $\t{g}_2^{[G]}=\t{g}_4^{[G]}=\t{v}_3^{[G]}=0$ in the approximation LPA1 for any values of the higher-derivative coupling $\t{Y}$.
As Fig. \ref{fig:y16_gfp} shows, in the approximation NLO1 we have found again the Gaussian FP to be
located at $\t{g}_2^{[G]}=\t{g}_4^{[G]}=\t{v}_3^{[G]}=0$ for $Z_\pe=Z_\pa=-1$. This means, that the equation
used to determine $\t{\sigma}_k$ becomes trivial, and thus no periodic
condensate is present in the Gaussian FP. Furthermore, the vanishing of the
periodic condensate also makes the $\beta_{\t{v}_3}$ vanishing. The
eigenvalues of the stability matrix in the Gaussian FP are the ones of the
ordinary $O(1)$ model extended with a zero, i.e., $-2,-1,0$.

\begin{figure}[htb]
\centerline{
\psfig{file=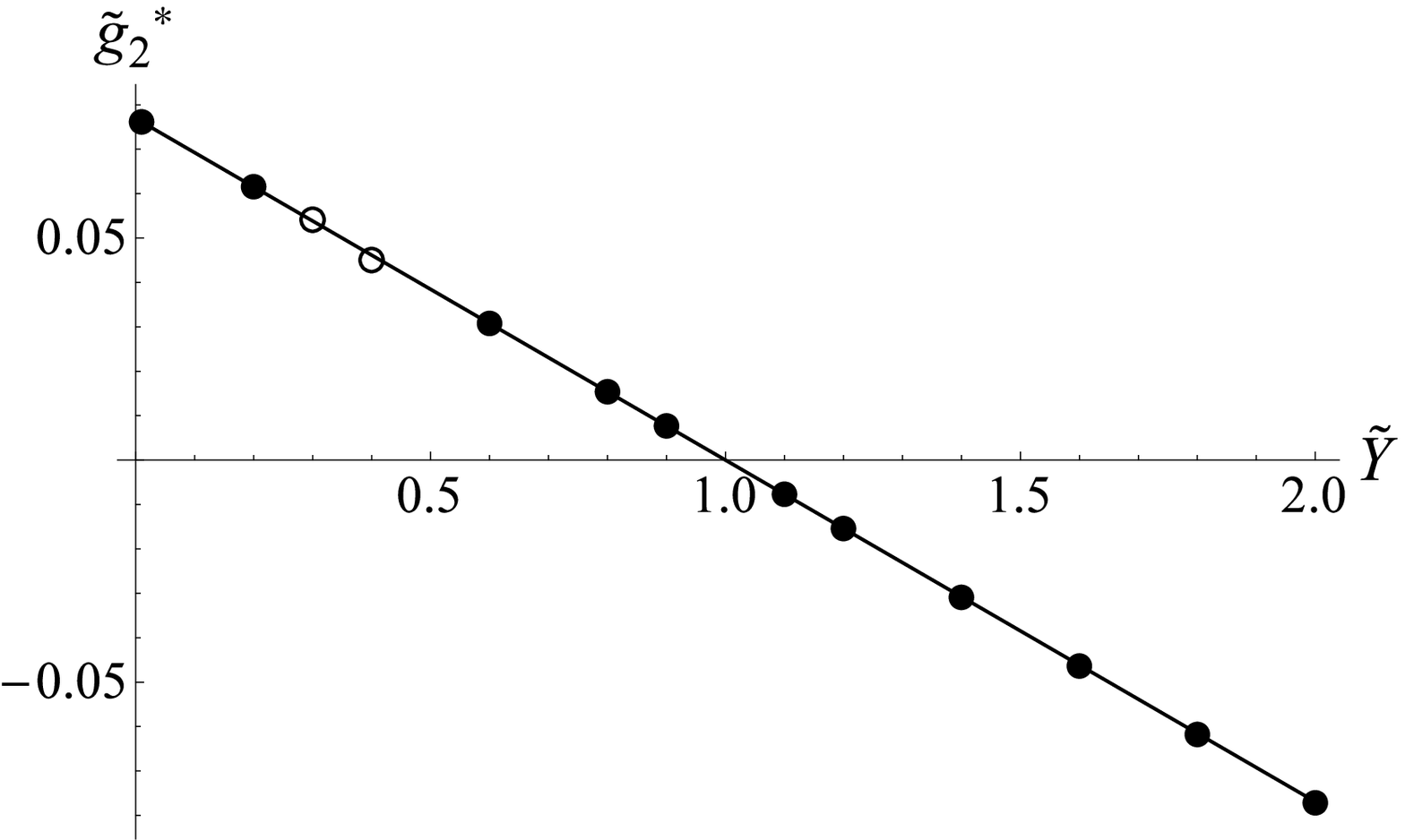,height=3.52cm,width=4.04cm,angle=0}
\psfig{file=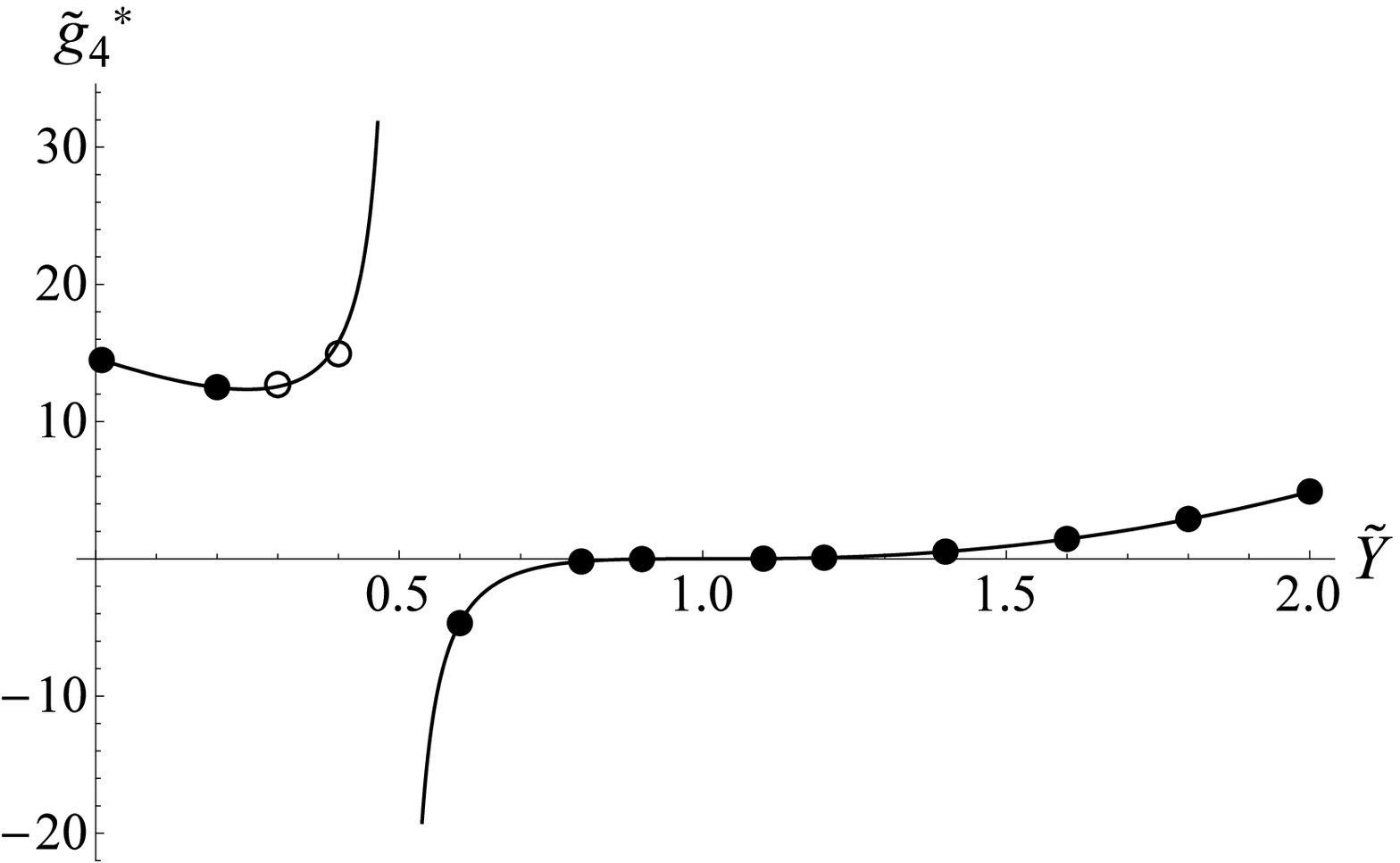,height=3.52cm,width=4.04cm,angle=0}
\psfig{file=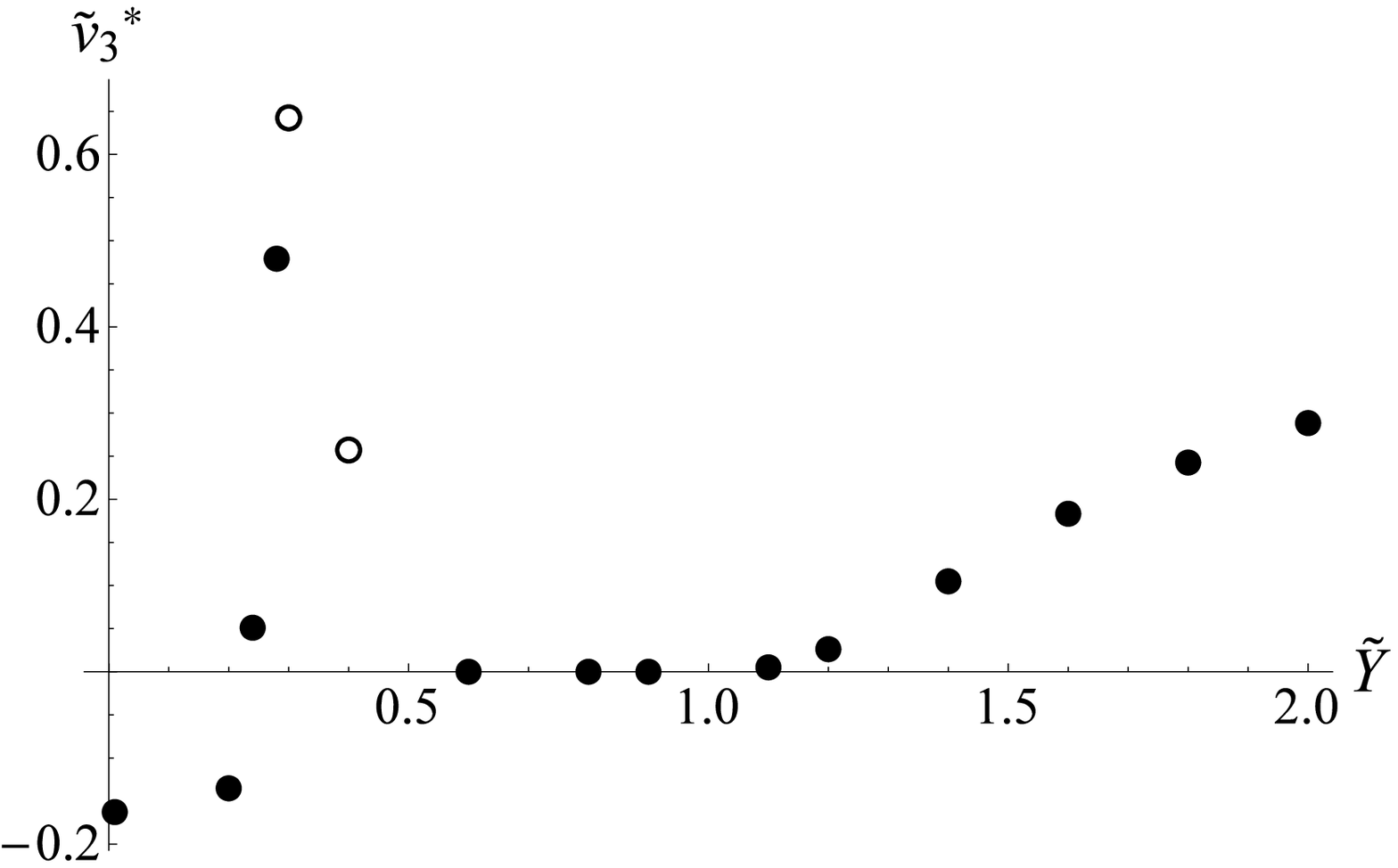,height=3.52cm,width=4.04cm,angle=0}}
\centerline{
\psfig{file=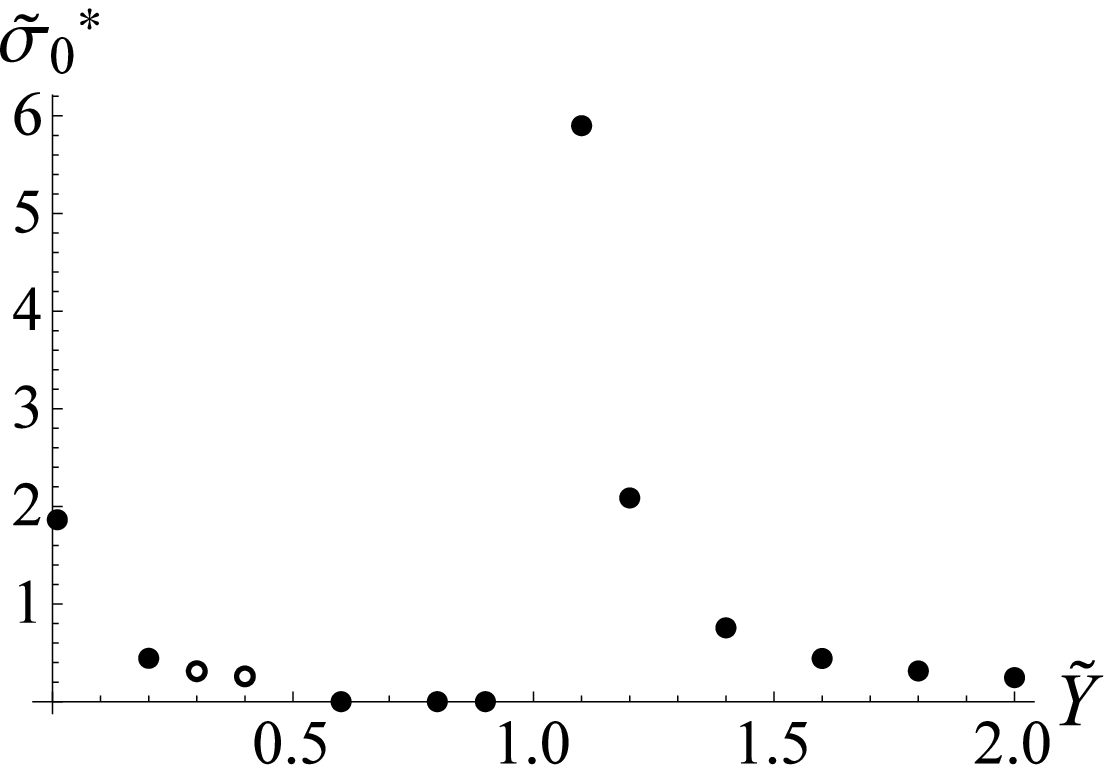,height=3.52cm,width=4.04cm,angle=0}
\psfig{file=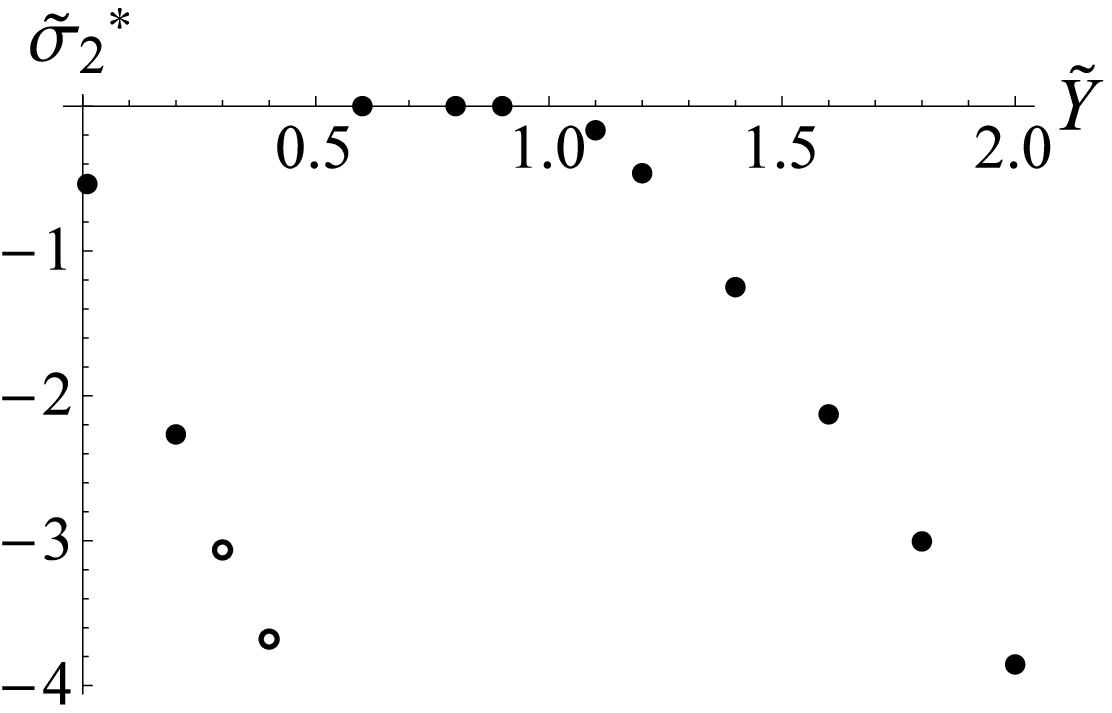,height=3.52cm,width=4.04cm,angle=0}}
\centerline{
\psfig{file=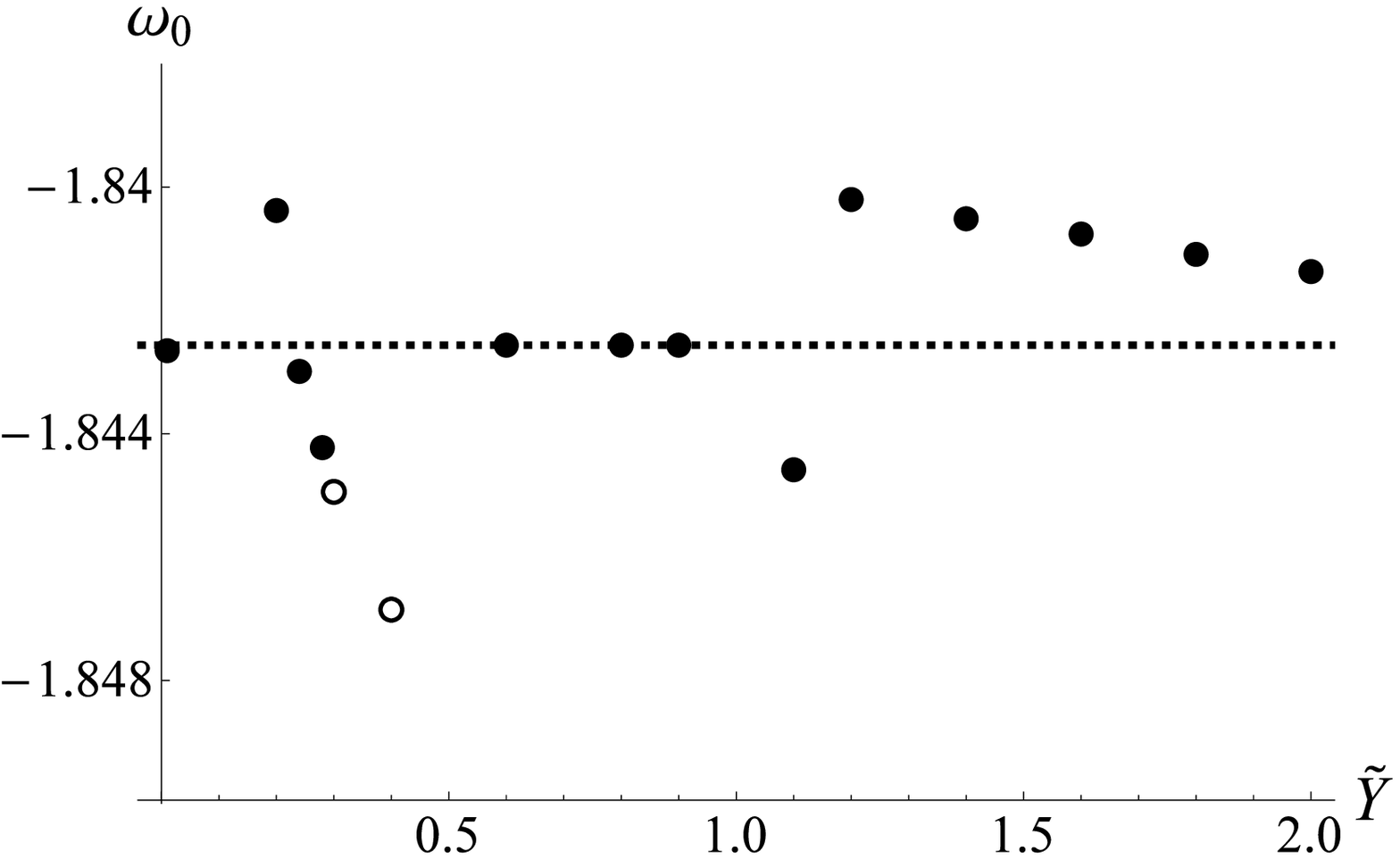,height=3.52cm,width=4.04cm,angle=0}
\psfig{file=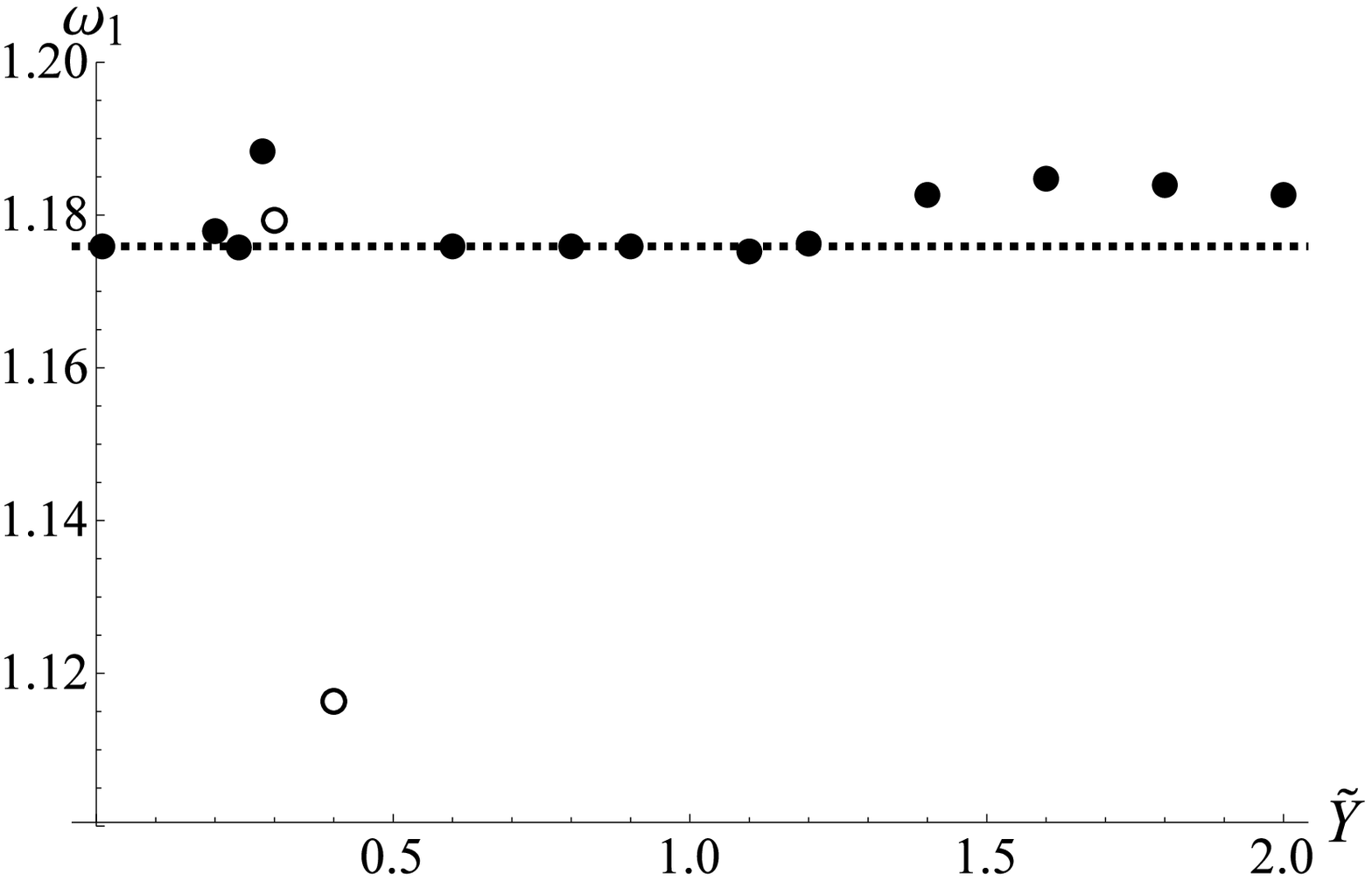,height=3.52cm,width=4.04cm,angle=0}
}
\caption{\label{fig:lpawf} The position of the WF FP, the coefficients of the amplitude of the periodic condensate at the WF FP, and the corresponding first two eigenvalues of the stability matrix $\u{\u{M}}$ vs. the higher-derivative coupling $\t{Y}$. The LPA0 and LPA1 results are indicated by the full lines and the full circles, respectively.   The empty circles represent the real parts of the corresponding parameters.}

\end{figure}  
The dependence of the position and the characteristics of the WF FP on the derivative coupling $\t{Y}$ have been determined in the approximations LPA0, LPA1, and NLO1. The results obtained in LPA0 and LPA1 are shown in  Fig. \ref{fig:lpawf}. As to the position of the WF FP in the parameter-plane $(\t{g}_2,\t{g}_4)$,
the approximations LPA0 and LPA1 yield the same result within the numerical accuracy, but there occurs a non-negligible induced potential in the approximation LPA1 (see the plots in the first row of Fig. \ref{fig:lpawf}). The second plot in the first row of Fig. \ref{fig:lpawf} shows the typical dependence of the FP value $\t{g}_4^*$ on the coupling $\t{Y}$ observed in both of the approximations LPA0 and LPA1, in accordance with the relations given in Eq. \eq{wffplpa0}. While in the ordinary scalar model (with $Z=+1$) $\t{g}_4^{*}$ is always positive for any non-negative $\t{Y}$, in the ghost model (with $Z\equiv \hf(Z_\pe+Z_\pa)\approx -1$)  $ \t{g}_4^{*}$ is positive if and only if $0\le \t{Y}\not \in [ 1/2, 1]$. For $\t{Y}\to 1/2$ from below (above) one gets $\t{g}_4^*\to +\infty$ $(-\infty)$ and
$\t{g}_2^*\to 1/26$.  
 The following conclusions can be made:
{\em (i)} For $1<\t{Y}_\Lambda\le 2$ there is a WF FP with  $ \t{g}_2^{*}<0$, $  \t{g}_4^{*}, ~\t{v}_3>0$, which moves towards the Gaussian FP at $\t{g}_2^{[G]}=\t{g}_4^{[G]}=\t{v}_3^{[G]}=0$ with decreasing $\t{Y}$ and merges into it for $\t{Y}=1$.  The dimensionless ordinary potential is a double-well potential opening upward. 
{\em (ii)} For $\t{Y}\in [1/2,1]$ there is no physically sensible fixed-point theory, because it holds  $\t{g}_4^{*}<0$ and the Euclidean action is unbounded from below. The dimensionless ordinary potential is a double-well potential opening downward.
In this interval the induced potential essentially vanishes with the accuracy $|\t{v}_3^*| < 10^{-8}$.
{\em (iii)} 
 For $\t{Y}\in (0, 1/2]$ the WF FP reappears with real parameters and with
$\t{g}_4^*>0$.  The critical theory is bounded from below again and describes
physics in  the strong-coupling regime  with positive mass squared and a convex dimensionless ordinary potential. 

The essential gain yielded by the one-mode approximation is the determination of the amplitude of the periodic condensate. The plots in the second row of Fig.
\ref{fig:lpawf} show that in the physically reliable cases (with Euclidean action bounded from below) the critical system corresponding to the WF FP exhibits a periodic condensate. Although the WF FP moves smoothly to the Gaussian FP when
$\t{Y}$ approaches the value $\t{Y}=1$, the $\t{Y}$-dependence of the amplitude $\t{\sigma}_0^*$ of the periodic condensate shows up a singularity: $\lim_{ \t{Y} -1 \to 0^+} \t{\sigma}_0^*=+\infty$ while $\lim_{ \t{Y} -1 \to 0^-} \t{\sigma}_0^*=0$.

\begin{figure}
\centerline{
\psfig{file=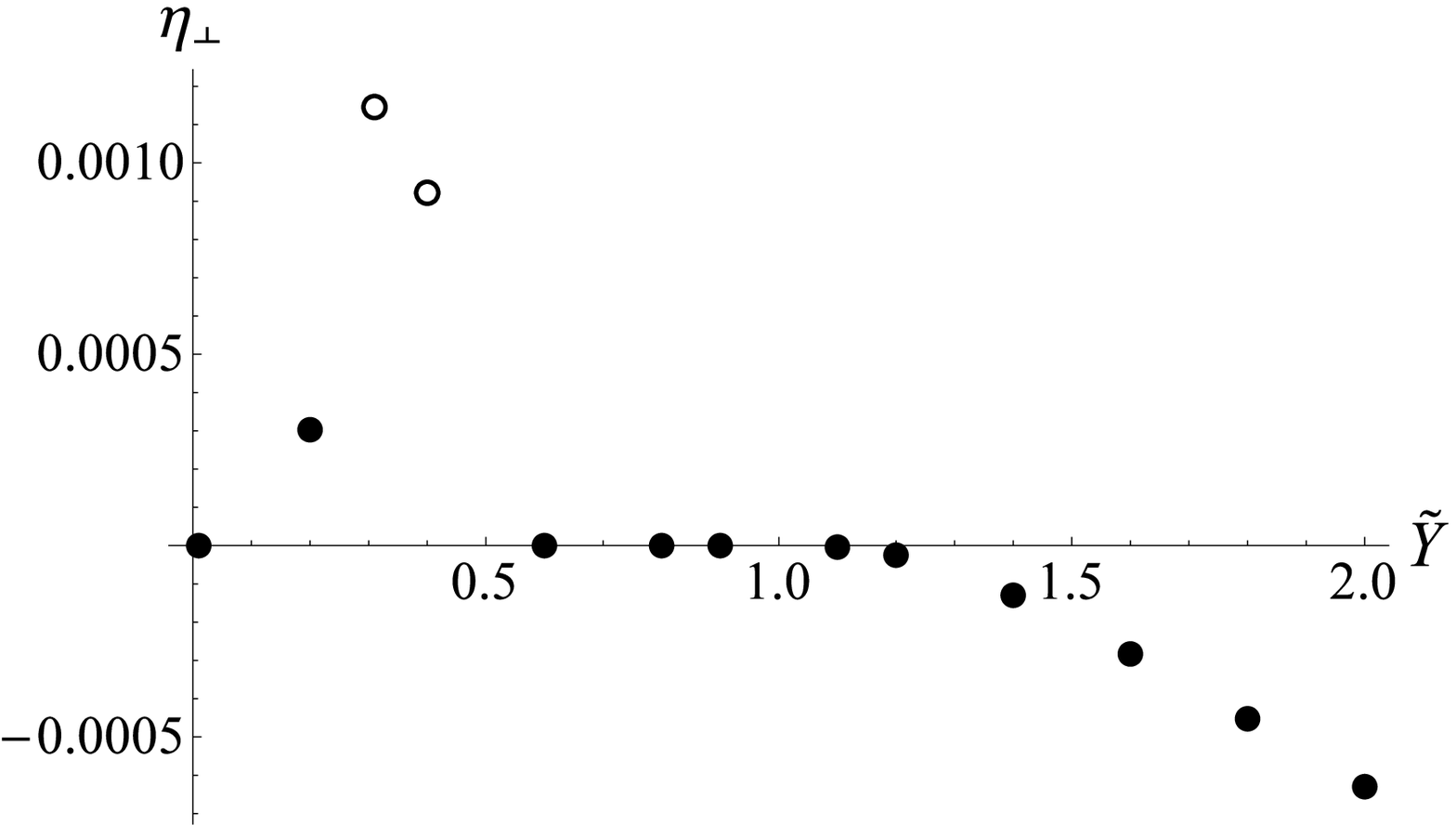,height=3.52cm,width=4.04cm,angle=0}
\psfig{file=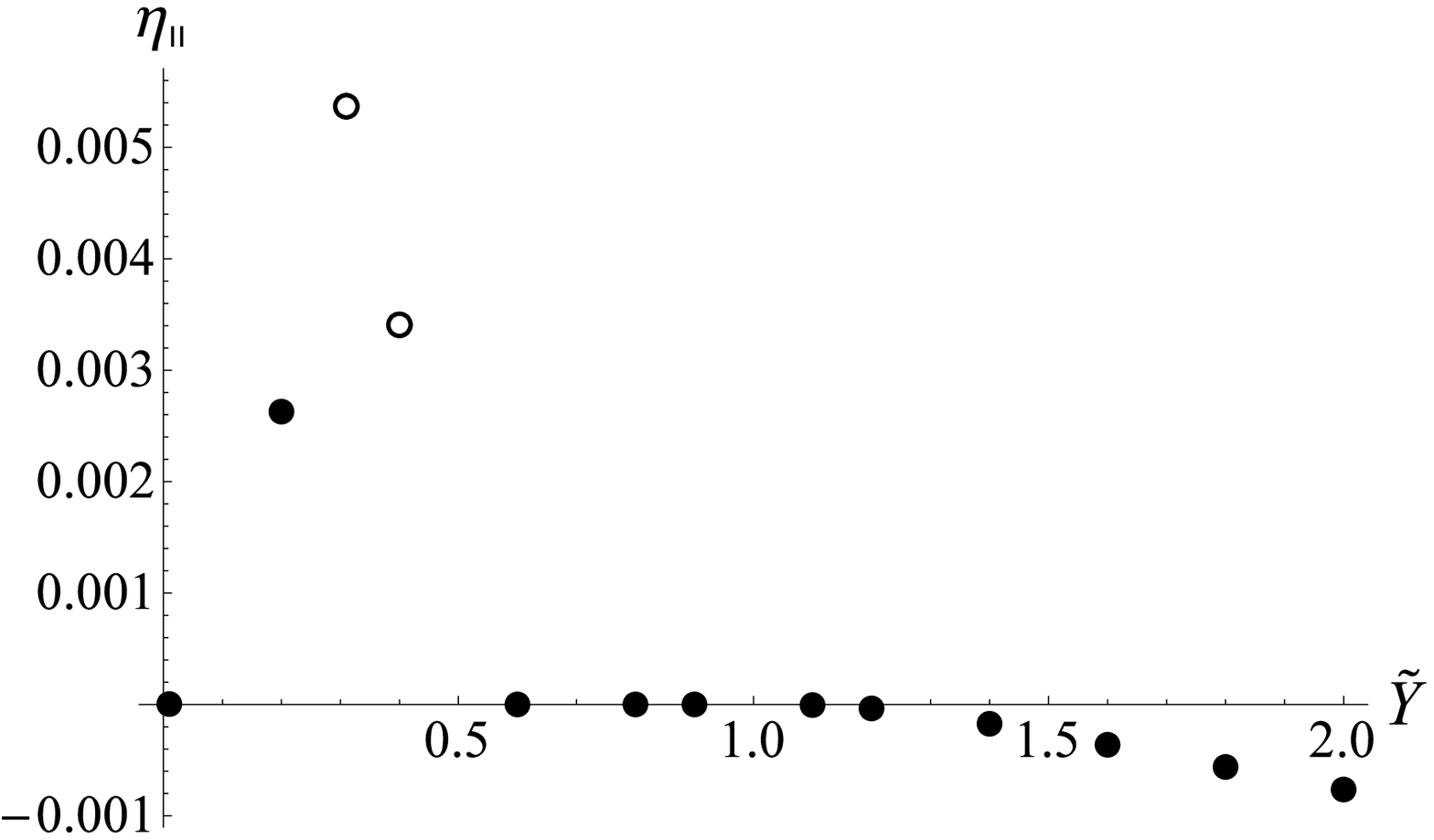,height=3.52cm,width=4.04cm,angle=0}}
\centerline{
\psfig{file=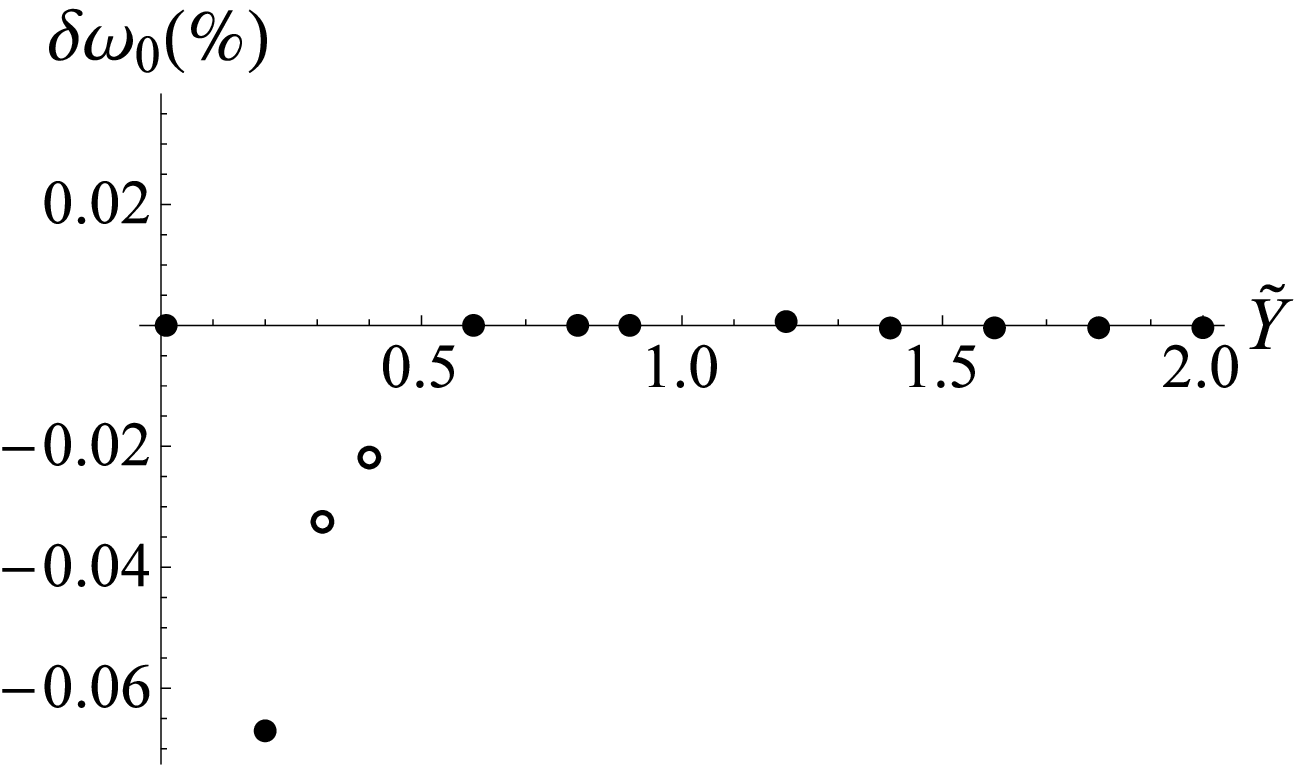,height=3.52cm,width=4.04cm,angle=0}
\psfig{file=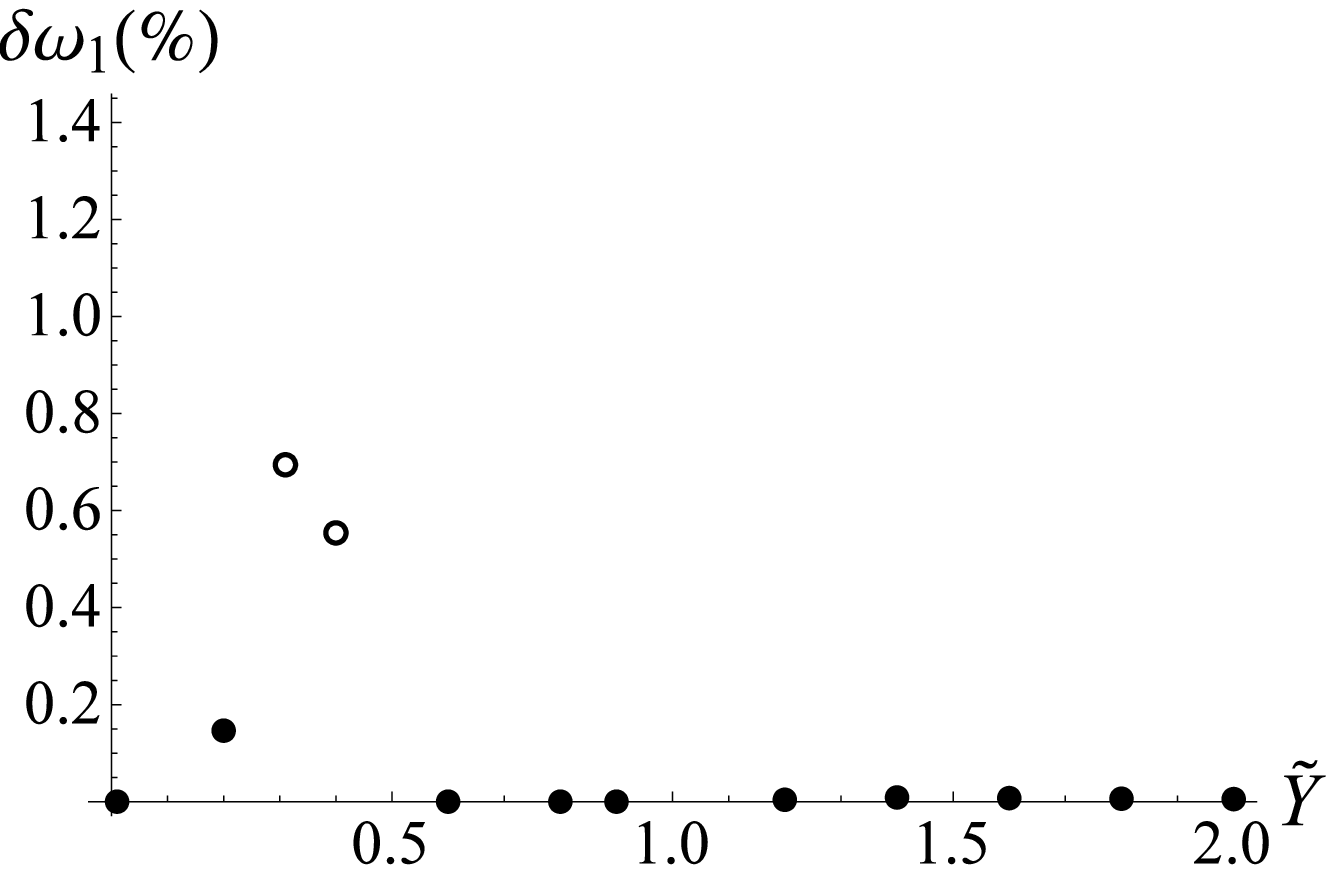,height=3.52cm,width=4.04cm,angle=0}}
\caption{\label{fig:nlocrexp} The anomalous dimensions $\eta_A$ $(A=\pe,~\pa)$
and the NLO effect $\delta \omega_i=100\times (\omega_i^{[NLO1]} -\omega_i^{[LPA1]})/\omega_i^{[LPA1]}$ on the critical exponents at the WF FP obtained in the approximation NLO1, for
 $Z_\pa^* = -1 + 2\times 10^{-3}$, $Z_\pe^* = -1+ 10^{-3}$ for $\t{Y} > 1$ and $Z_\pa^* = -1 - 2\times 10^{-3}$, $Z_\pe^* = -1 - 10^{-3}$ for $\t{Y} < 1$. }
\end{figure}

The first two eigenvalues $\omega_0=-1/\nu<0$ and $\omega_1$ of the stability matrix $M_{ij} =\partial \beta_{\t{c}_i}/\partial \t{c}_j$ at the WF FP have also been determined (the third row of plots in Fig. \ref{fig:lpawf}), where $\nu$ is the correlation length's critical exponent and $\t{c}_j$ stands for the various couplings $\t{g}_2,~\t{g}_4,~\t{v}_3$.  It is found that the WF FP should be a crossover point with the eigenvalues $\omega_0<0$ and $\omega_1>0$.  In the approximation LPA0 the eigenvalues $\omega_0$ and $\omega_1$ are independent of $\t{Y}$, while they seem to exhibit a slight  $\t{Y}$-dependence in the approximation LPA1.  One has to note, however, that the  strong truncation of the potentials possibly makes the values of the eigenvalues and their $\t{Y}$-dependence introduced by the one-mode approximation rather tentative. The strong truncation of the potentials may be the reason of another observed peculiarity. In the approximation LPA1 we have found, on the one hand,  that the FP values $\t{v}_3^*$ acquire a non-negligible imaginary parts for $\t{Y} \in (\sim 0.26, 1/2]$ (see the data illustrated by empty circles in Figs. \ref{fig:lpawf} and  \ref{fig:nlocrexp}.    On the other hand,  solving the flow equations we found qualitatively the same phase structure and quite similar crossover scaling regions for any values of $\t{Y}\in (0,1/2]$. 

As it was mentioned at the end of Sect. \ref{onnum},  we could not find an educated guess of the roots satisfying the full set of the FP equations including also the vanishing of the beta-functions of the wave function renormalizations $Z_A$ $(A=\pe,~\pa)$. This is connected with the observation that the flow of the
wave function renormalizations $Z_A$ does not exhibit the typical crossover region,  `the plateau' in the vicinity of the WF FP  (see Sect. \ref{phstruc}),
 just as in the case of the ordinary $O(N)$ models \cite{Peli2018}.
 Moreover, using the parameters of the WF FP obtained in LPA1 as initial conditions and starting to solve the full set \eq{setflow} of the flow equations in NLO1 towards the UV scale $\Lambda$, we could not arrive at $Z_A(\Lambda)=-1$ $(A=\pe,~\pa)$  which are supposed to hold in the bare theory. 
It is possible, however, to find the WF FP for all values of $\t{Y}\in [0,2]$ numerically in the restricted parameter space $(\t{g}_2,\t{g}_4,\t{v}_3)$ for given values of $Z_A$ $(A=\pe,~\pa)$ with the same root-finder method which was applied in the approximation LPA1.  Nevertheless, we could  estimate  the values $Z_A^*$ at the WF FP  in the following manner. We have taken through various test values of  $Z_A$, determined the corresponding positions  $(\t{g}_2^*(Z_A),
\t{g}_4^*(Z_A),\t{v}_3^*(Z_A)) $ of the WF FP in the 
restricted parameter space, took these test values as initial conditions 
  at $k=10^{-4}$ for solving the flow equations \eq{setflow} in the approximation NLO1 towards the UV scale $\Lambda=1$. The couplings $Z_A$ retained a constant value $Z_{A~\Lambda}$ in the entire UV scaling region.
 Finally, we identified the WF FP as
$(\t{g}_2^*(Z_A^*),\t{g}_4^*(Z_A^*),\t{v}_3^*(Z_A^*), Z_A^*) $ for those $Z_A=Z_A^*$ for which both differences $|Z_{A~\Lambda}-(-1)|$  became smaller than $\ord{10^{-4}}$. Thus we get $Z_\pa^* = -1 + 2\times 10^{-3}$, $Z_\pe^* = -1+ 10^{-3}$ for $\t{Y} > 1$ and $Z_\pa^* = -1 - 2\times 10^{-3}$, $Z_\pe^* = -1 - 10^{-3}$ for $\t{Y} < 1$. Concluding, it seems to exist a 2-dimensional critical surface in approximation NLO1, and the point identified on it as the WF FP is the one which attracts the trajectories corresponding to bare theories with $|Z_{A~\Lambda}-(-1)|\le 0.0001$.

Due to the fact that the FP values of the wave function renormalizations
are slightly different of their bare values $-1$, the position of the WF FP
and the FP values  $\t{\sigma}_0^*$ and $\t{\sigma}_2^*$ determining the amplitude of the periodic condensate in the critical point differ only slightly from those obtained in the approximation LPA1: a change  $|\Delta Z_A^*| \approx 10^{-3}$ modifies  $\t{g}_4^*$, $\t{g}_2^*$, and $\t{v}_3^*$   in their second, third, and fourth valuable digits, respectively. As compared to the approximations LPA0 and LPA1, the approximation NLO1 provides information on the critical dimensions $\eta_A =- Z_A^{*-1} \beta_{Z_A}( \{ \t{c}_i^{[WF]} \})$ $(A=\pe,~\pa)$ in the vicinity of the WF FP. The plots in the first row in Fig.
\ref{fig:nlocrexp} illustrate the $\t{Y}$-dependence of the anomalous dimensions $\eta_A$  $(A=\pe,~\pa)$. For $\t{Y}$ decreasing from 2 towards 1, $\eta_A$ are negative for the ghost model and the magnitudes $|\eta_A|$ decrease strictly monotonically to zero.  The scaling of $Z_A$ in the UV regime - which is essentially constant - in the vicinity of the Gaussian FP yields $\eta_A^{[G]}=0$. For $\t{Y}$ decreasing from $1/2$ towards $ 0$ the anomalous dimensions  $\eta_A$ are positive and  show a decreasing trend. The anomalous dimension $\eta_\pa$ for the longitudinal mode  is generally a few times larger in magnitude than $\eta_\pe$ for the perpendicular modes. Also the first two eigenvalues, $\omega_i$ $(i=0,1)$  of the stability matrix at the WF FP have been determined in NLO1
restricting the matrix to the reduced parameter space $(\t{g}_2,\t{g}_4,\t{v}_3)$. The plots in the second row in Fig. \ref{fig:nlocrexp} illustrate the effect of the run of the wave function renormalizations for various values of the higher-derivative coupling $\t{Y}$. It is found that the NLO effect on the critical exponents $\omega_i$ does occur only for $\t{Y}\le 1/2$. The differences between the results obtained in the approximations NLO1 and LPA1 and shown in Fig.
 \ref{fig:nlocrexp} are rather small, because the  change $|\Delta Z_A^*| \approx 10^{-3}$ modifies the values of $\eta$ and $\omega_i $ $(i=0, )$ only in their third and fifth valuable digits, respectively. This is in agreement with the findings in Refs. \cite{Reuter2000,Reuter2013} that the non-trivial saddle point of the path integral, and correspondingly here the non-trivial inhomogeneous minimum of the EAA determines the low-energy  behaviour of the system and the role of quantum fluctuations is small.

\subsection{Phase structure and flow of the various couplings}\label{phstruc}

The phase structure of the model has been investigated in the approximations
LPA0, LPA1, and NLO1, while the results obtained in LPA0 were helpful to find typical trajectories corresponding to the various phases of the model in the other approximations. It has been established that the approximations used provide qualitatively the same phase structure as well as qualitatively  similar scaling
laws characterizing the particular phases. The smallness of the quantitative differences is again in favour of the findings of the authors of Refs. \cite{Reuter2000,Reuter2013} mentioned at the end of the previous Section.  Therefore we discuss in details below the results obtained  in the approximation NLO1. It is a peculiar feature of the RG evolution of the potentials of the ghost $O(1)$ model in the LPA that there is only tree-level scaling for $\t{Y}=\hf Z\equiv -\frac{1}{4} (Z_\pe+ Z_\pa)=\hf$ without any quantum corrections.
This happens because the usage of Litim type regulators results in occurring the factor $Z+2\t{Y}$ in front of the terms describing the loop corrections in Eqs. \eq{ulpadiml} and \eq{potlpadiml}. We shall see that in NLO1 with the approximations used by us the scale-dependence of the wave function renormalizations $Z_{\pe~k}$ and $Z_{\pa~k}$ is rather weak, a few tenths of per cents. Therefore one may expect that the phase structure  on one side of the hypersurface $\t{Y}\approx 1/2$  is different of that on the other side. Since the approximations used the coupling $\t{Y}$ does not run, we have mapped out the phase structure
in various sections of the parameter space taken at given values of $\t{Y}$.
Taking several test trajectories in the approximation NLO1, it has been checked numerically  that the inclusion of the term $\rho_k(p_\pe^2, p_\pa^2)$, given in Eq. \eq{rhonlo},  in the regulator's scale-derivative  has a negligible effect  on the run of the RG trajectories, so that in the detailed analysis the term $
 \rho_k(p_\pe^2, p_\pa^2)$ has been neglected.

\begin{figure}
\centerline{
\psfig{file=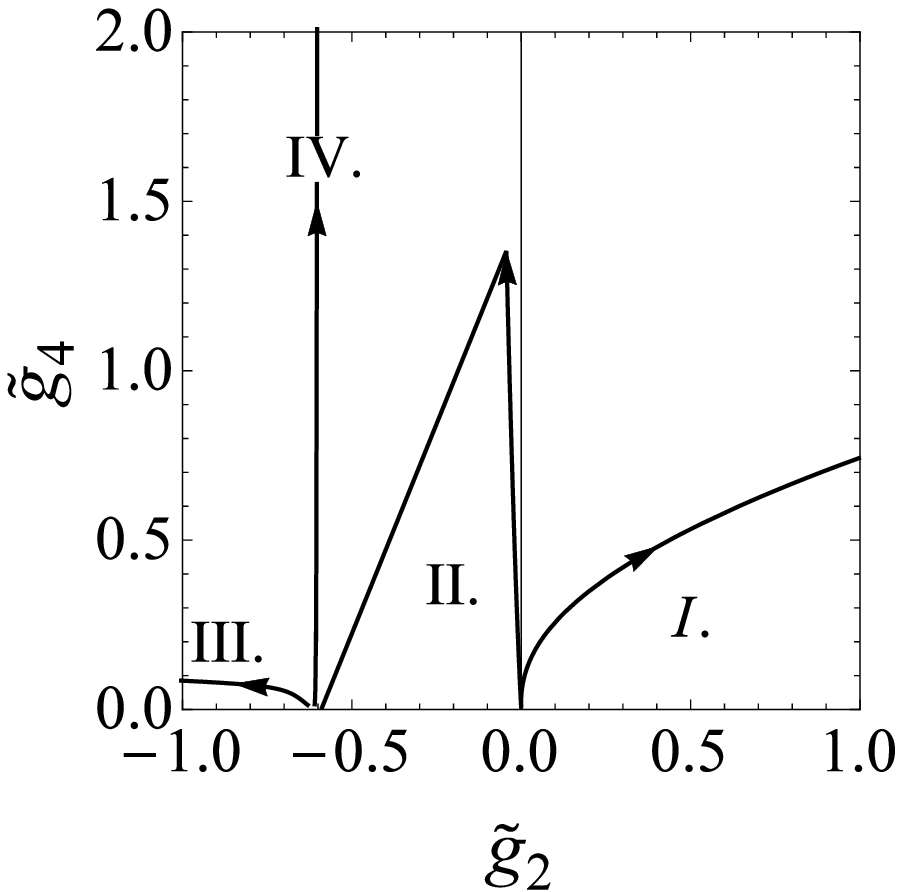,height=3.52cm,width=4.04cm,angle=0}
\psfig{file=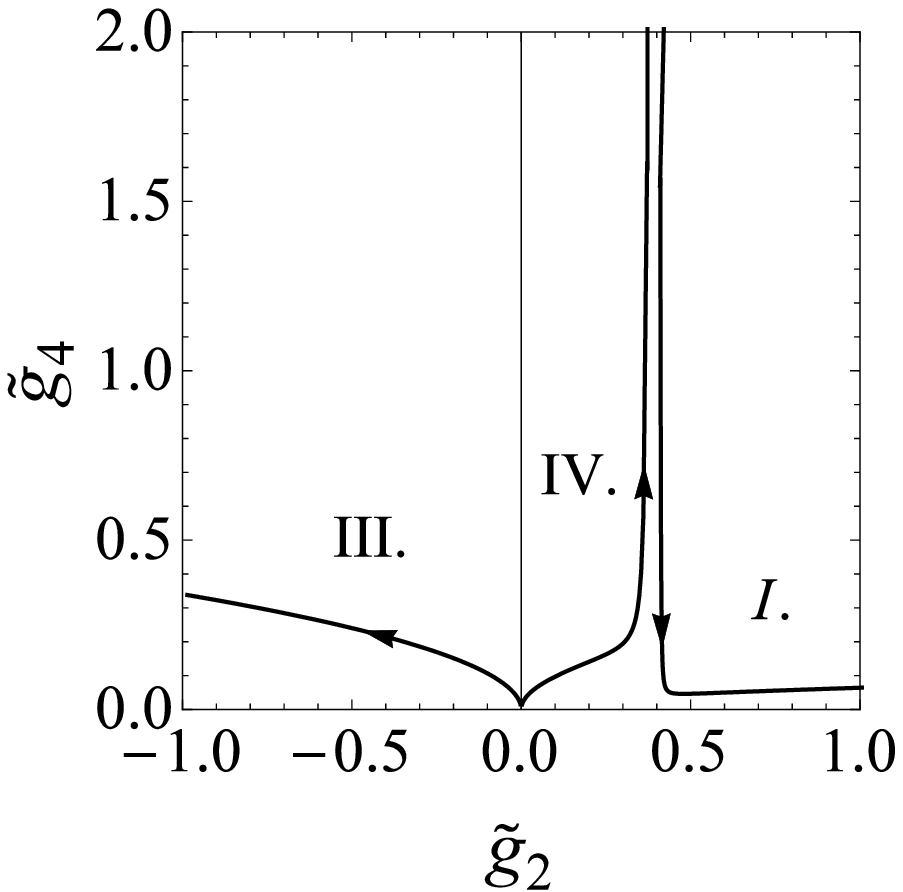,height=3.52cm,width=4.04cm,angle=0}
\psfig{file=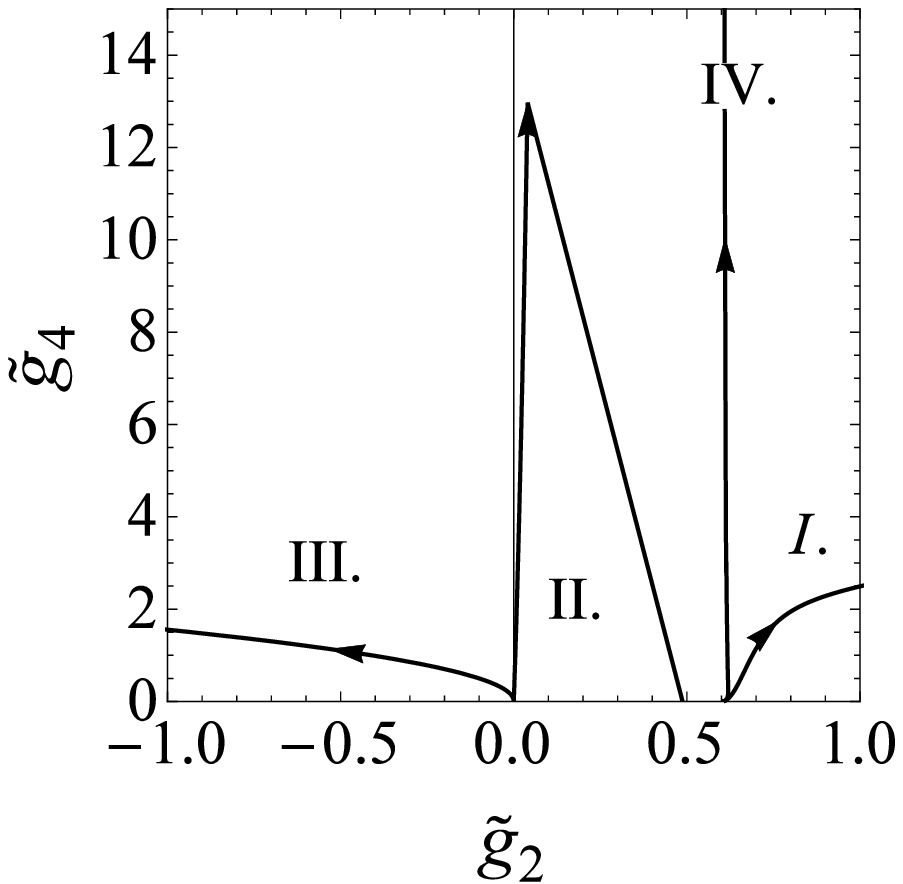,height=3.52cm,width=4.04cm,angle=0}
}
\caption{\label{fig:phdnlo} Various sections of the phase diagram of the ghost $O(1)$ model  for $\t{Y}=1.6$, $0.6$, and $0.4$ in the approximation  NLO1. The various phases are indicated by typical RG trajectories.}
\end{figure}
Typical RG trajectories were determined numerically for  various constant values of $\t{Y}$ and projected onto the 2-dimensional plane $\c{P}=(\t{g}_2,\t{g}_4)$
of the parameter space. The corresponding projected sections of the phase diagram are shown in Fig. \ref{fig:phdnlo} obtained in the approximation NLO1. The $\t{Y}$ values were chosen from the intervals $I_+\equiv (1,2]$, $I_0\equiv (1/2, 1)$, and $I_-\equiv (0,1/2)$ in which the WF FP occurs at positive, negative, and again positive values of the quartic coupling $\t{g}_4^*$ of the ordinary potential, respectively (as shown on the second plot in the first row  in Fig. \ref{fig:lpawf}). On these projected sections the Gaussian FP is in the origin of the plane $\c{P}$, there is a line of singularity  at $\t{g}_2=-Z -\t{Y}\approx 1-\t{Y}$, and also the WF FP is present for $\t{Y}\in I_\pm$. 
The parameter regions belonging to the same phase on the different projected sections were identified by the close similarity of the IR scaling laws of the various
couplings, found to be qualitatively the same in all of the used approximations LPA0, LPA1, and NLO1. Finally the existence of the following phases have been
established.
\begin{itemize}
\item 
\begin{figure}[htb]
\centerline{
\psfig{file=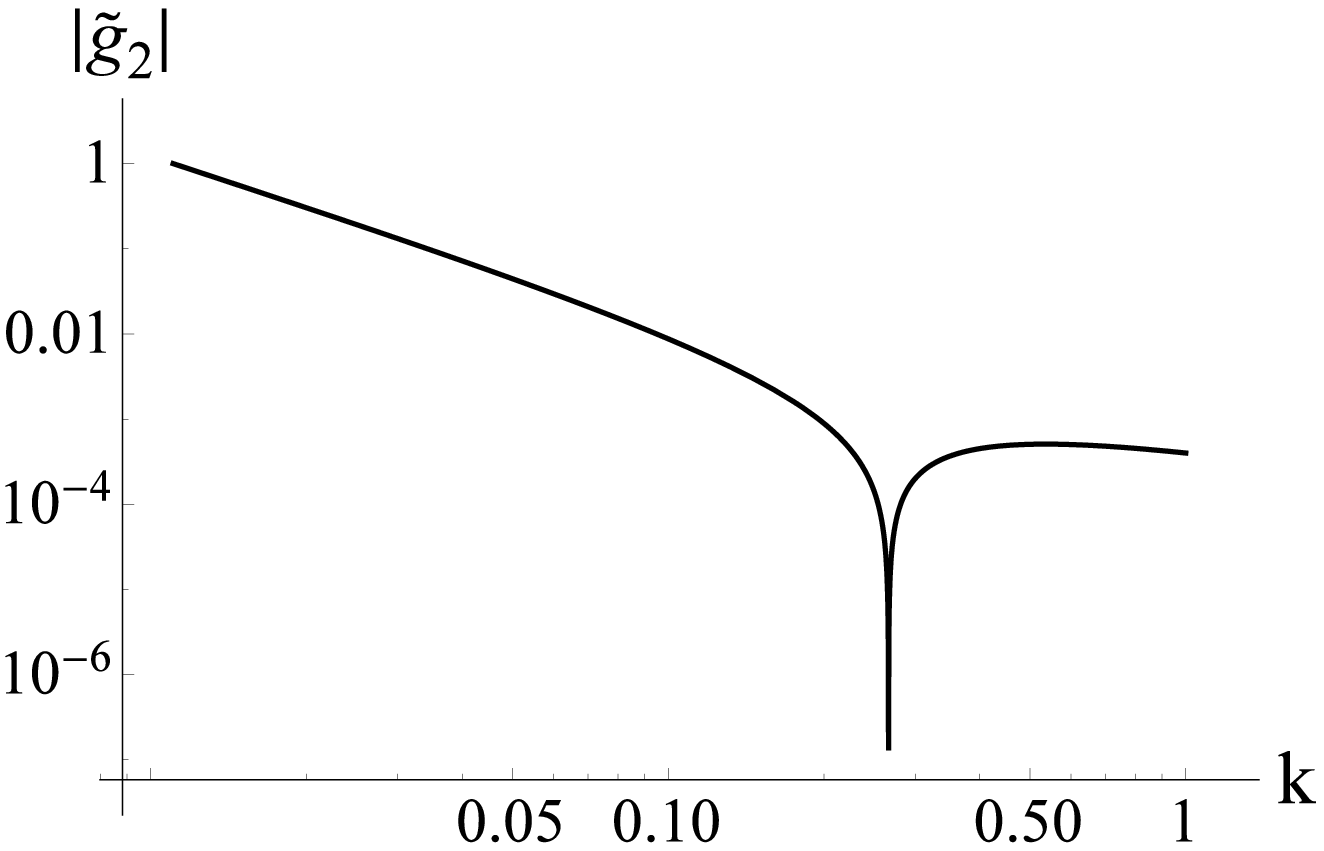,height=3.52cm,width=4.04cm,angle=0}
\psfig{file=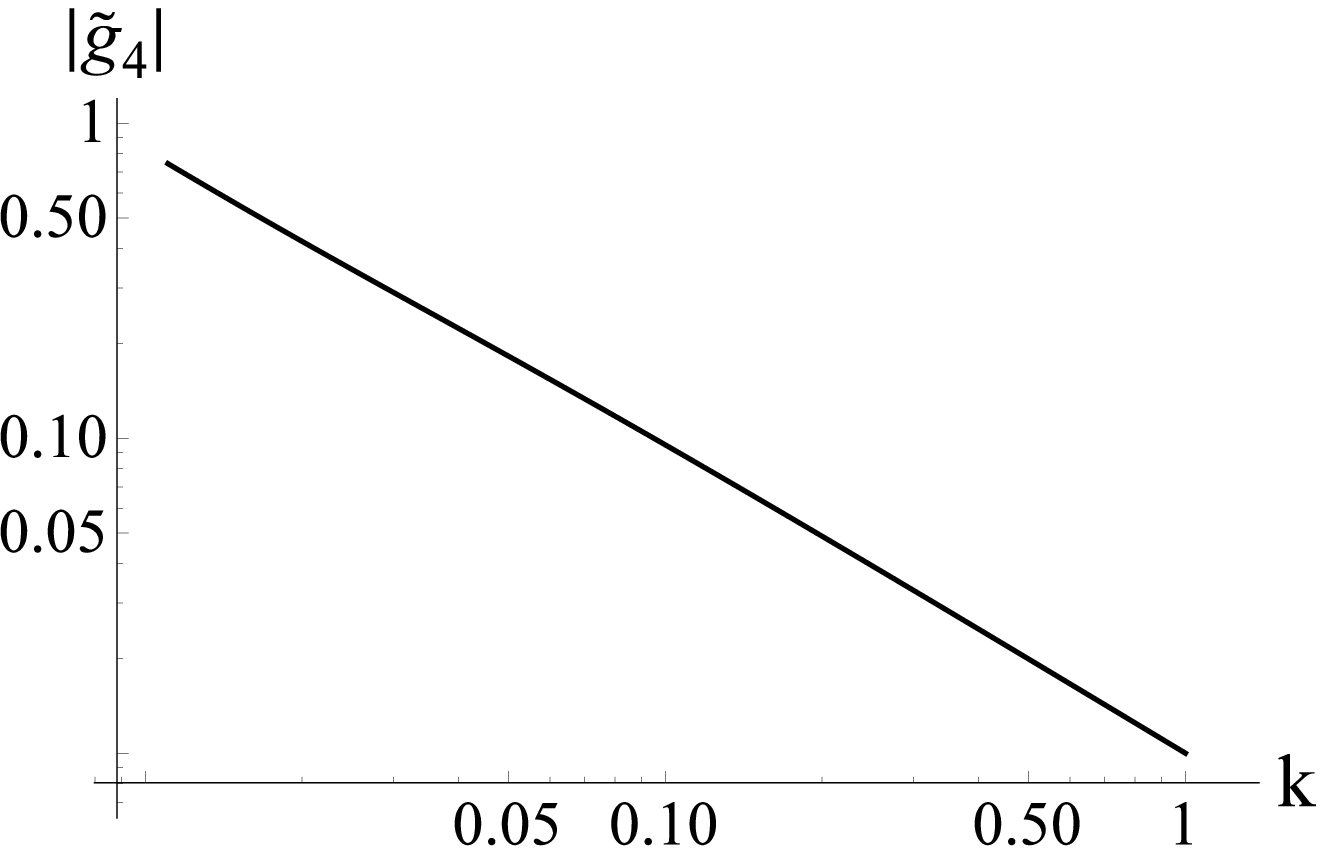,height=3.52cm,width=4.04cm,angle=0}
\psfig{file=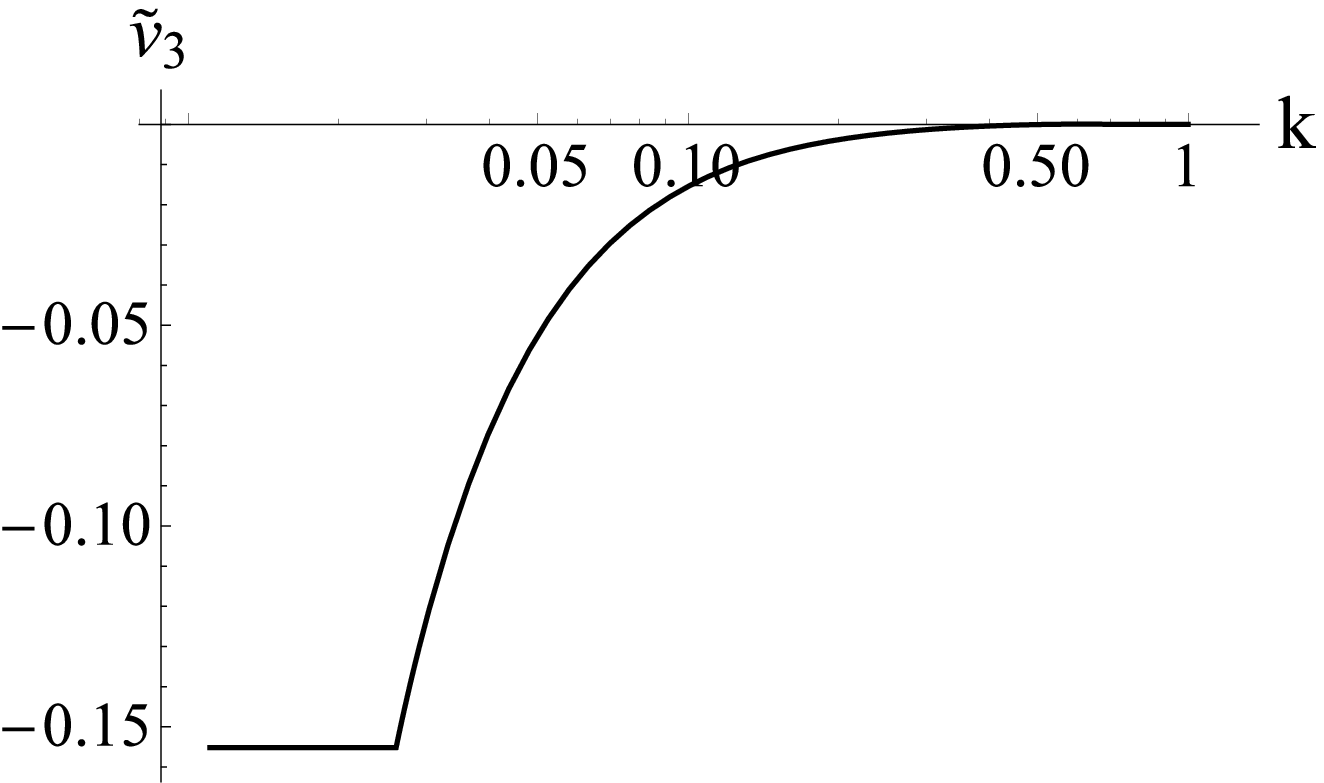,height=3.52cm,width=4.04cm,angle=0}}
\centerline{
\psfig{file=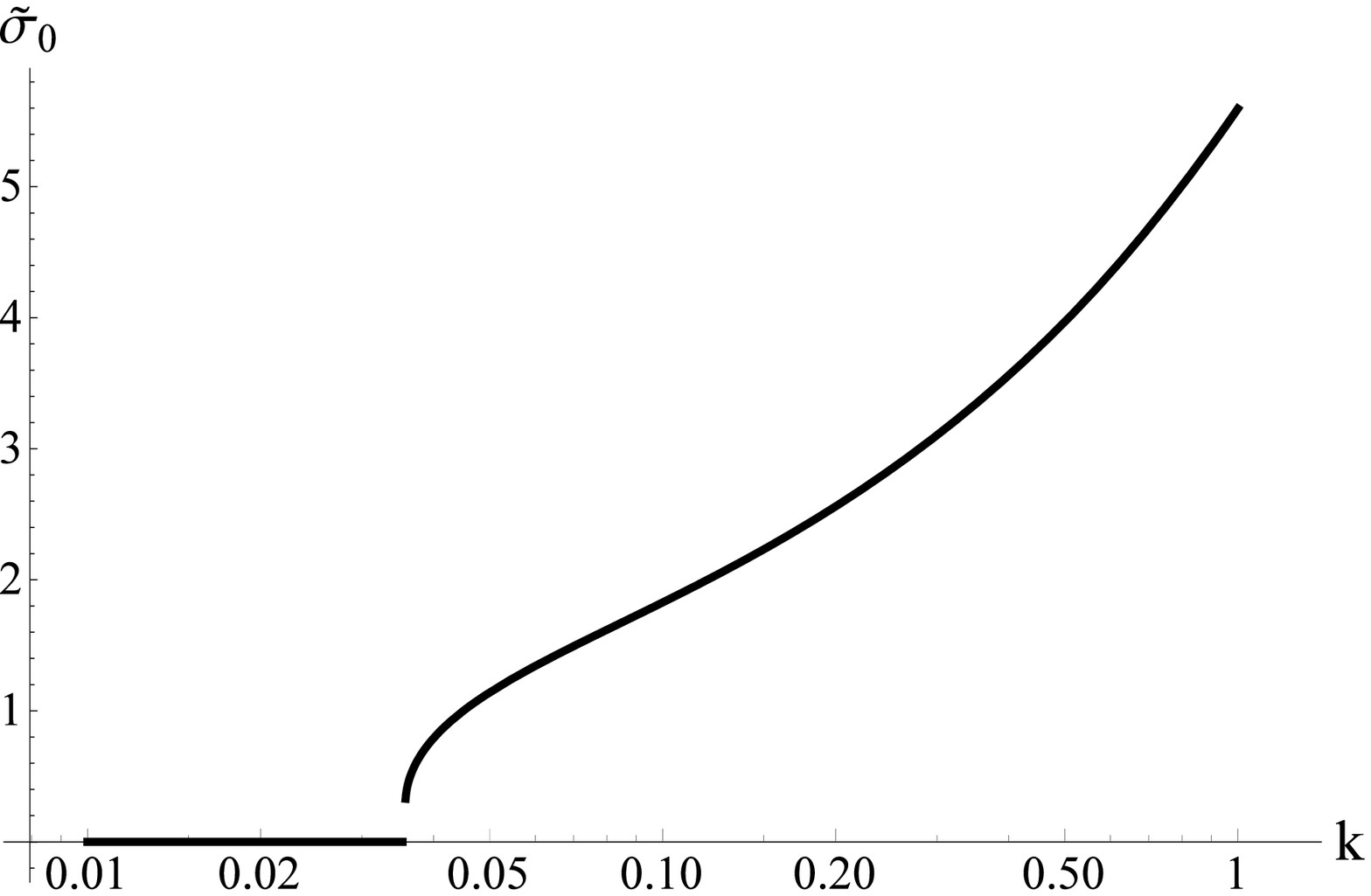,height=3.52cm,width=4.04cm,angle=0}
\psfig{file=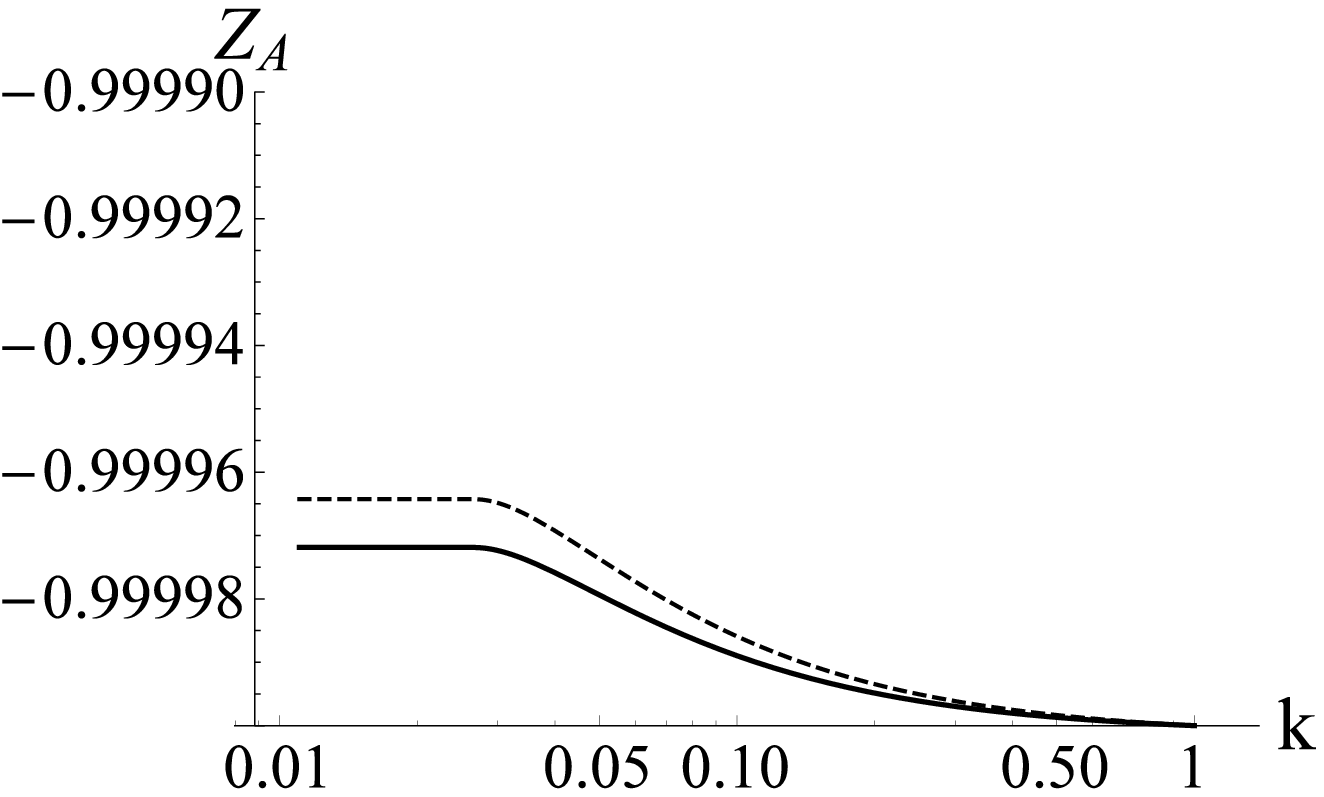,height=3.52cm,width=4.04cm,angle=0}
\psfig{file=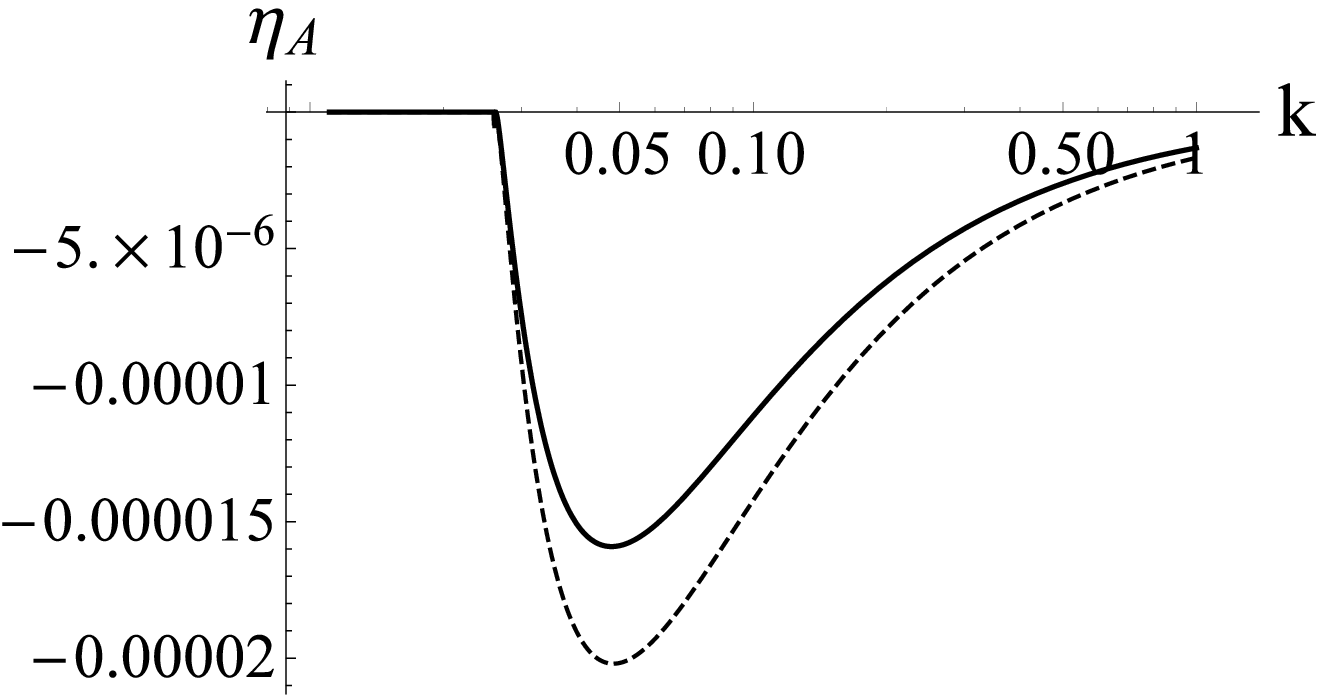,height=3.52cm,width=4.04cm,angle=0}
}
\caption{\label{fig:NLO_phase1_y16} Flows of the dimensionless couplings $\t{g}_2$, $\t{g}_4$, $\t{v}_3$, the amplitude of the periodic condensate $\t{\sigma}_0$, the wave functions renormalizations and the corresponding anomalous dimensions  $Z_\pe$, $\eta_\pe$ (full lines) and $Z_\pa$, $\eta_\pa$ (dashed lines)  for $\t{Y}=1.6$ along trajectories belonging to Phase I, obtained in the approximation NLO1.
}
\end{figure}  
 Phase I occurs for all values $\t{Y}\in (0,2]$. The typical scaling of the various couplings are shown in Fig. \ref{fig:NLO_phase1_y16} for $\t{Y}=1.6$. The dimensionless couplings of the ordinary potential exhibit tree-level IR scaling laws, $\t{g}_2\sim c_2 k^{-2}$ and $\t{g}_4\sim c_4k^{-1}$ with the constants $c_2,~c_4>0$,  so that the dimensionful couplings take finite
positive IR limits, $g_2(0)=c_2>0$ and $g_4(0)=c_4>0$. The dimensionful ordinary  potential is convex in the IR limit, having its   trivial minimum at $\Phi_*=0$.  The UV scaling region is generally rather short and somewhat different for
the various intervals $I_+$, $I_0$, and $I_-$. For $\t{Y}\in I_+$ the trajectories run out from the Gaussian FP, while for $\t{Y}\in I_0$ and  $\t{Y}\in I_-$
they move off the singularity line for decreasing scale $k$. In the UV region generally occurs the periodic condensate with a non-vanishing amplitude $\t{\sigma}_0$ for $\Phi=0$, but it dies out when the IR scaling sets on leaving behind a
non-vanishing dimensionless induced potential (the non-vanishing value of $\t{v}_3$) as a relict. Therefore no periodic condensate is present in the IR limit.
Thus Phase I can be identified with the symmetric phase of the 3-dimensional ghost $O(1)$ model. The wave function renormalizations $Z_A$ $(A=\pe,~\pa)$ exhibit monotonic scale-dependence in the UV region and become constant in the IR region where $\t{\sigma}_0=0$, but the overall change is rather small, $|\lim_{k\to 0}Z_{A~k} -(-1)|<10^{-4}$. Correspondingly, the anomalous dimensions $\eta_A$ vanish in the IR region. For  $\t{Y}\in I_+$ the RG trajectories belonging to Phase I emanate from the Gaussian FP, so that the flow of the couplings is perturbative in the `potential piece' of the reduced propagator, but the amplitude $\t{\sigma}_0$ behaves non-perturbatively for $\t{g}_4\to 0$. For $\t{Y}\in I_0$ the RG trajectories start from the singularity $\t{g}_2\to -Z-\t{Y}$, $\t{g}_4\to \infty$, with a sudden decrease of $\t{g}_4$ approach the IR singular point
 $\t{g}_2\to -Z-\t{Y}$, $\t{g}_4=0$ which acts as a crossover point and are repulsed from it in the IR region. For $\t{Y}\in I_-$ the typical RG trajectories approach the singular point  $\t{g}_2\to -Z-\t{Y}$, $\t{g}_4=0$ in the UV limit. Nevertheless, the IR scaling laws are in each of these cases qualitatively identical.

\item
\begin{figure}[htb]
\centerline{
\psfig{file=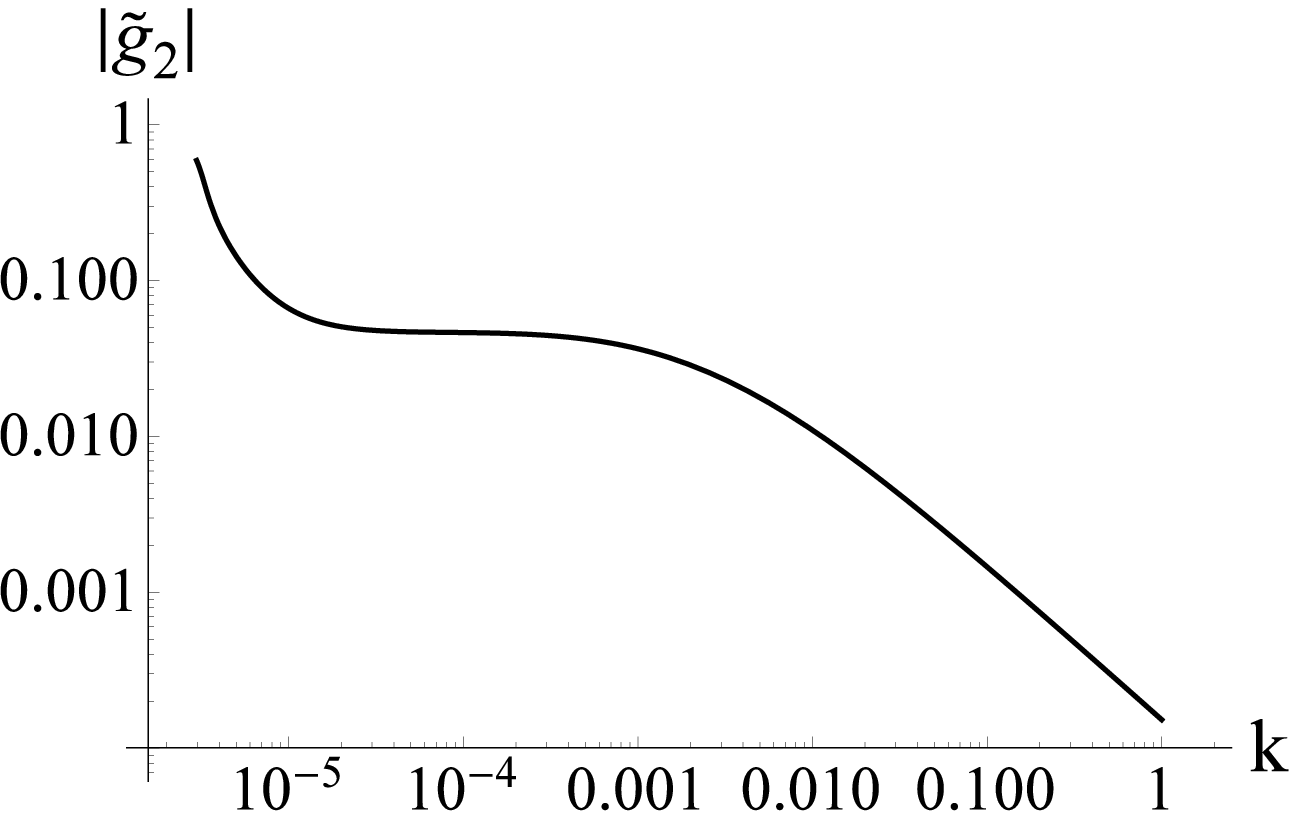,height=3.52cm,width=4.04cm,angle=0}
\psfig{file=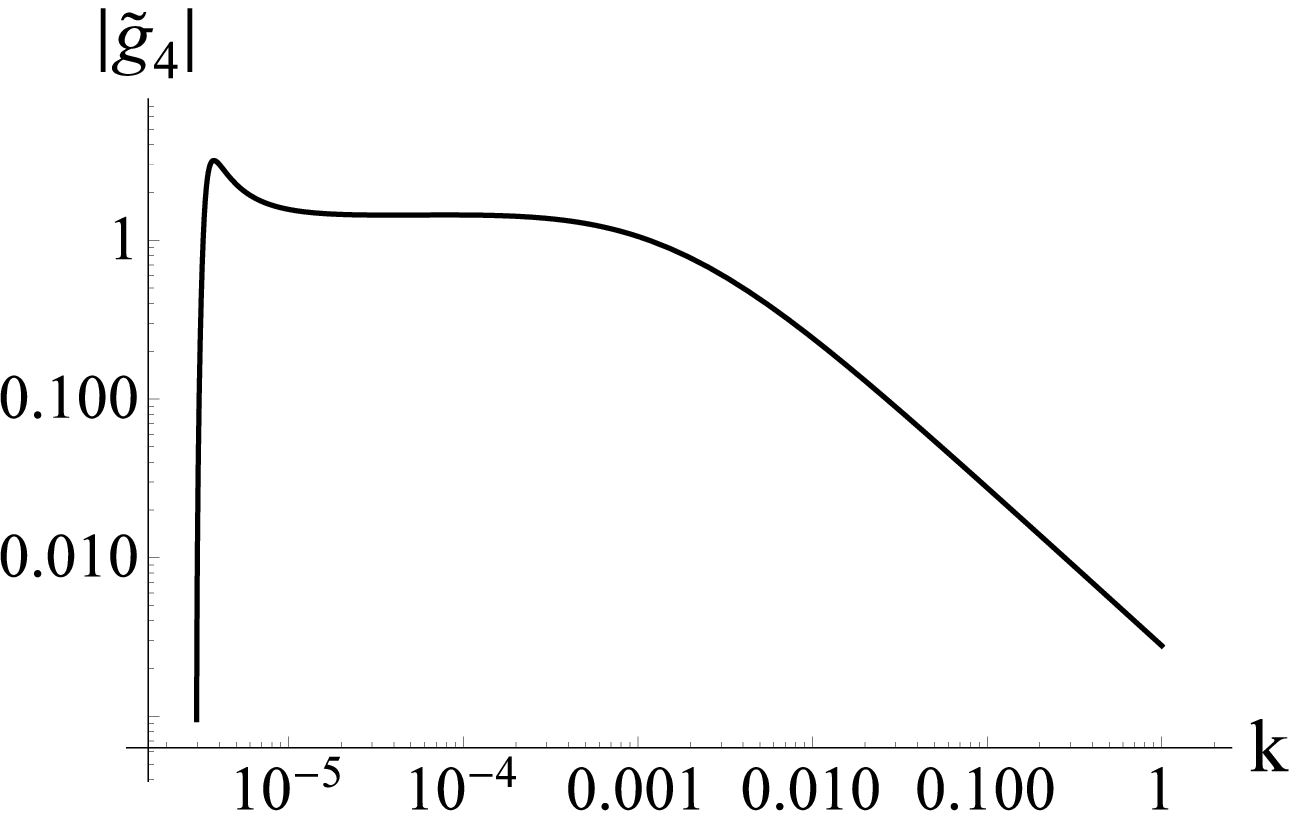,height=3.52cm,width=4.04cm,angle=0}
\psfig{file=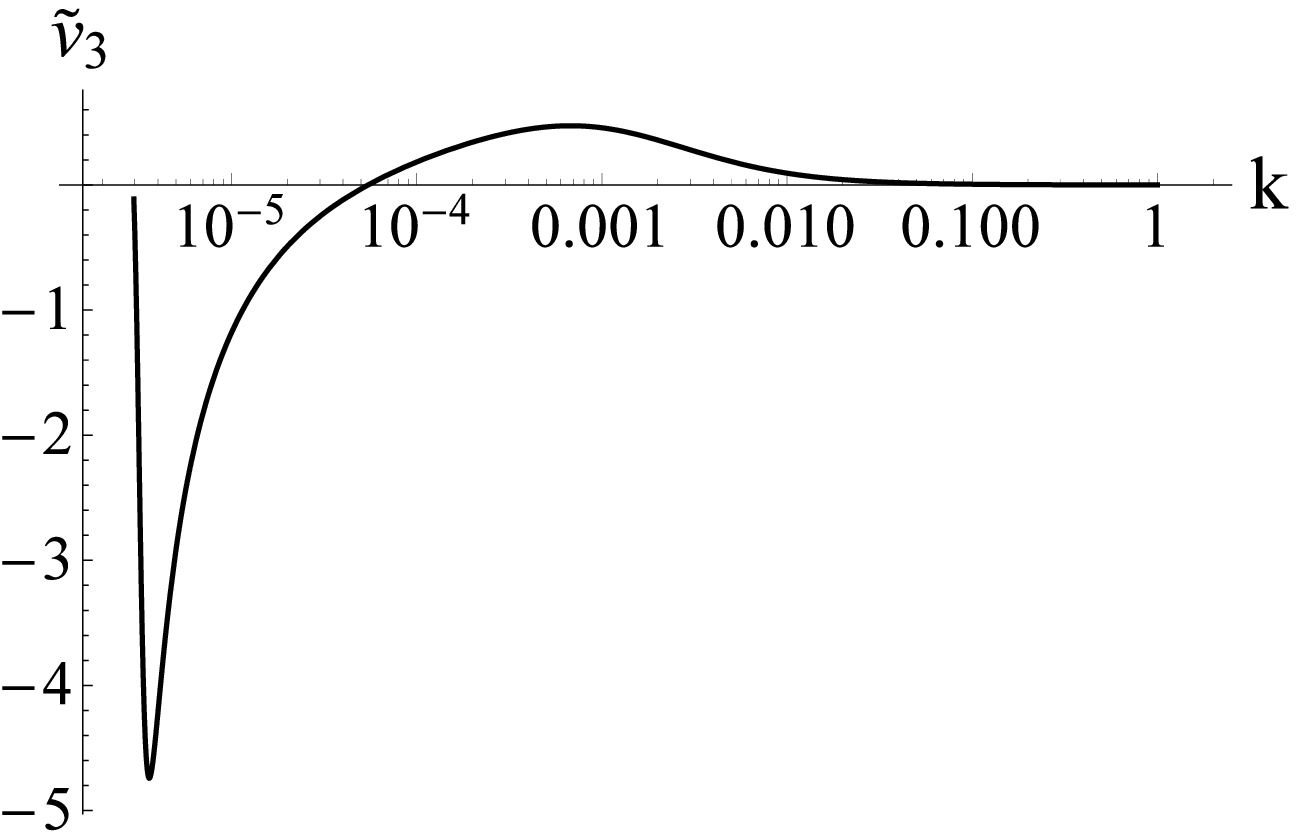,height=3.52cm,width=4.04cm,angle=0}}
\centerline{
\psfig{file=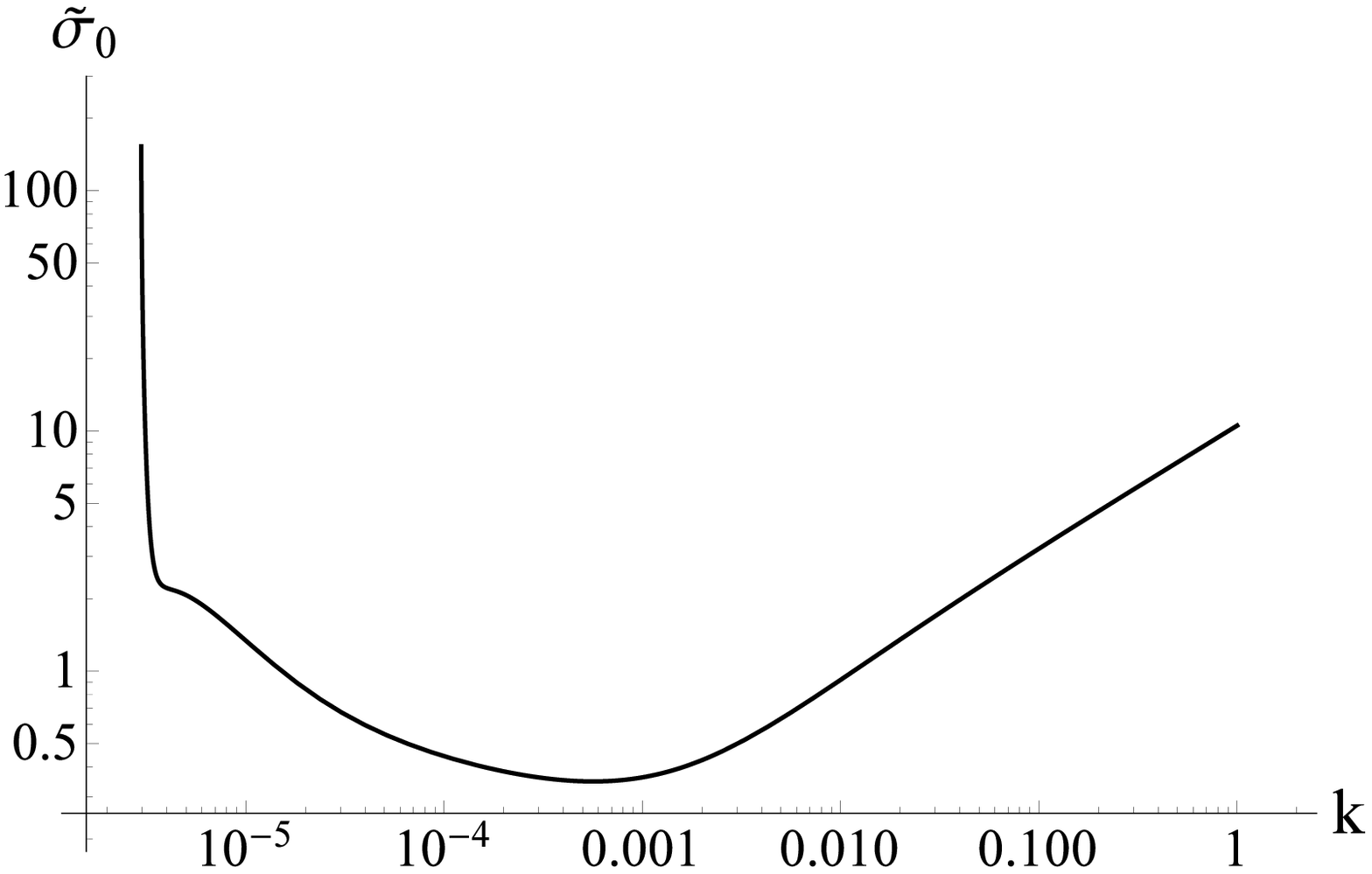,height=3.52cm,width=4.04cm,angle=0}
\psfig{file=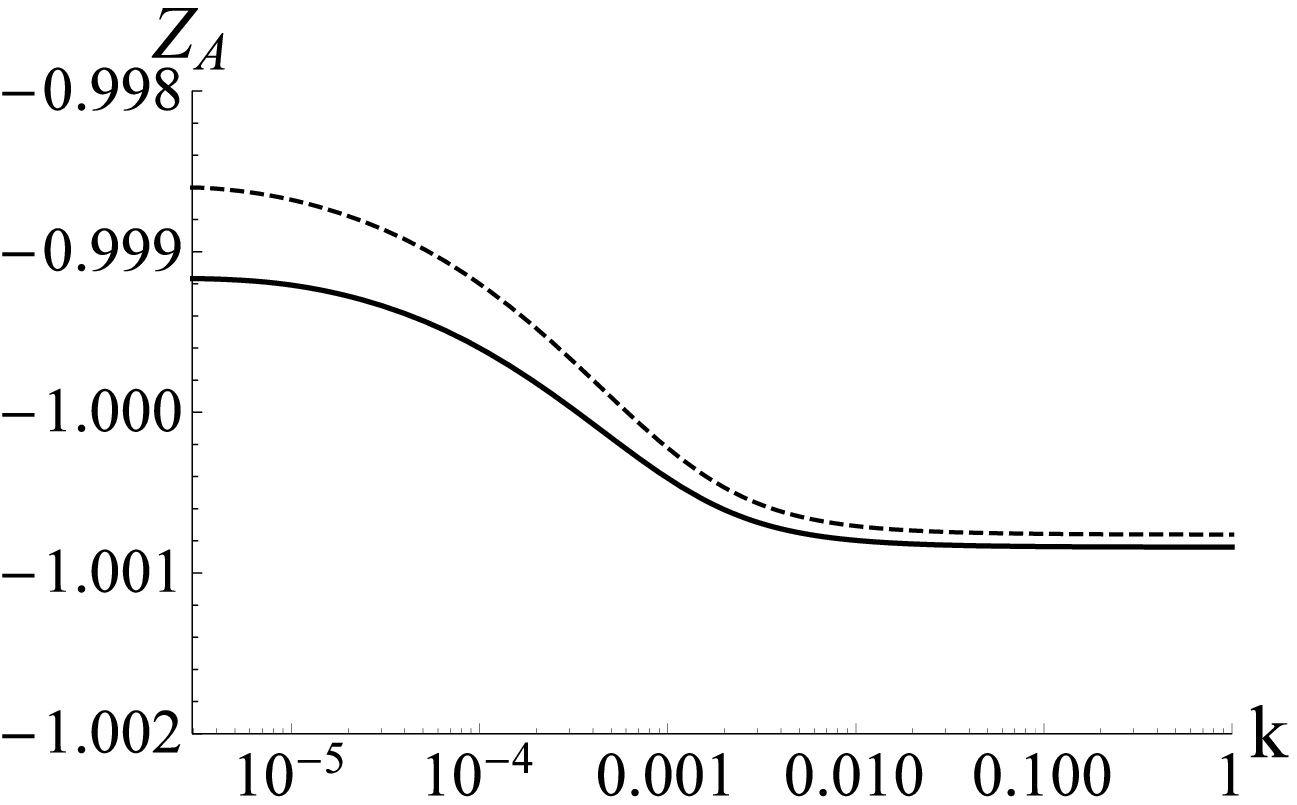,height=3.52cm,width=4.04cm,angle=0}
\psfig{file=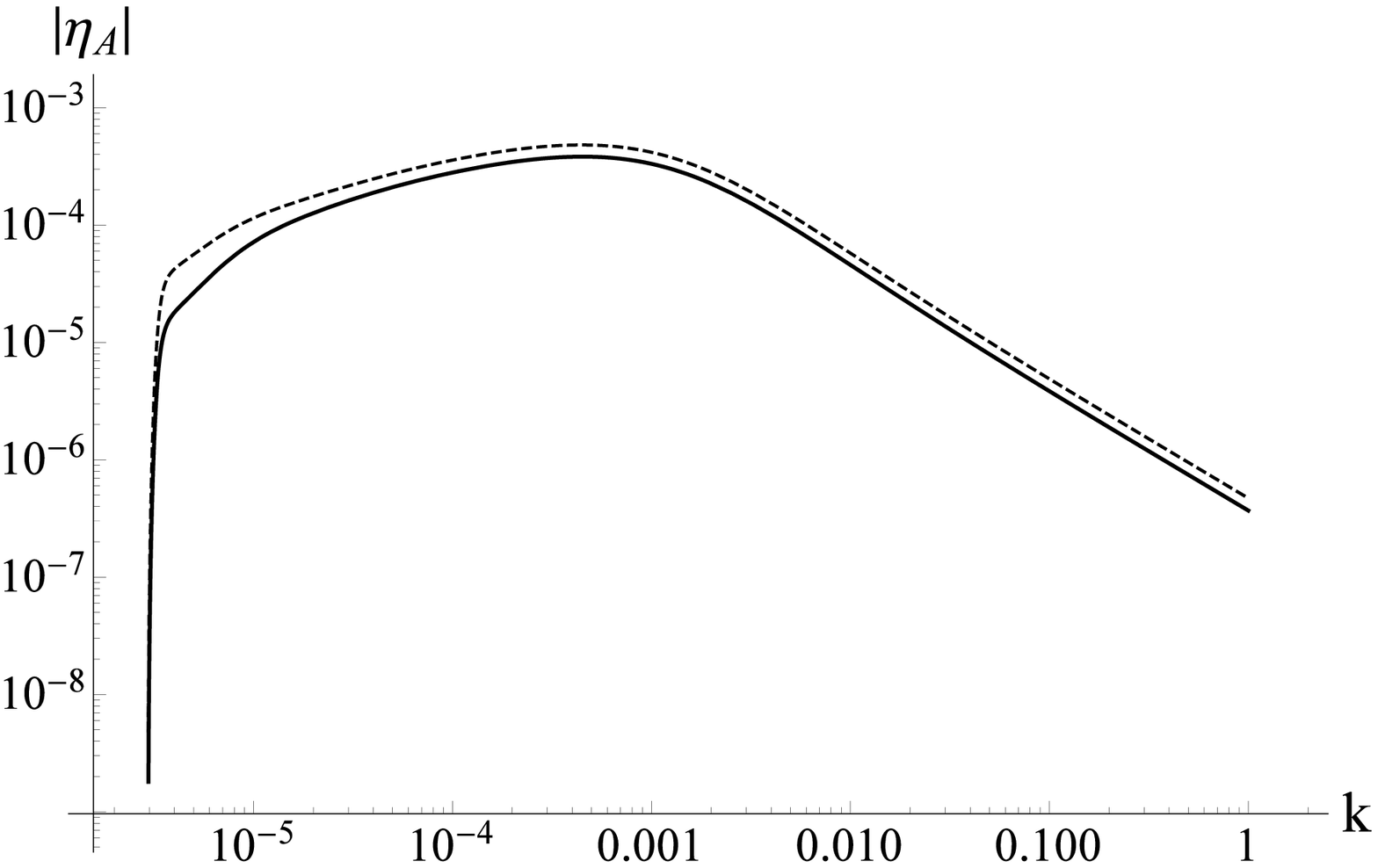,height=3.52cm,width=4.04cm,angle=0}
}
\caption{\label{fig:NLO_phase2_y16} Flows of the dimensionless couplings $\t{g}_2$, $\t{g}_4$, $\t{v}_3$, the amplitude of the periodic condensate $\t{\sigma}_0$, the wave functions renormalizations and the corresponding anomalous dimensions  $Z_\pe$, $\eta_\pe$ (full lines) and $Z_\pa$, $\eta_\pa$ (dashed lines) for $\t{Y}=1.6$ along almost critical trajectories belonging to Phase II, obtained in the approximation NLO1.}
\end{figure}  
Phase II occurs for $\t{Y}\in I_+$ and $\t{Y}\in I_-$ when there is a WF FP with
positive quartic coupling $\t{g}_4^*>0$. The typical scaling of the various couplings along the almost critical trajectories are shown in Fig. \ref{fig:NLO_phase2_y16} for $\t{Y}=1.6$.  The almost critical trajectories start
from the Gaussian FP, pass by the WF FP and approach the singularity at $\t{g}_2
\approx 1-\t{Y}$ in the IR region. Correspondingly the scaling of the couplings  $\t{g}_2$ and $\t{g}_4$ exhibit an UV scaling region, which  is followed by the WF crossover region where $\t{g}_2$ and $\t{g}_4$ are nearly constant. Finally, a rather short IR region occurs when the flow approaches the singularity.  The 
presence of the `plateau' of the WF crossover region and the very good agreement of the plateau values with those computed from the FP equations  are quite similar to that in the symmetry broken phase of the ordinary 3-dimensional $O(1)$ model.
 In the ghost $O(1)$ model, however, there occurs  the periodic condensate with the amplitude $\t{\sigma}_0$ decreasing with decreasing scale $k$. In the crossover region $\t{\sigma}_0$ becomes nearly independent of the scale $k$ and it starts to increase suddenly when the trajectory approaches the singularity. The coupling $\t{v}_3$ of the induced potential shows up a slight scale-dependence in the UV region, but it increases significantly in magnitude for the crossover scales and suddenly dies out near the singularity. The scale-dependence of the wave function renormalizations is weak, but an order of magnitude stronger than in Phase I. The anomalous dimensions do not show up any plateau in the crossover region. One should, however, mention that the fine-tuning of the RG trajectories in order to get closer to the separatrix becomes rather involved in the 6-dimensional parameter-space.
Therefore the nearly critical trajectories where generated starting the flow
in the close neighbourhood of the WF FP and following it both with decreasing and increasing scales. This explains why the UV values $\t{Z}_{A~\Lambda}$ are different of $-1$ at the UV scale $k=\Lambda=1$ on the 2. plot in the second row of Fig. 
\ref{fig:NLO_phase2_y16}.  In Phase II the dimensionless ordinary potential is unversal in the IR limit $\t{g}_2\to -\hf (Z_{\pe~0} +Z_{\pa~0})-\t{Y}\approx 1-\t{Y}$, $\t{g}_4\to 0$ and the dimensionful ordinary  potential flattens out. There occurs a  periodic condensate in the crossover region, and its amplitude explodes in the IR limit (due to the factors $1/\t{g}_4$ in Cardano's formulas). It should be mentioned that quite similar scaling laws have been found numerically for $\t{Y}=0.4$ and $\t{Y}=0.2$, reflecting the crossover character of the WF FP.
This indicates that our finding that $\t{v}_3^*$ acquires imaginary part at the WF FP is rather the consequence of the strong truncation of the induced potential than a sign of some physics. The trajectories belonging to  Phase II emanate from the Gaussian FP, therefore the flow equations can be treated perturbatively in the UV region (the loop integrals can be expanded in powers of the potential piece of the reduced propagator), but the amplitude $\t{\sigma}_0$ of the periodic condensate behaves non-perturbatively for $\t{g}_4\to 0$.

\item 
\begin{figure}
\centerline{
\psfig{file=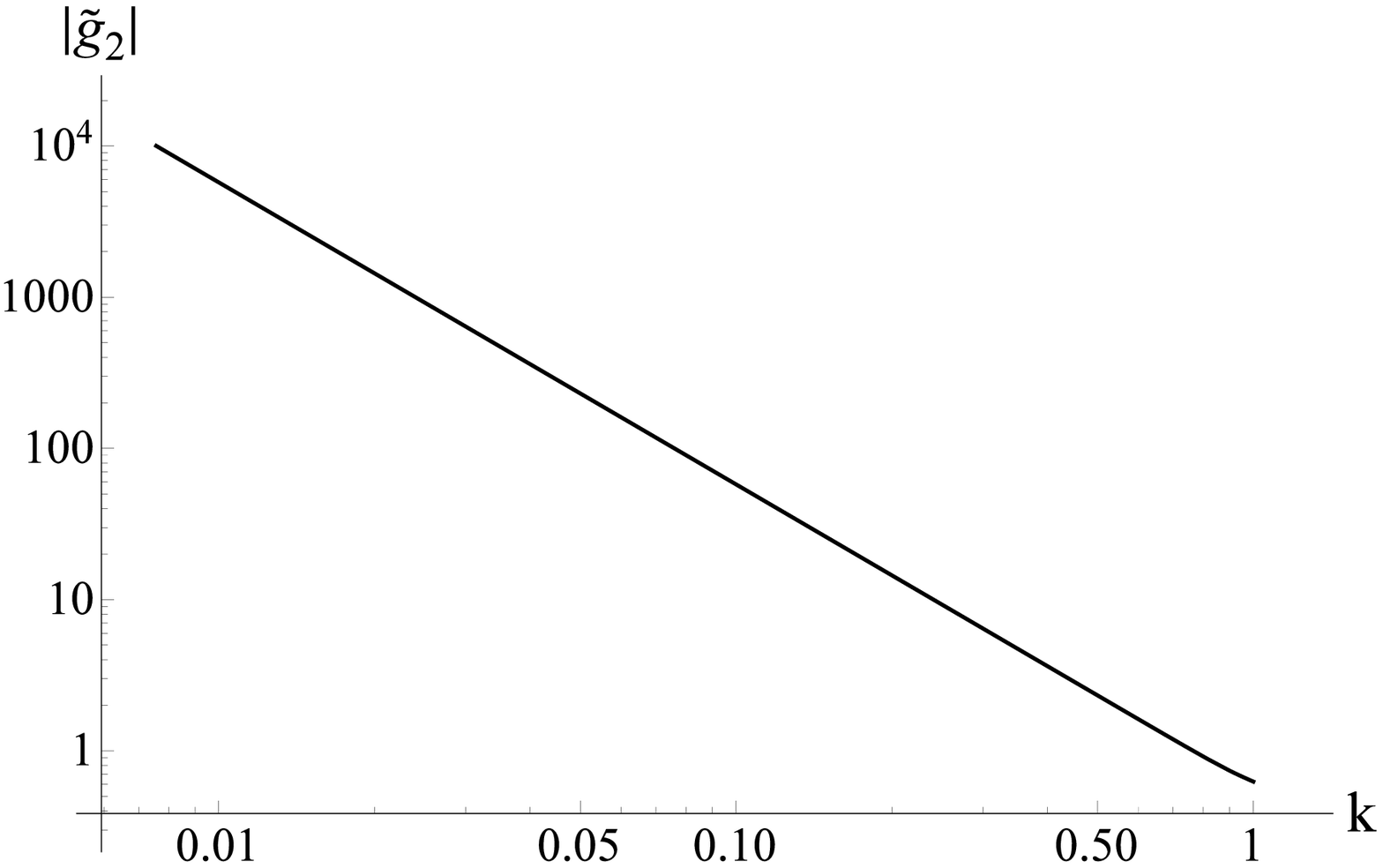,height=3.52cm,width=4.04cm,angle=0}
\psfig{file=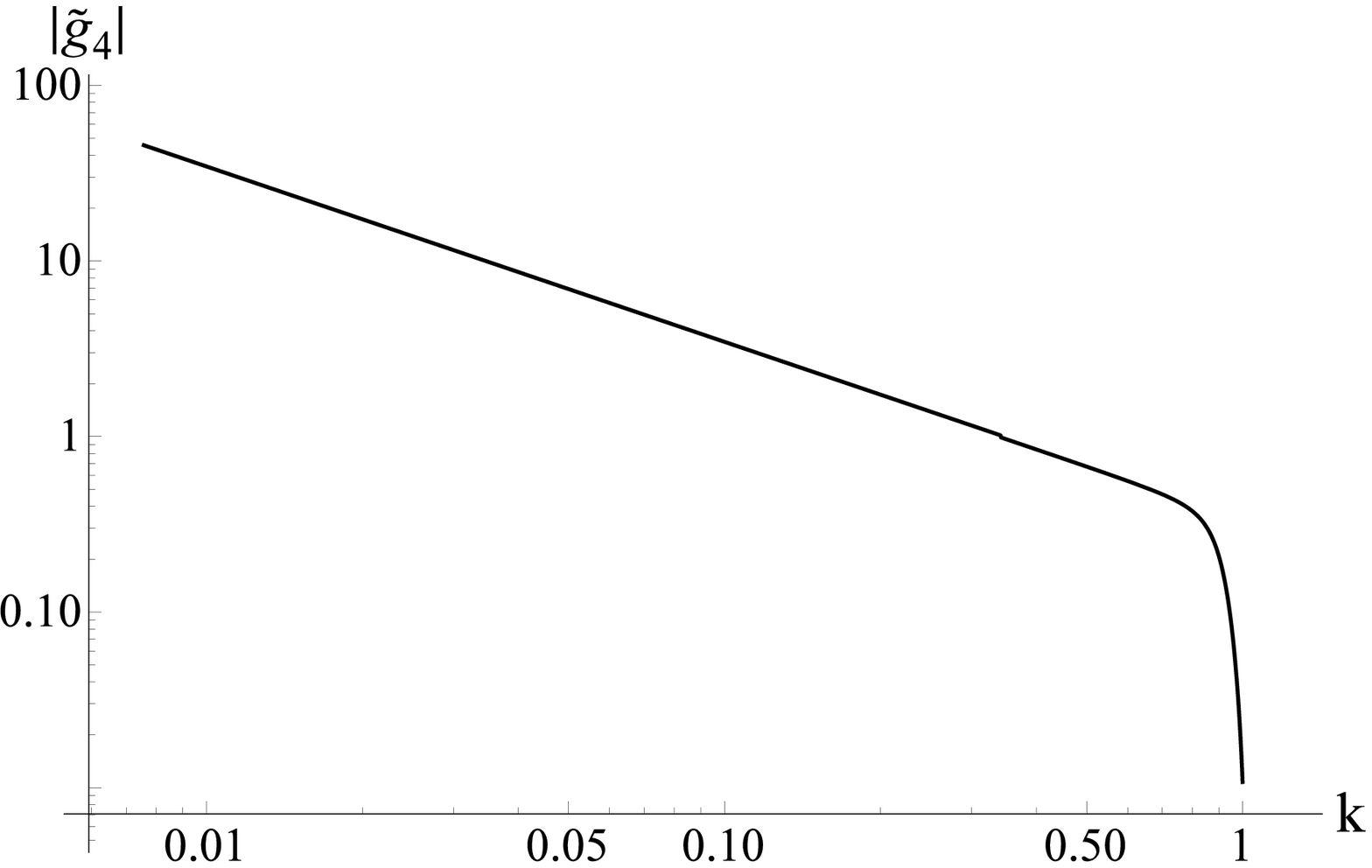,height=3.52cm,width=4.04cm,angle=0}
\psfig{file=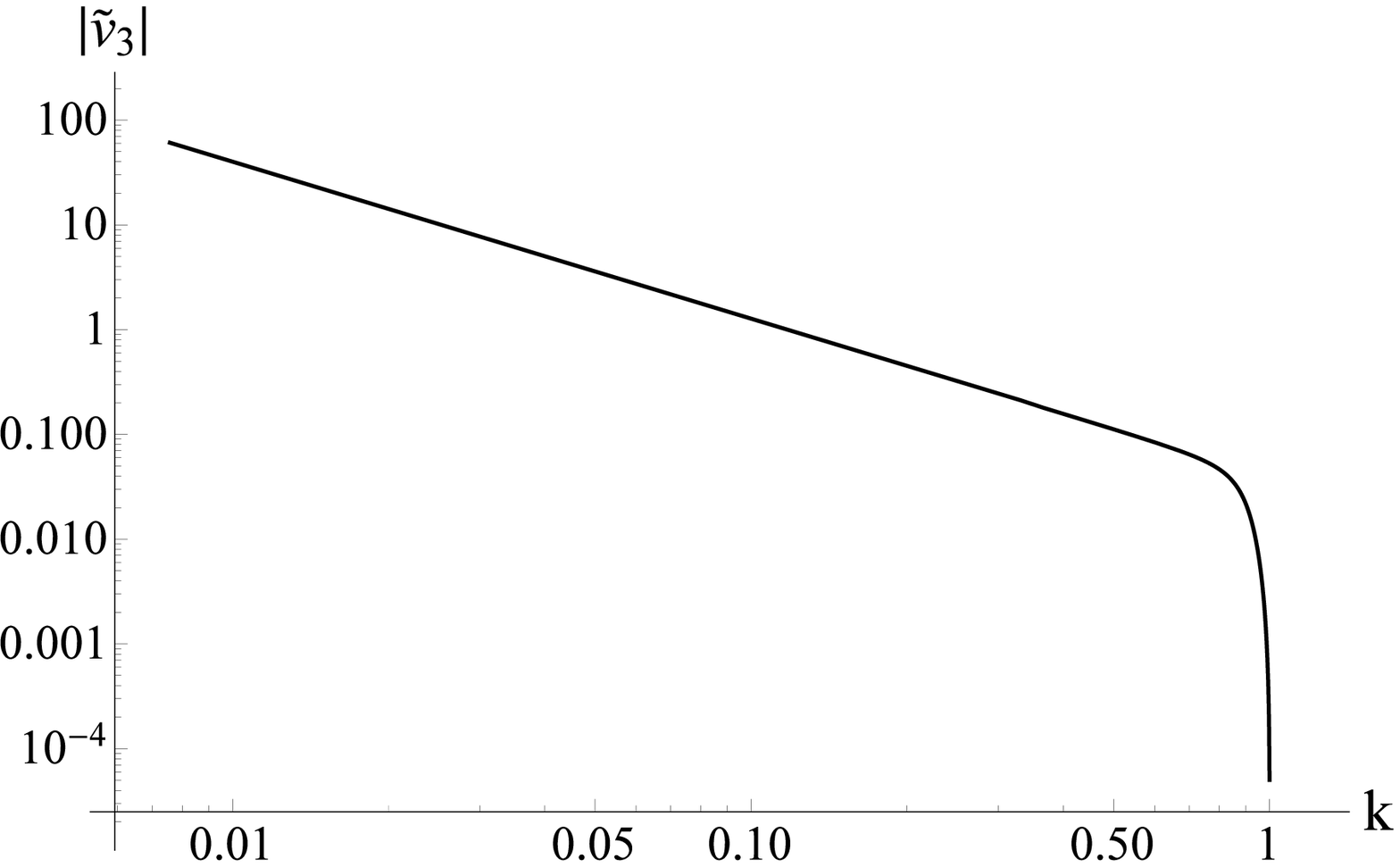,height=3.52cm,width=4.04cm,angle=0}}
\centerline{
\psfig{file=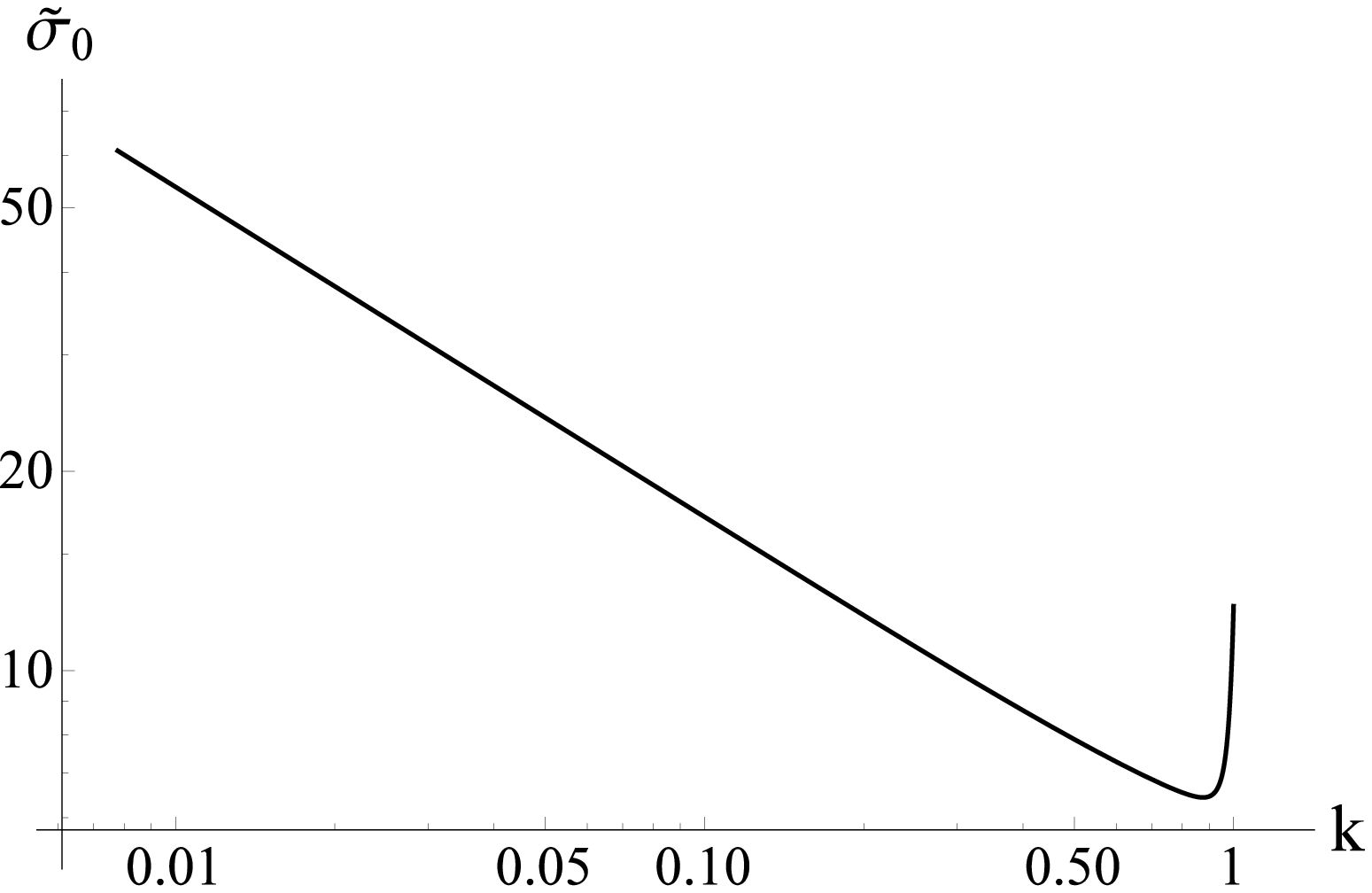,height=3.52cm,width=4.04cm,angle=0}
\psfig{file=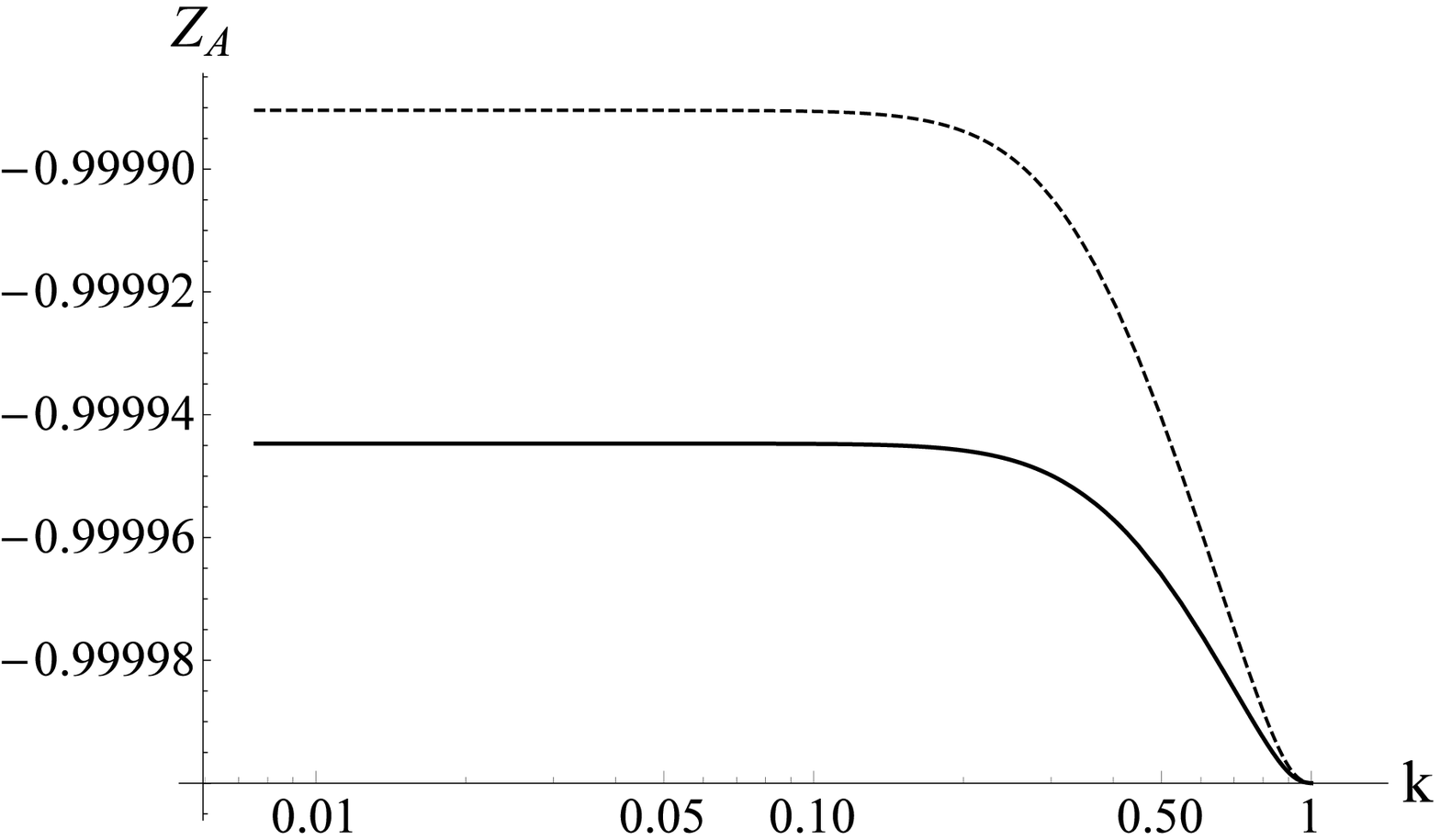,height=3.52cm,width=4.04cm,angle=0}
\psfig{file=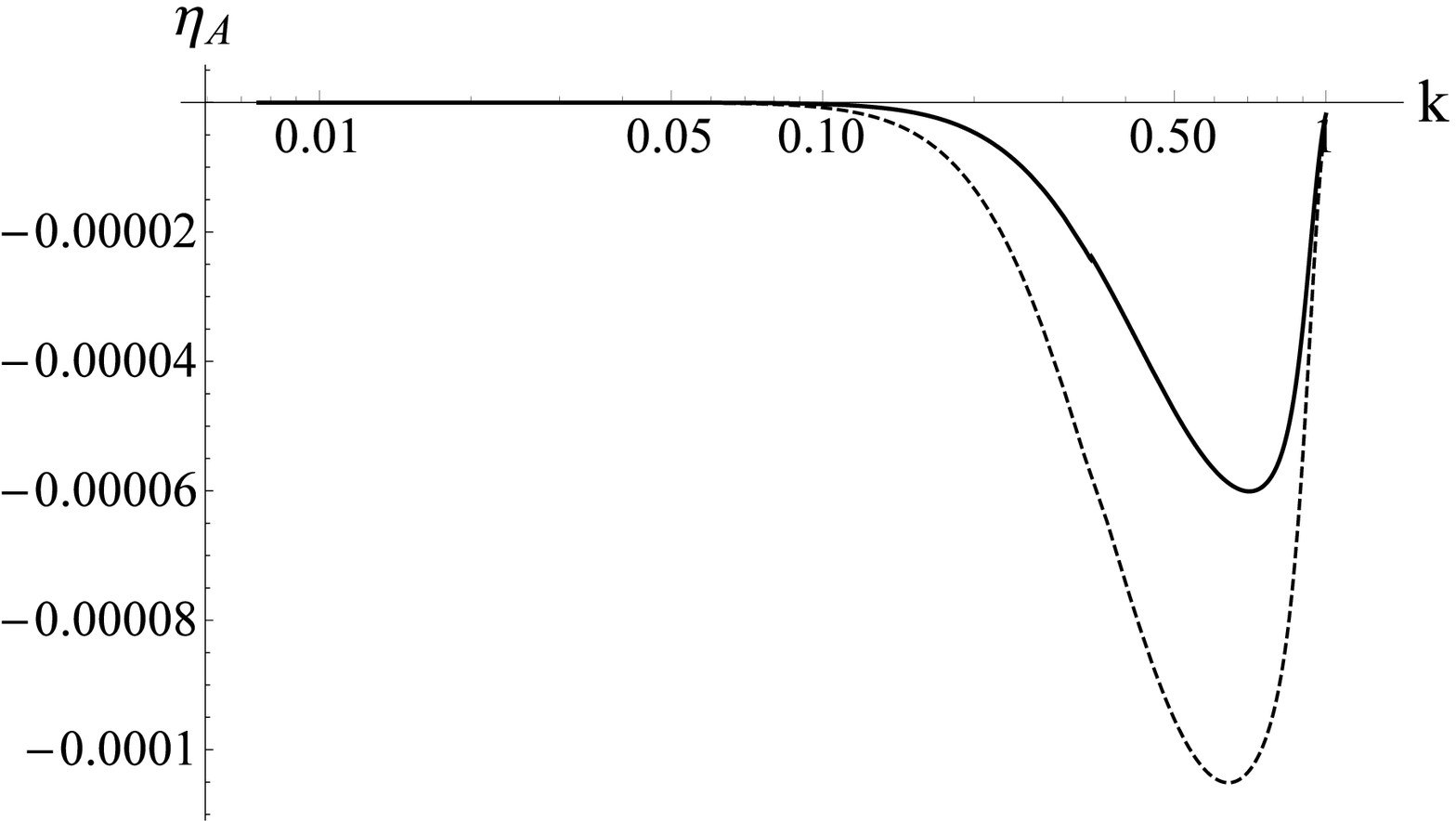,height=3.52cm,width=4.04cm,angle=0}
}
\caption{\label{fig:NLO_phase3_y16}
Flows of the dimensionless couplings $\t{g}_2$, $\t{g}_4$, $\t{v}_3$, the amplitude of the periodic condensate $\t{\sigma}_0$, the wave function renormalizations and the corresponding anomalous dimensions  $Z_\pe$, $\eta_\pe$ (full lines) and $Z_\pa$, $\eta_\pa$ (dashed lines)  for $\t{Y}=1.6$ along the RG trajectories belonging to Phase III, obtained in the approximation  NLO1.}
\end{figure}  
Phase III occurs for all values $\t{Y}\in(0,2]$. For $\t{Y}\in I_+$ the RG trajectories are emanated from the point of singularity $\t{g}_2= -\hf (Z_\pe+ Z_\pa) -\t{Y} =1-\t{Y}$, $\t{g}_4=0$, while for $\t{Y}\in I_0$ and $\t{Y}\in I_-$ they are emanated from the Gaussian FP (see Fig. \ref{fig:phdnlo}). The typical scaling of the various parameters of the model are shown in Fig. \ref{fig:NLO_phase3_y16}. A very short UV scaling region is followed by the IR scaling region, where the dimensionless couplings $\t{g}_2$ and $\t{g}_4$ of the ordinary potential as well as the coupling $\t{v}_3$ of the induced potential and the amplitude $\t{\sigma}_0$ of the periodic condensate show up tree-level scaling, so that the corresponding dimensionful quantities tend to constant non-vanishing values in the IR limit. In the approximations used the dimensionful ordinary potential remains a double-well potential (because $g_2(0)<0$  and $g_4(0)>0$) facing upwards. The wave function renormalizations  $Z_A$ $(A=\pe,~\pa)$ vary monotonically in a short UV region and become constant in the IR region, their overall change does not exceed a few hundredths of per cents. Correspondingly, the anomalous dimensions vanish in the IR limit.
  
\item
\begin{figure}
\centerline{
\psfig{file=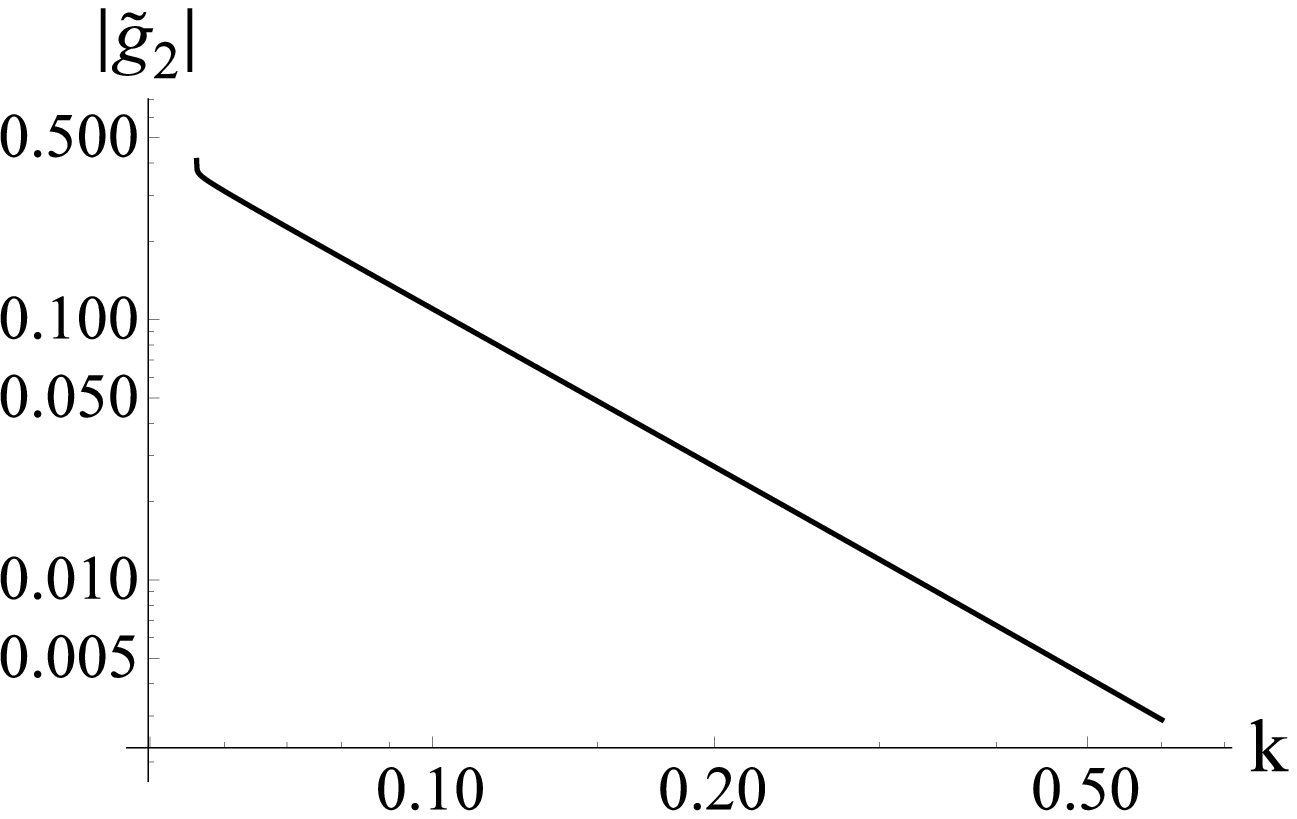,height=3.52cm,width=4.04cm,angle=0}
\psfig{file=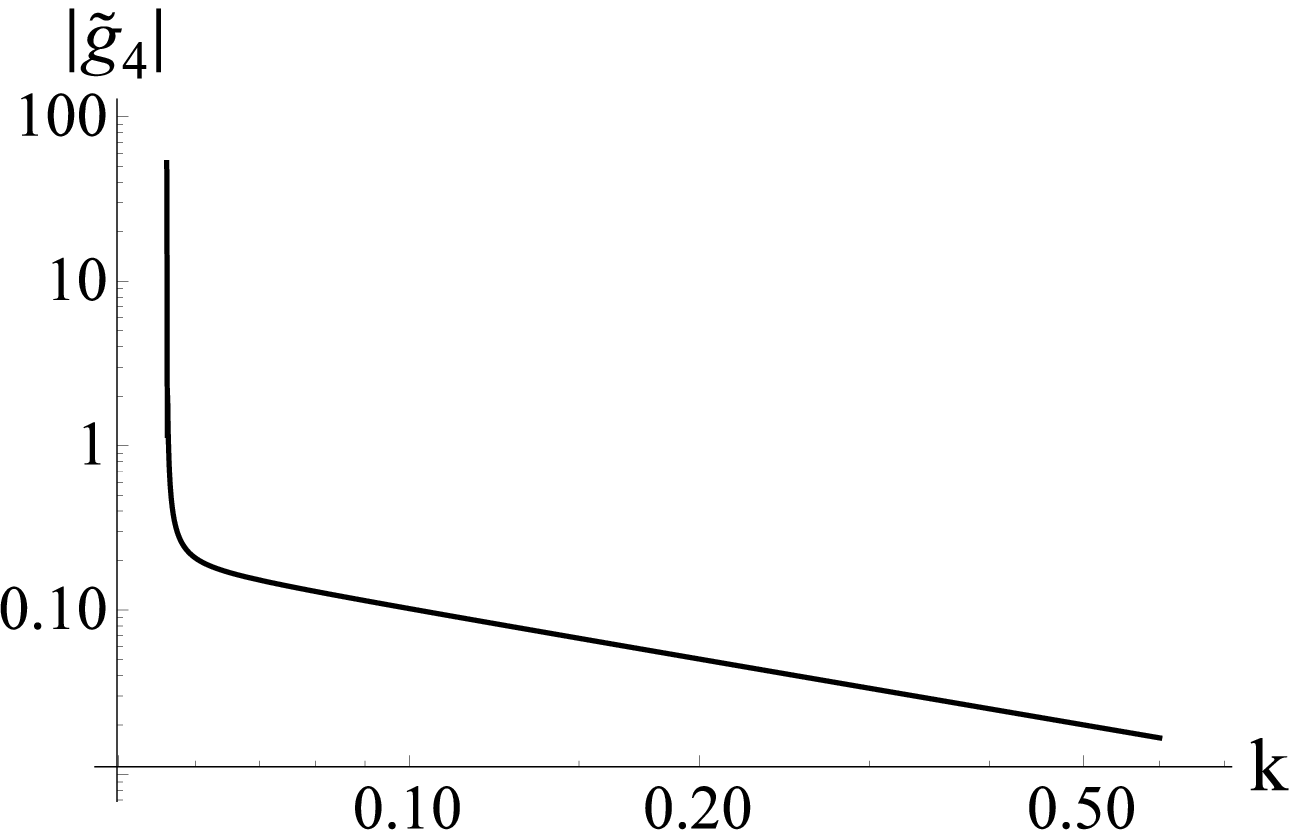,height=3.52cm,width=4.04cm,angle=0}
\psfig{file=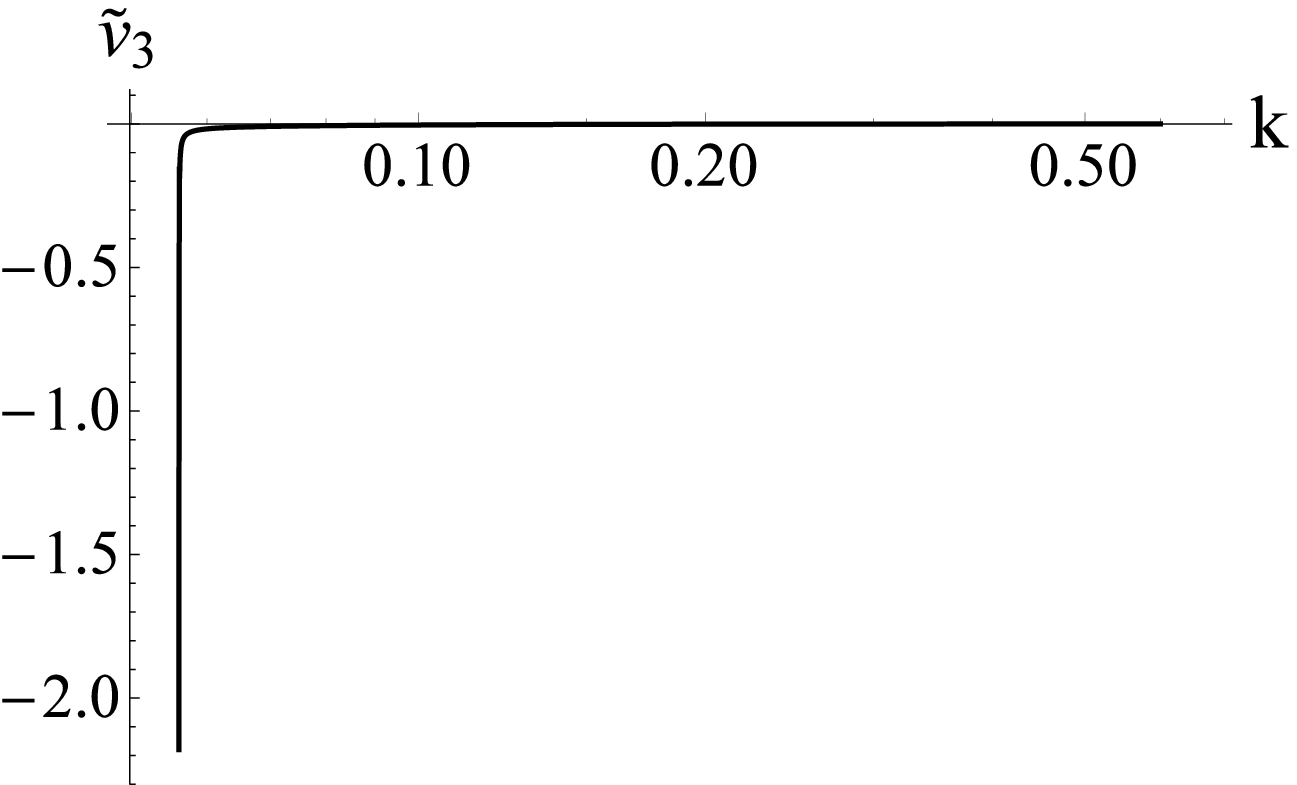,height=3.52cm,width=4.04cm,angle=0}}
\centerline{
\psfig{file=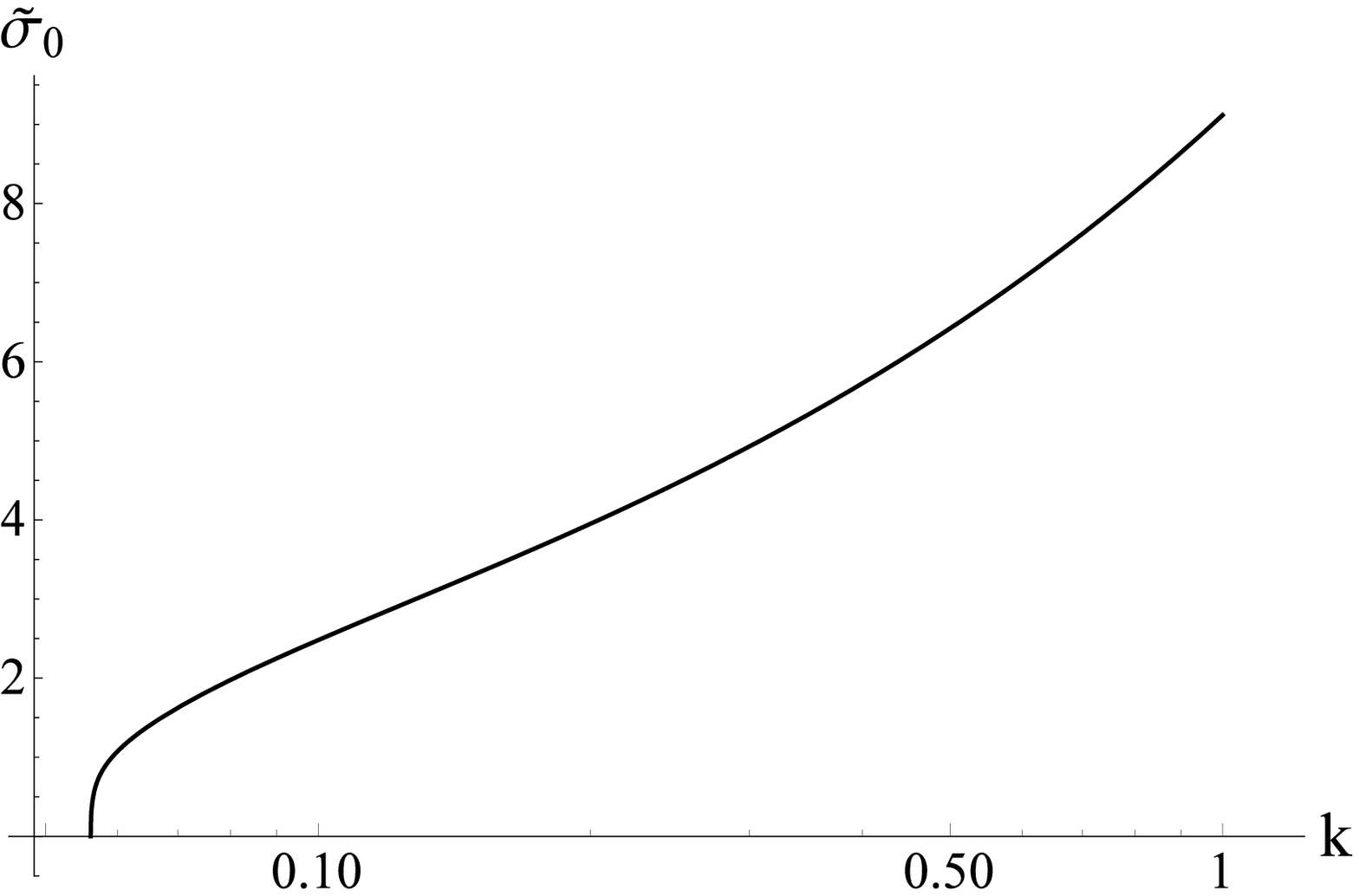,height=3.52cm,width=4.04cm,angle=0}
\psfig{file=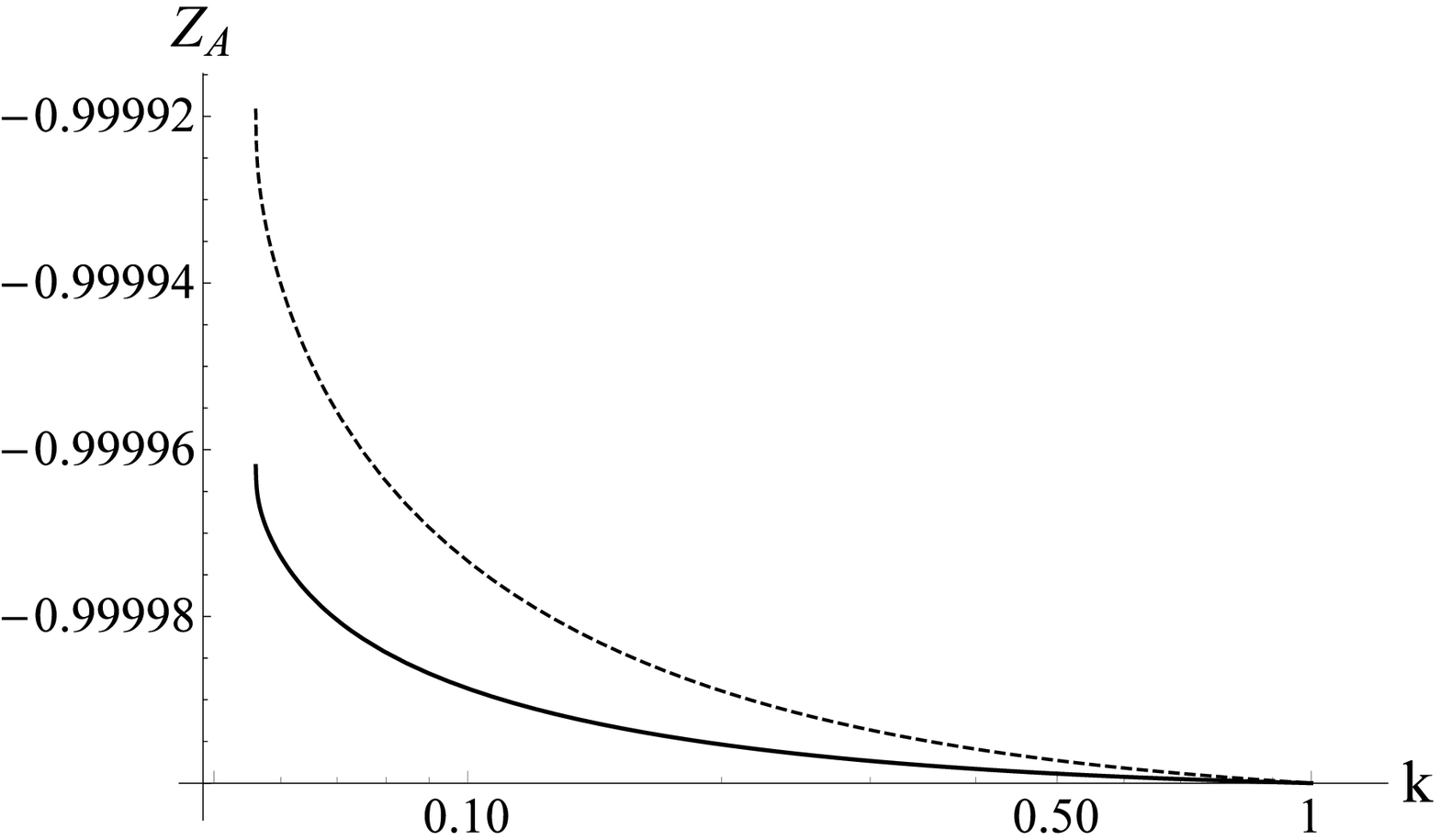,height=3.52cm,width=4.04cm,angle=0}
\psfig{file=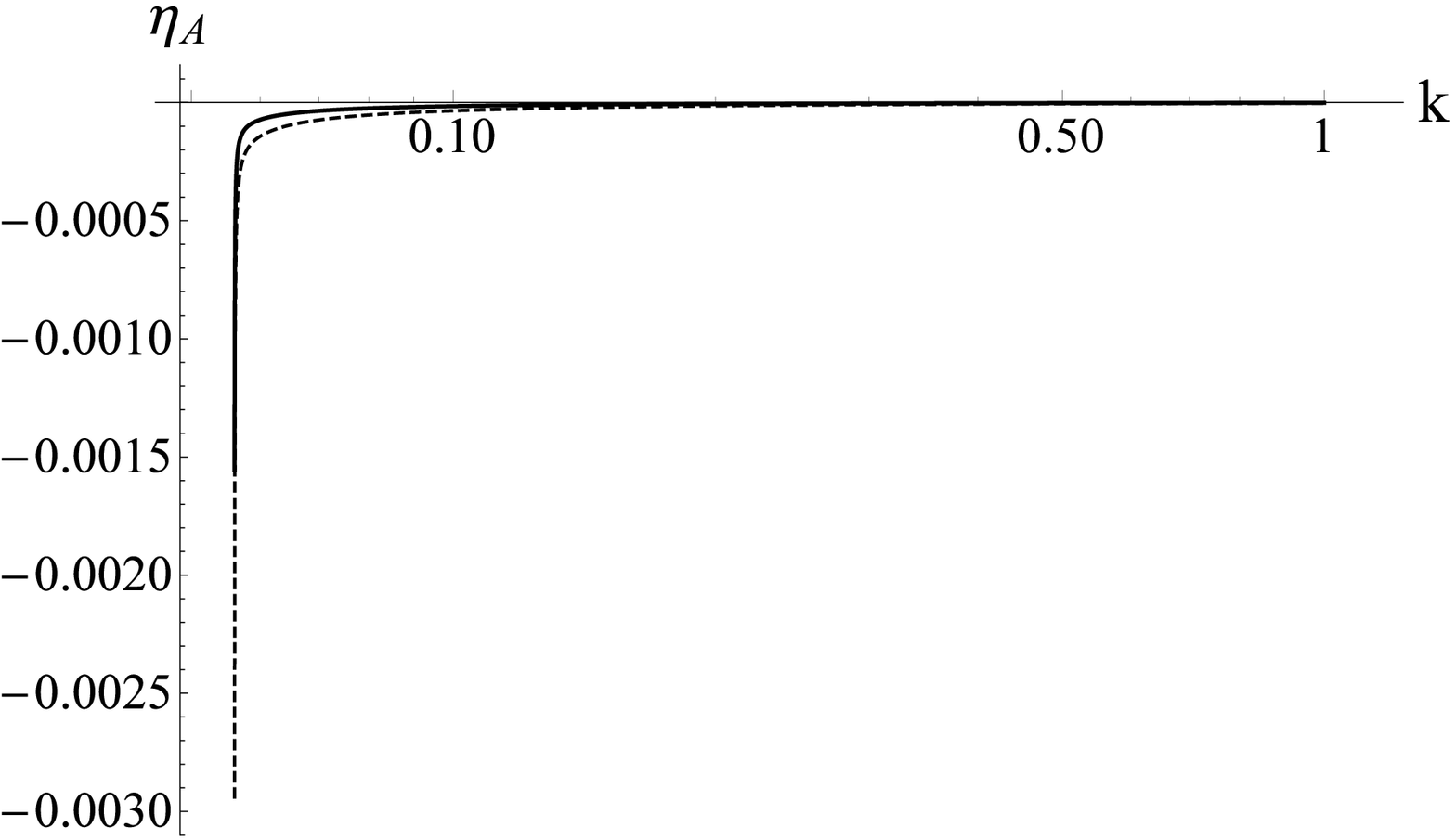,height=3.52cm,width=4.04cm,angle=0}
}
\caption{\label{fig:NLO_phase4_y06}
Flows of the dimensionless couplings $\t{g}_2$, $\t{g}_4$, $\t{v}_3$, the amplitude of the periodic condensate $\t{\sigma}_0$, the wave function renormalizations and the corresponding anomalous dimensions  $Z_\pe$, $\eta_\pe$ (full lines) and $Z_\pa$, $\eta_\pa$ (dashed lines) for $\t{Y}=0.6$ along the RG trajectories belonging to Phase IV, obtained in the approximation  NLO1.
}
\end{figure}  
For $\t{Y}\in I_0$ Phase IV is spanned by the RG trajectories which are emanated from the Gaussian FP and after a short tree-level UV scaling region approach
the line $\t{g}_2=-\hf(Z_\pe+Z_\pa) -\t{Y}$ where $\t{g}_4$ as well as its dimensionful counterpart explode rather suddenly  (see Figs.  \ref{fig:phdnlo} and 
\ref{fig:NLO_phase4_y06}). Similar behaviour is shown up by the coupling $\t{v}_3$ of the induced potential, the wave function renormalizations $Z_A$, and the
anomalous dimensions $\eta_A$. When the RG trajectories approach the line
 $\t{g}_2=-\hf(Z_\pe+Z_\pa) -\t{Y}$, the amplitude $\t{\sigma}_0$ of the periodic condensate becomes vanishing. For $\t{Y}\in I_+$ and $\t{Y}\in I_-$ Phase IV represents on the parameter plane $\c{P}$ a rather narrow region restricted to the very close neighbourhood of the line $\t{g}_2=-\hf(Z_\pe+Z_\pa) -\t{Y}$.

\end{itemize}

\section{Summary and outlook}\label{summary}

A modified version of the EAA RG method,  called by us the Fourier-Wetterich RG scheme has been developed in order to study kinetic condensates with spatial periodicity on the example of the 3-dimensional, Euclidean ghost $O(1)$ model, in the action of which the usual $\ord{\partial^2}$ kinetic term appears with the `wrong' sign and a positive definite quartic $\ord{\partial^4}$ derivative term have been included. It is assumed that the  condensate can be approximated with a truncated Fourier-series containing the number of $N_m$ modes, among them zero-mode is the homogeneous background $\Phi$ and the first mode of the scale-dependent wave number $P_k$ and the amplitude $\sigma_k(\Phi)$ is the fundamental mode. The scheme outlined enables one to take into account upper harmonics, but their role is possibly suppressed for decreasing  gliding scale $k$. The parameters of the  periodic condensate are determined by minimizing the EAA at each scale $k$ during the solution of the RG flow equations. The periodic condensate is assumed to reveal periodicity in one spatial direction, so that the system exhibits cylindrical symmetry. Therefore the direction parallel to the condensate's
axis and those perpendicular to it should be distinguished in the kinetic terms of the EAA. In the NLO of the GE this means that in principle  wave function renormalizations $Z_\pa$ and $Z_\pe$, differing in the longitudinal and transverse directions may occur. In order to close the set of the flow equations for the truncation $N_m$ of the Fourier series, additional potentials and additional derivative terms are included into the ansatz for the EAA, which represent  terms induced by the various Fourier-modes of the periodic condensate. The zero-mode approximation ($N_m=0$) corresponds to the usual EAA RG approach when the periodic condensate is not taken into account. The proposed Fourier-Wetterich RG scheme enables one to study when the periodic condensate occurs and how does it evolve during the RG flow and to decide in which of the phases does it survive the IR limit.

Numerical work was done in order to determine the characteristics of the WF FP and the phase structure of the model in the LPA in the zero- (LPA0,  $N_m=0$) and one-mode  (LPA1, $N_m=1$), and in the NLO of the GE in the one-mode approximation (NLO1, $N_m=1$). In all of these approximations the higher-derivative couplings do not evolve and keep the bare value $\t{Y}$. Therefore various sections of the phase diagram have been investigated at constant values of $\t{Y}\in(0,2]$ and projected onto the plane $\t{g}_2$ and $\t{g}_4)$ corresponding to the dimensionless  mass squared and the quartic coupling of the ordinary potential, respectively. The approximations LPA0, LPA1, and NLO1 provided qualitatively identical  results. The WF FP with positive quartic coupling $\t{g}_4^*$ was found for $1<\t{Y}\le 2$ and $0<\t{Y}<1/2$, corresponding to a physically reliable critical theory (with the EAA bounded from below). Four different phases of the system were identified according to their  IR behaviours.
Phase I occurs to be the symmetric one where no periodic condensate survives the IR limit and it is present for all values $\t{Y}\in (0,2]$. Phase II occurs only when the WF FP with positive quartic coupling is present. Phase II seems to be the counterpart of the symmetry broken phase of the ordinary 3-dimensional Euclidean $O(1)$ model, 
in the IR limit it is characterized by a universal dimensionless ordinary potential,
the flattening out the dimensionful ordinary potential, but it exhibits a periodic condensate with infinite amplitude. It is a phase in which $Z_2$ symmetry, rotational symmetry in space, and translational symmetry along the axis of the periodic condensate is spontaneously broken. Phase III is present for all values  $\t{Y}\in (0,2]$, characterized by a  finite non-vanishing amplitude of the periodic condensate and a non-convex, dimensionful double-well ordinary potential. Finally, Phase IV seems to appear as the phase replacing Phase II for $1/2 <\t{Y}<1$ when no WF FP with $\t{g}_4^*>0$ exists. In the IR limit  the periodic condensate does not survive, the dimensionless potential becomes universal and the dimensionful potential describes infinitely strong self-interaction.

 In all of the phases the flow of the wave function renormalizations $Z_\pe$ and $Z_\pa$ as well as its effect on the whole dynamics was found almost negligible, although one would expect a rather severe role of the flow of the derivative terms in the dynamics of the kinetic condensate. This feature of our numerical results may be connected with the
various approximations used by us.   In order to make the explicit forms of the flow equations as simple as possible, additional approximations have been involved: {\em (i)} a modified version of Litim's regulator adjusted to the cylindrical symmetry of the system has been used, {\em (ii)} the kinetic term induced by the fundamental Fourier mode of the periodic condensate has been neglected, {\em (iii)}  the wave function renormalizations (and the couplings of the quartic gradient terms) have been assumed to be field-independent, {\em (iv)}  the potentials have been expanded at vanishing value $\Phi=0$ of the homogeneous background, {\em (v)} the potentials have been truncated at their quartic terms. Approximation {\em (i)} kills the dependence of the reduced propagators on the loop-momentum, however, the dynamical role of the periodic condensate leading to the shift of the longitudinal momenta is still kept in the traces on the right-hand sides of the flow equations.  As discussed at the end of Sect. \ref{regul}, approximation {\em (ii)} kills the momentum-dependence of the induced vertex $\mf{V}$  which is responsible for the interaction of the quantum fluctuations of the field with the periodic condensate, and together with    approximation {\em (iv)} result in disappearing this interaction entirely on the right-hand sides of the flow equations for the wave function renormalizations.
 Therefore, one may suggest that the removal of approximations {\em (ii)} and {\em (iv)}
should be crucial in taking aim at physically  more reliable results. The expansion in powers of the homogeneous background $\Phi$  at the  minimum of the ordinary potential
provides much more reliable numerical results in the case of the ordinary $O(1)$ model, although it introduces an explicit symmetry breaking into the form of the EAA. Similarly, the inclusion of the induced wave function renormalization $\c{E}_k$ breaks $Z_2$ symmetry explicitly (see our discussion before Eq. \eq{we}). Nevertheless, this should be the way to avoid getting rid of the main dynamics involved in the induced vertex.
 Also the removal of approximation {\em (iii)} and  allowing for at least quadratic dependence of the wave function renormalization on the field may make improvements, especially regarding the flow in the symmetric phase, Phase I.  Similarly, the usage of polynomials higher than of quartic order may give more accurate results, like in the case of ordinary $O(N)$ models. We believe that the above mentioned possibilities of the  improvement of the proposed Fourier-Wetterich RG framework are worthy of future investigations.

\section*{Acknowledgements}

S. Nagy acknowledges financial support from a J\'anos Bolyai Grant of the
Hungarian Academy of Sciences, and the Hungarian National Research,
Development and Innovation Office NKFIH (Grant Nos. K112233, KH126497).

\appendix
\section{Functional derivatives}\label{fderiv}

In the one-mode approximation, i.e., for the truncation $N_m=1$ in Eq. \eq{eaaon}, 
the first four functional derivatives of the rEAA are given as
\bea\label{fdgam1}
  \fd{\Gamma_k}{\phi_x} &=&   \c{Z}_k(-\Delta_\pe, -\partial_\pa^2)\phi_x
+ \hf \cos(Px_\pa)  \c{E}_k(-\Delta_\pe, -\partial_\pa^2)\phi_x
+ \hf \c{E}_k(-\Delta_\pe, -\partial_\pa^2) \lbrack \phi_x \cos(Px_\pa)\rbrack
\nn
&&
~~~+ U_k'(\phi_x) + V_k'(\phi_x) \cos(P x_\pa),\nn
\eea
\bea\label{fdgam2}
 \fdd{\Gamma_k}{\phi_x}{\phi_y} &=& \c{Z}_k(-\Delta_\pe, -\partial_\pa^2)\delta_{x,y}
+ \hf \cos(Px_\pa)\c{E}_k(-\Delta_\pe, -\partial_\pa^2) \delta_{x,y}
+\hf \c{E}_k(-\Delta_\pe, -\partial_\pa^2)  \lbrack\delta_{x,y} \cos(P x_\pa)\rbrack
\nn
&&
~~~+ \biggl( U_k''(\phi_x) + V_k''(\phi_x)  \cos(Px_\pa)\biggr)\delta_{x,y},
\eea
\bea\label{fdgam3}
 \fddd{\Gamma_k}{\phi_z}{\phi_y}{\phi_x} &=& 
\biggl( U_k'''(\phi_x) + V_k'''(\phi_x) \cos(P x_\pa)\biggr)\delta_{x,y}\delta_{x,z},
\eea
and
\bea\label{fdgam4}
 \fdddd{\Gamma_k}{\phi_u}{\phi_z}{\phi_y}{\phi_x} &=&
\biggl( U_k''''(\phi_x) + V''''_k(\phi_x) \cos(P x_\pa)\biggr)\delta_{x,y}\delta_{x,z}\delta_{x,u}.
\eea

\section{Left-hand side of the WE}\label{lhswe}

In the one-mode approximation  the left-hand side of the WE, 
\bea
{\dot \Gamma}_k[\phi]&=& \hf \int_x  \phi_x
\biggl(
 {\dot{\c{Z}}}_k(-\Delta_\pe,-\partial^2_\pa)
+ \cos (P_kx_\pa){\dot{\c{E}}}_k(-\Delta_\pe, -\partial_\pa^2)\biggr)\phi_x
+ \int_x\biggl(  {\dot U}_k(\phi_x) + {\dot V}_k (\phi_x) \cos(P_k x_\pa)
\biggr)
\eea
with $\phi_x=\phi_{Bx}+\eta_x= \Phi+ 2\sigma \cos(P x_\pa)+\eta_x$ is evaluated in the momentum representation. Then we write
\bea
 &&\phi_{BQ}= \int_x e^{iQx} \phi_{Bx}
= \Phi\delta_{Q,0} + \sigma \sum_\tau \delta_{Q+\tau P_k e,0},~~~~
\eta_Q= e^{iQx}\eta_x,
\eea
where we assume that $\eta_Q$ vanishes for $Q_\pa=0,\pm  P_k$. In order to evaluate the pieces of ${\dot \Gamma}_k[\phi_B+\eta]$ containing the potentials and keeping the terms up to the order $\ord{\eta^2}$, first, one has to Fourier-expand
$(\phi_B+\eta)^n$ and $(\phi_B+\eta)^n\cos(P_kx_\pa)$ with $n=1,2,3,4$ keeping the modes up to $\cos(P_kx_\pa)$, but during the calculation of  $(\phi_B+\eta)^n\cos(Px_\pa)$ one has to keep also the mode $\cos(2P_kx_\pa)$ of the factor   $(\phi_B+\eta)^n$
because  the terms containing the product  $\cos(2P_k x_\pa)\cos(P_k x_\pa)$ contribute  to the mode   $\cos(P_k x_\pa)$. As to the next, one performs the 
 3-dimensional volume integrals in the Euclidean space over the cube $V=L^3$ with $L=N\frac{2\pi}{P_k}$ and $N\to \infty$. Then it is straightforward to evaluate the integrals $\int_x U_k(\phi_{Bx}+\eta_x)$ and $\int_x V_k(\phi_{Bx}+\eta_x)\cos Px_\pa$ for the truncation $M=4$ of the potentials. Finally, one re-expresses the result  in terms of the ordinary potential  $U_k(\Phi)$ and the induced potential $V_k(\Phi)$ and their derivatives with respect to the homogeneous background field $\Phi$.

In order to evaluate the derivative pieces of ${\dot \Gamma}_k[\phi_B+\eta]$ in the truncation $N_m=1$, one performs the volume integrals in the same way as
it has been done for the terms of the potentials. Making use of the Fourier-integral representations of $\phi_{Bx}$, $\eta_x$ and the Dirac-delta, one finds 
\bea
\hf \int_x (\phi_{Bx}+\eta_x) \c{Z}_k(-\Delta_\pe,-\partial_\pa^2) (\phi_{Bx} +\eta_x )
&=&
V\sigma^2\c{Z}_k(0,P^2)  + \sigma \c{Z}_k(0,P^2) (\eta_{P}+\eta_{-P})
+ \hf\int_Q  \c{Z}_k(Q_\pe^2,Q_\pa^2)\eta_Q\eta_{-Q}\nn
\eea
and
\bea
\hf \int_x  \cos(Px_\pa)(\phi_{Bx}+\eta_x) \c{E}_k(-\Delta_{\pe},-\partial^2_{\pa})
(\phi_{Bx}+\eta_x)
&=&
V\hf \Phi\sigma \c{E}_k(0,P^2)
+ \frac{1}{4}\Phi\c{E}_k(0,P^2)(\eta_{P}+\eta_{-P})
\nn
&&
+\frac{1}{4}  \sum_{\tau=\pm 1}\int_{Q,Q'} \delta_{Q+Q'+\tau Pn_1,0}
 \c{E}_k(Q_\pe^2,Q_\pa^2)\eta_Q\eta_{Q'}.
\eea
Finally we obtain the expressions given in Eqs. \eq{gammadot}-\eq{gam21} for the scale derivative ${\dot \Gamma}_k[\phi_B+\eta]$ of the rEAA.

\section{Closed form of the full propagator}\label{a:fullprop}

In this section we show that in the one-mode approximation the full propagator  $\c{G}_{p,q}$ can be obtained in a closed form by the inversion of the matrix
\bea
  \c{G}^{-1}_{p,q} &=& G^{-1}(p_\pe^2,p_\pa^2,\Phi)  \delta_{p+q,0}
+ \mf{V}_{p,q},
\eea
given by Eq. \eq{fullpro1},
where $\mf{V}_{p,q}$ is the induced vertex given in Eq. \eq{indverm}. For the sake of simplicity the $\Phi$-dependences are not indicated below in this appendix.
Formally the matrix of the full propagator $\c{G}$ is given by the Neumann-series via the matrix equation
\bea\label{geomexp}
\c{G}\equiv  [ G^{-1}+\mf{V} ]^{-1}&=& \biggl( 1+ G\mf{V} \biggr)^{-1} G =
\sum_{n=0}^\infty (-G\mf{V})^nG =
\sum_{n=0}^\infty \biggl\lbrack 
(-G\mf{V})^{2n} G+ (-G\mf{V})^{2n+1} G\biggr\rbrack ,
\eea
where we have separated the terms containing even and odd powers of the induced vertex. It is worthwhile mentioning that the reduced propagator $G_{p,q}$ and the induced vertex ${\mf{V}}(p_\pe^2,p_\pa^2,q_\pe^2,q_\pa^2)$ are symmetric matrices in the momentum space, but they do not commute, $\lbrack G,{\mf{V}} \rbrack\not=0$, therefore nor the product $G\mf{V}$ neither the full propagator $\c{G}$
 are symmetric matrices.

Now we resum the series in the right-hand side of Eq. \eq{geomexp}.
One finds easily the terms with $n=1$ of the series,
\bea
 (G\mf{V})_{p,q} &=&
 G(p_\pe^2,p_\pa^2) {\mf{V}}(p_\pe^2,p_\pa^2,q_\pe^2,q_\pa^2)
\sum_{\tau=\pm 1} \delta_{p+q+\tau e,0} ,
\eea
and
\bea
[(G\mf{V})^2]_{p,q}&= & 
G(p_\pe^2,p_\pa^2)\sum_{\tau=\pm 1} G(p_\pe^2,(p_\pa+\tau P)^2)
\times \mf{V}^2(p_\pe^2,p_\pa^2,p_\pe^2,(p_\pa+\tau P)^2) \delta_{p+q,~0}.
\eea
In order to determine the terms with $n>1$ of the series in the right-hand side of Eq. \eq{geomexp} one has to take the matrix products in accordance with the one-mode approximation, i.e., neglect all contributions when in the intermediate steps longitudinal momenta shifted by  more than one times of the elementary shift $\pm P$ occur. It is  straightforward to show by mathematical induction that one gets
\bea
[(G\mf{V})^{2n}]_{p,q} &=&
 G^n(p_\pe^2,p_\pa^2)\biggl(\sum_{\tau=\pm 1} G(p_\pe^2,(p_\pa+\tau P)^2) \mf{V}^2(p_\pe^2,p_\pa^2,p_\pe^2,(p_\pa+\tau P)^2)\biggr)^n\delta_{p+q,~0}
\eea
and
\bea
[(G\mf{V})^{2n+1}]_{p,q} &=& 
G^{n+1}(p_\pe^2,p_\pa^2)\biggl(\sum_{\tau'=\pm 1} G(p_\pe^2,(p_\pa+\tau' P)^2)\mf{V}^2(p_\pe^2,p_\pa^2,p_\pe^2,(p_\pa+\tau' P)^2)\biggr)^n
\nn
&&
~~~\times {\mf{V}}(p_\pe^2,p_\pa^2,q_\pe^2,q_\pa^2)
\sum_{\tau =\pm 1}\delta_{p+q+\tau Pe,0}
\eea
in the one-mode approximation. Then one finds
\bea\label{prsum1}
&&\sum_{n=0}^\infty [(G\mf{V})^{2n}G]_{p,q}= 
\sum_{n=0}^\infty
G^{n+1}(p_\pe^2,p_\pa^2)\biggl(\sum_{\tau=\pm 1} G(p_\pe^2,(p_\pa+\tau P)^2)
\mf{V}^2(p_\pe^2,p_\pa^2,p_\pe^2,(p_\pa+\tau P)^2)\biggr)^n\delta_{p+q,0}
\nn
&=&
~~~G(p_\pe^2,p_\pa^2)\biggl( 1- G(p_\pe^2,p_\pa^2)\sum_{\tau=\pm 1} G(p_\pe^2,(p_\pa+\tau P)^2)
\times \mf{V}^2(p_\pe^2,p_\pa^2,p_\pe^2,(p_\pa+\tau P)^2)\biggr)^{-1}\delta_{p+q,0} 
\nn
&&
~~~\equiv\c{G}(p_\pe^2,p_\pa^2)\delta_{p+q,0}
\eea
and
\bea\label{prsum2}
&& -\sum_{n=0}^\infty [ (G\mf{V})^{2n+1}G]_{p,q} =  
-\sum_{n=0}^\infty G^{n+1}(p_\pe^2,p_\pa^2)\biggl(\sum_{\tau'=\pm 1} G(p_\pe^2,(p_\pa+\tau' P)^2)
\mf{V}^2(p_\pe^2,p_\pa^2,p_\pe^2,(p_\pa+\tau' P)^2)\biggr)^n 
\nn
&&
~~~\times{\mf{V}}(p_\pe^2,p_\pa^2,q_\pe^2,q_\pa^2)\sum_{\tau=\pm 1}\delta_{p+q+\tau Pe,0}   G(q_\pe^2,q_\pa^2)
\nn
&=&
~~~-\sum_{n=0}^\infty G(p_\pe^2,p_\pa^2)\biggl( 1- G(p_\pe^2,p_\pa^2)\sum_{\tau=\pm 1} G(p_\pe^2,(p_\pa+\tau P)^2)
\mf{V}^2(p_\pe^2,p_\pa^2,p_\pe^2,(p_\pa+\tau P)^2)\biggr)^{-1} 
\nn
&&
~~~\times G(q_\pe^2,q_\pa^2){\mf{V}}(p_\pe^2,p_\pa^2,q_\pe^2,q_\pa^2)\sum_{\tau'=\pm 1}\delta_{p+q+\tau' Pe,0}
=
- \c{G}(p_\pe^2,p_\pa^2) G(q_\pe^2,q_\pa^2){\mf{V}}(p_\pe^2,p_\pa^2,q_\pe^2,q_\pa^2)\sum_{\tau'=\pm 1}
\delta_{p+q+\tau' Pe,0}.
\eea
The last equation in \eq{prsum1} defines the function $\c{G}(p_\pe^2,p_\pa^2)$ also exhibiting  $\Phi$-dependence.
Making use of the sums \eq{prsum1} and \eq{prsum2} we obtain from Eq. \eq{geomexp} the explicit form of the full propagator in the one-mode approximation
given by Eqs. \eq{fullpro2} and \eq{resumg}.

\section{Evaluation of the traces on the right-hand side of the WE in the one-mode approximation}\label{traces}
We evaluate the traces \eq{Ts0}-\eq{Ts2C} in momentum space. In the notation  the dependences on the scale $k$,  the momentum-dependence of ${\dot R}_k(p_\pe^2,
p_\pa^2)$ as well as the $\Phi$-dependences of the propagators and the induced vertices  are suppressed for the sake of simplicity.

\subsection{Trace of the matrix independent of $\eta$}\label{trace0}

In order to evaluate the trace \eq{Ts0} explicitly, one inserts the series expansion \eq{geomexp} of the IR regulated full propagator $\c{G}_{p,q}$ into the right-hand side. Making use of the sums \eq{prsum1} and \eq{prsum2} it is straightforward to find the traces,
\bea\label{Ts0ex}
T_{0I}&=&\sum_{n=0}^\infty\Tr  [(G\mf{V})^{2n}G{\dot R}]=
\int_p  \c{G}(p_\pe^2, p_\pa){\dot R}
\eea
and
\bea
T_{0II}&=&-\sum_{n=0}^\infty\Tr [ (G\mf{V})^{2n+1}G{\dot R}] =
= 
-\int_p\c{G}(p_\pe^2, p_\pa) G(p_\pe^2,p_\pa^2) {\mf{V}}(p_\pe^2,p_\pa^2,p_\pe^2,p_\pa^2)
\sum_{\tau=\pm 1}\delta_{\tau Pe,0} {\dot R}=0. \nn
\eea
The  trace $T_{0II}$ vanishes because the characteristic momentum $P$ is non-vanishing. Then we obtain $T_0=T_{0I}$ given by Eq. \eq{Ts0ex}.

\subsection{Trace of the matrix linear in $\eta$}\label{trlineta}

Inserting the explicit forms \eq{bpqr} and \eq{fullpro2} of the matrices $B_{p,q,r}$ and $\c{G}_{p,q}$, respectively, into the right-hand side of Eq. \eq{Ts1}, one gets 
\bea
 T_1 &=&-\int_Q \eta_Q \int_{p}
\c{G}(p_\pe^2,p_\pa)\biggl(b_0\c{G}_{p-Q,-p}+ b_1 \sum_{\tau} \c{G}_{p-Q+\tau Pe,-p}\biggr) {\dot R}\nn
&&
+ \int_Q \eta_Q  \int_{p}\sum_{\tau_1} b_0 G(p_\pe^2,(p_\pa+\tau_1 P)^2)
{\mf{V}}(p_\pe^2,p_\pa^2,p_\pe^2,(p_\pa+\tau_1 P)^2 ))
\c{G}_{p+\tau_1 Pe  -Q,-p}{\dot R}
\nn
&&
+ \int_Q \eta_Q  \int_{p}\sum_{\tau_1,\tau_2} b_1 G(p_\pe^2,(p_\pa+\tau_1 P )^2)
{\mf{V}}(p_\pe^2,p_\pa^2,(p_\pa+\tau_1 P)^2 ))
\c{G}_{p+\tau_1 Pe -Q+\tau_2 Pe,-p}{\dot R}
\eea 
after some algebra. In the one-mode approximation one has to keep only the terms with $\tau_2=-\tau_1$ in the sum  $\sum_{\tau_1,\tau_2}$ in the last line. Then one finds
\bea
  T_1&=& \sum_{A=I,II,III} T_{1A}
\eea
with
\bea
  T_{1I} &=&
- \int_Q \eta_Q \int_{p}
\c{G}(p_\pe^2,p_\pa)\biggl\lbrack
  b_0\c{G}_{p-Q,-p}+ b_1 \sum_{\tau} \c{G}_{p-Q+\tau Pe,-p}
 \biggr\rbrack {\dot R},\nn
T_{1II}
&=&
\int_Q \eta_Q  \int_{p}\sum_{\tau} b_0 G(p_\pe^2,(p_\pa+\tau P)^2)
{\mf{V}}(p_\pe^2,p_\pa^2, p_\pe^2,(p_\pa+\tau P)^2 )
\c{G}_{p+\tau Pe   -Q,-p}{\dot R}\nn
T_{1III}
&=&
2 \int_Q \eta_Q  \int_{p}\sum_{\tau} b_1 G(p_\pe^2,(p_\pa+\tau P)^2)
{\mf{V}}(p_\pe^2,p_\pa^2, p_\pe^2,(p_\pa+\tau P)^2 )
\c{G}_{p -Q,-p}{\dot R}.
\eea 
Here once again one makes use of the expression \eq{fullpro2} of the matrix $\c{G}_{p,q}$  and after neglecting 
 all terms containing propagators with shifted longitudinal momenta $p_\pa\pm 2P$ or $p_\pa-Q_\pa\pm 2P$ one gets expressions for the $T_{1A}$s in which either $\eta_0$ or $\eta_{\pm Pe}$ occurs. Therefore the trace $T_1$ vanishes when one chooses the fluctuating field with vanishing modes $\eta_0=0$ and  $\eta_{\pm Pe}=0$, as it was supposed.

\subsection{Traces of the matrices quadratic in $\eta$}\label{trquadeta}

The explicit expressions for the traces $T_{2B}$ and $T_{2C}$ have been derived
similarly to the evaluation of the trace $T_1$. Namely, the expressions \eq{bpqr} and \eq{cpqrs} were inserted first into the right-hand sides of Eqs. \eq{Ts2B} and \eq{Ts2C}, then the matrices $\c{G}_{p,q}$
subsequently replaced by their expressions \eq{fullpro2} followed always by the removal of the terms in which the longitudinal momenta shifted by $\pm 2 P$ occurred. In this manner a lengthy but straightforward calculation provided the explicit expressions looked for. It has been found that the trace $T_{2C}$ given by Eq. \eq{Ts2C} contains 
terms only proportional to the quadratic forms  either $\int_{Q,Q'}\eta_Q\eta_{Q'} \delta_{Q+Q',0}$  or  $\int_{Q,Q'}\eta_Q\eta_{Q'} \delta_{Q+Q'\pm Pe,0}$, so that there occurs no contribution from $T_{2C}$ to the flow equations for $\c{Z}(Q_\pe^2, Q_\pa^2)$ and $\c{E}(Q_\pe^2, Q_\pa^2)$. This is the consequence of the usage of 
field-independent derivative couplings, otherwise the field-derivatives of $\c{Z}$ and $\c{E}$ would occur with $Q$-dependences in the integrands in Eq. \eq{Ts2C}.
 The various terms contributing to the trace $T_{2B}$ given by Eq. \eq{Ts2B} can be sorted according to the power  $n$ on which the induced vertex $\mf{V}$ explicitly appears in them,
\bea
 T_{2B}&= & \sum_{n=0}^3T_{2B}^{[n],m}
\eea
where
\bea
T_{2B}^{[0],0}&=&
b_0^2 \int_{Q,Q',p}\eta_Q\eta_{-Q}\delta_{Q+Q',0}
 \c{G}^2(p_\pe^2,p_\pa)\c{G}((p_\pe-Q_\pe)^2,p_\pa-Q_\pa){\dot R} \nn
&&+b_1^2\int_{Q,Q',p} \eta_Q\eta_{-Q}\delta_{Q+Q',0}
\c{G}^2(p_\pe^2,p_\pa)\sum_{\tau=\pm} \c{G}((p_\pe-Q_\pe)^2,p_\pa-Q_\pa+\tau P)
{\dot R},
\eea

\bea
T_{2B}^{[0],1}&=&
b_0b_1 \sum_{\tau=\pm}\int_{Q,Q',p}  \eta_Q\eta_{Q'}\delta_{-Q-Q'+ \tau Pe,0}\c{G}^2(p_\pe^2,p_\pa)
\biggl\lbrack \c{G}((p_\pe-Q_\pe)^2,p_\pa-Q_\pa)
+\c{G}((p_\pe-Q_\pe)^2,p_\pa-Q_\pa+\tau P) 
\biggr\rbrack{\dot R}, \nn
\eea

\bea
&& T_{2B}^{[1],0}=
- b_0b_1\int_{Q,Q',p} \eta_Q\eta_{-Q}\delta_{Q+Q',0}\c{G}(p_\pe^2,p_\pa)\nn
&&\times
\biggl\lbrack
\c{G}(p_\pe^2,p_\pa)  \sum_{\tau=\pm}
\c{G}((p_\pe-Q_\pe)^2,p_\pa-Q_\pa)G((p_\pe-Q_\pe)^2,(p_\pa-Q_\pa-\tau P)^2)\nn
&&~~~~\times
  {\mf{V}}((p_\pe-Q_\pe)^2,(p_\pa-Q_\pa)^2,(p_\pe-Q_\pe)^2,(p_\pa-Q_\pa-\tau P)^2)
\nn
&&+ G(p_\pe^2,p_\pa^2) \sum_{\tau=\pm}
\c{G}(p_\pe^2,p_\pa+\tau P)  {\mf{V}}((p_\pe^2,(p_\pa+\tau P)^2,p_\pe^2,p_\pa^2)
\nn
&& \times \c{G}((p_\pe-Q_\pe)^2,p_\pa-Q_\pa)
\nn
&&+\c{G}(p_\pe^2,p_\pa)\sum_{\tau=\pm}G(p_{\pe}^2,(p_{\pa}+\tau P)^2) 
 {\mf{V}}(p_\pe^2,p_\pa^2,p_{\pe}^2,(p_{\pa}+\tau P)^2)
\nn
&& \times \c{G} ( (p_\pe-Q_\pe)^2, p_\pa-Q_\pa+ \tau P ) \nn
&&+
 G( p_\pe^2, p_\pa^2)\sum_{\tau=\pm} \c{G}(p_\pe^2,p_\pa+\tau P)
\mf{V}( p_\pe^2,(p_\pa+\tau P)^2, p_\pe^2, p_\pa^2)\nn
&& \times \c{G}((p_\pe-Q_\pe)^2,p_\pa-Q_\pa+\tau P)\nn
&&
+ \c{G}(p_\pe^2,p_\pa)\sum_{\tau=\pm} 
\c{G}((p_\pe-Q_\pe)^2,p_\pa-Q_\pa+\tau P)\nn
&&~~~~\times G((p_\pe-Q_\pe)^2,(p_\pa-Q_\pa)^2)  
\nn
&& \times {\mf{V}}((p_\pe-Q_\pe)^2,(p_\pa-Q_\pa+\tau P)^2,(p_\pe-Q_{\pe})^2,(p_\pa-Q_{\pa})^2)\nn
&&
+\c{G}(p_\pe^2,p_\pa)\biggl(\sum_{\tau=\pm}
G(p_\pe^2,(p_\pa+\tau P)^2) 
{\mf{V}}(p_\pe^2,p_\pa^2,p_\pe^2,(p_\pa+\tau P)^2) \biggr)\nn
&& \times \c{G}( (p-Q)_\pe^2, p_\pa-Q_\pa)\biggr\rbrack{\dot R}, 
\eea

\bea
T_{2B}^{[1],1}&=&
-b_0^2\sum_{\tau=\pm} \int_{Q,Q',p}\eta_Q\eta_{Q'}\delta_{-Q-Q'+\tau Pe,0}
\c{G}(p_\pe^2,p_\pa)\nn
&&\times
\biggl\lbrack
G(p_\pe^2,p_\pa^2)
 {\mf{V}}(p_\pe^2,(p_\pa-\tau P)^2,p_\pe^2,p_\pa^2)\c{G}(p_\pe^2,p_\pa-\tau P)\nn
&&
\times\c{G}((p_\pe-Q_\pe)^2,p_\pa-Q_\pa)\nn
&&
+ \c{G}(p_\pe^2,p_\pa)
\c{G}((p_\pe-Q_\pe)^2,p_\pa-Q_\pa)\nn
&& 
\times G((p_\pe-Q_\pe)^2,(p_\pa-Q_\pa+\tau P)^2) \nn
&&~~~~\times
 {\mf{V}}((p_\pe-Q_\pe)^2,(p_\pa-Q_\pa)^2,(p_\pe-Q_\pe)^2,(p_\pa-Q_\pa+\tau P)^2)\nn
&&
+\c{G}(p_\pe^2,p_\pa)
G(p_\pe^2,(p_\pa+\tau P)^2)  {\mf{V}}(p_\pe^2,p_\pa^2,p_\pe^2,(p_\pa+\tau P)^2)\nn
&&
\times\c{G}((p_\pe-Q_\pe)^2,p_\pa-Q_\pa+\tau P)
\biggr\rbrack{\dot R}\nn
&&- b_1^2\sum_{\tau}\int_{Q,Q',p} \eta_Q\eta_{Q'}\delta_{-Q-Q'+\tau Pe,0}\c{G}(p_\pe^2,p_\pa)
\nn
&&\times
\biggl\lbrack
G(p_\pe^2, p_\pa^2)
  \c{G}( p_\pe^2, p_\pa-\tau P) \mf{V}(p_\pe^2, (p_\pa-\tau P)^2,p_\pe^2, p_\pa^2)\nn
&&
\times\biggl( \sum_{\tau'}\c{G}((p_\pe-Q_\pe)^2,p_\pa-Q_\pa+\tau' P)\biggr)
\nn
&&
+
 \sum_{\tau'=\pm} 
\c{G}^2((p_\pe-Q_\pe)^2,p_\pa-Q_\pa+\tau' P)G((p_\pe-Q_{\pe})^2,(p_{\pa}-Q_\pa)^2) \nn
&&~~~~\times
 {\mf{V}}((p_\pe-Q_\pe)^2, (p_\pa-Q_\pa+\tau' P)^2,(p_\pe-Q_{\pe})^2,(p_\pa-Q_\pa)^2)
\nn
&&
+\c{G}(p_\pe^2,p_\pa)
 \biggl(\sum_{\tau'=\pm}
 G(p_{\pe}^2,(p_{\pa}-\tau' P)^2) {\mf{V}}(p_\pe^2,p_\pa^2,p_{\pe}^2,(p_{\pa}-\tau' P)^2)\biggr)\nn
&& 
\times\c{G}( (p-Q)_\pe^2,p_\pa-Q_\pa)
\biggr\rbrack{\dot R},
\eea

\bea
&&  T_{2B}^{[2],0}=
b_0^2 \int_{Q,Q',p}\eta_Q\eta_{-Q}\delta_{Q+Q',0}\c{G}(p_\pe^2,p_\pa)\nn
&&\times
\biggl\lbrack
G(p_\pe^2,p_\pa^2)\sum_{\tau=\pm}\c{G}(p_\pe^2,p_\pa+\tau P)
 {\mf{V}}(p_\pe^2,(p_\pa+\tau P)^2,p_\pe^2,p_\pa^2)
\nn 
&& \times\c{G}((p_\pe-Q_\pe)^2,p_\pa-Q_\pa)
 G((p_\pe-Q_\pe)^2,(p_\pa-Q_\pa+\tau P)^2) \nn
&&{\mf{V}}((p_\pe-Q_\pe)^2,(p_\pa-Q_\pa)^2,(p_\pe-Q_\pe)^2,(p_\pa-Q_\pa+\tau P)^2)\nn
&&
+ G(p_\pe^2,p_\pa^2)\sum_{\tau=\pm} \c{G}(p_\pe^2,p_\pa+\tau P)
G(p_\pe^2,(p_\pa+\tau P)^2) \nn
&&\times {\mf{V}}^2(p_\pe^2,p_\pa^2,p_\pe^2,(p_\pa+\tau P)^2)
\c{G}((p_\pe-Q_\pe)^2,p_\pa-Q_\pa+\tau P)\nn
&&
+\c{G}(p_\pe^2,p_\pa)\sum_{\tau=\pm} 
G(p_\pe^2,(p_\pa+\tau P)^2)  {\mf{V}}(p_\pe^2,p_\pa^2,p_\pe^2,(p_\pa+\tau P)^2)
\nn
&&\times
\c{G}((p_\pe-Q_\pe)^2,p_\pa-Q_\pa+\tau P)G((p_\pe-Q_{\pe})^2,(p_\pa-Q_\pa)^2) \nn
&&\times
 {\mf{V}}((p_\pe-Q_\pe)^2,(p_\pa-Q_\pa+\tau P)^2,(p_\pe-Q_{\pe})^2,(p_\pa-Q_\pa)^2)
\biggr\rbrack{\dot R}\nn
&&
+b_1^2\int_{Q,Q',p} \eta_Q\eta_{-Q}\delta_{Q+Q',0} \c{G}(p_\pe^2,p_\pa)\nn
&&\times
\biggl\lbrack
 G(p_\pe^2,p_\pa^2)
 \sum_{\tau,\tau'=\pm} 
\c{G}(p_\pe^2,p_\pa+\tau'P) {\mf{V}}(p_\pe^2, (p_\pa+\tau'P)^2, p_\pe^2,p_\pa^2)
\nn
&&\times
\c{G}((p_\pe-Q_\pe)^2,p_\pa-Q_\pa+\tau P)G((p_\pe-Q_{\pe})^2,(p_{\pa}-Q_\pa)^2) \nn
&&\times
 {\mf{V}}((p_\pe-Q_\pe)^2, (p_\pa-Q_\pa+\tau P)^2,(p_\pe-Q_{\pe})^2,(p_\pa-Q_\pa)^2)
\nn
&&
+ G( p_\pe^2, p_\pa^2) 
\biggl( \sum_{\tau=\pm}
 G(p_{\pe}^2,(p_{\pa}-\tau P)^2) {\mf{V}}(p_\pe^2,p_\pa^2,p_{\pe}^2,(p_{\pa}-\tau P)^2)  \biggr)^2\nn
&& \times\c{G}( (p_\pe-Q_\pe)^2,p_\pa-Q_\pa)   
\nn
&&
+\c{G}(p_\pe^2,p_\pa)
 \biggl(\sum_{\tau=\pm}
 G(p_{\pe}^2,(p_{\pa}-\tau P)^2) {\mf{V}}(p_\pe^2,p_\pa^2,p_{\pe}^2,(p_{\pa}-\tau P)^2)  \biggr)\nn
&&\times
\biggl(\sum_{\tau'}
G( (p_\pe-Q_\pe)^2, (p_\pa-Q_\pa-\tau'P)^2)\nn
&& \times \mf{V} (  (p_\pe-Q_\pe)^2, (p_\pa-Q_\pa)^2 , (p_\pe-Q_\pe)^2,(p_\pa-Q_\pa-\tau'P)^2 )
\biggr)\nn
&&\times \c{G}( (p_\pe-Q_\pe)^2,p_\pa-Q_\pa)
\biggr\rbrack{\dot R},
\eea

\bea
T_{2B}^{[2],1}&=&
 b_0b_1 \sum_{\tau=\pm}\int_{Q,Q',p}  \eta_Q\eta_{Q'}\delta_{-Q-Q'+ \tau Pe,0} \c{G}(p_\pe^2,p_\pa)\nn
&&
\biggl\lbrack 
G(p_\pe^2,p_\pa^2)  \sum_{\tau'=\pm}
\c{G}(p_\pe^2,p_\pa+\tau' P)
{\mf{V}}((p_\pe^2,(p_\pa+\tau' P)^2,p_\pe^2,p_\pa^2)\nn
&&\c{G}((p_\pe-Q_\pe)^2,p_\pa-Q_\pa)\nn
&&\times
G((p_\pe-Q_\pe)^2,(p_\pa-Q_\pa+\tau P)^2) \nn
&&\times{\mf{V}}((p_\pe-Q_\pe)^2,(p_\pa-Q_\pa)^2,(p_\pe-Q_\pe)^2,(p_\pa-Q_\pa+\tau P)^2)\nn
&&
+\c{G}(p_\pe^2,p_\pa)G(p_{\pe}^2,(p_{\pa}+\tau P)^2) 
 {\mf{V}}(p_\pe^2,p_\pa^2,p_{\pe}^2,(p_{\pa}+\tau P)^2)
\nn
&&\times\c{G} ( (p_\pe-Q_\pe)^2, p_\pa-Q_\pa+ \tau P )
 G((p_\pe-Q_\pe)^2, (p_\pa-Q_\pa)^2)\nn
&&\times
\mf{V}(  (p_\pe-Q_\pe)^2,(p_\pa-Q_\pa+ \tau P)^2, (p_\pe-Q_\pe)^2,
 (p_\pa-Q_\pa)^2)\nn
&&
+G(p_\pe^2,p_\pa^2)\biggl(\sum_{\tau'=\pm}
\c{G}(p_\pe^2,p_\pa+\tau'P) {\mf{V}}( p_\pe^2,(p_\pa+\tau'P)^2,p_\pe^2,p_\pa^2)
\biggr)\nn
&&\times
G(p_{\pe}^2,(p_{\pa}+\tau P)^2) 
{\mf{V}}(p_\pe^2,p_\pa^2,p_{\pe}^2,(p_{\pa}+\tau P)^2)\nn
&&\times\c{G}( (p_\pe-Q_\pe)^2, p_\pa-Q_\pa+\tau P)\nn
&&
+G(p_\pe^2,p_\pa^2) \sum_{\tau'}
\mf{V}( p_\pe^2, (p_\pa-\tau P)^2,p_\pe^2,p_\pa^2)
\c{G}( p_\pe^2, p_\pa-\tau P)\nn
&&\times
\c{G}((p_\pe-Q_\pe)^2,(p_\pa-Q_\pa+\tau' P) G((p_\pe-Q_{\pe})^2,(p_\pa-Q_\pa)^2)  \nn
&&\times
{\mf{V}}((p_\pe-Q_\pe)^2,(p_\pa-Q_\pa+\tau' P)^2,(p_\pe-Q_\pe)^2,(p_\pa-Q_\pa)^2)
\nn
&&
+G( p_\pe^2, p_\pa^2) \sum_{\tau'=\pm}
G(p_\pe^2,(p_\pa+\tau' P)^2) 
{\mf{V}}(p_\pe^2,p_\pa^2,p_\pe^2,(p_\pa+\tau' P)^2)\nn
&&\times
\c{G} ( p_\pe^2, p_\pa-\tau P)
\mf{V}( p_\pe^2,( p_\pa-\tau P)^2,  p_\pe^2, p_\pa^2) 
\c{G}( (p_\pe-Q_\pe)^2, p_\pa-Q_\pa) \nn
&&
+ \c{G}(p_\pe^2,p_\pa)\sum_{\tau'=\pm}
G(p_\pe^2,(p_\pa+\tau' P)^2) 
{\mf{V}}(p_\pe^2,p_\pa^2,p_\pe^2,(p_\pa+\tau' P)^2)\nn
&&\times\c{G}( (p_\pe-Q_\pe)^2, p_\pa-Q_\pa)
 G((p_\pe-Q_\pe)^2,( p_\pa-Q_\pa+\tau P)^2)\nn
&&\times\mf{V}( (p_\pe-Q_\pe)^2, (p_\pa-Q_\pa)^2,(p_\pe-Q_\pe)^2, (p_\pa-Q_\pa+\tau P)^2)
\biggr\rbrack {\dot R},\nn
\eea

\bea
 T_{2B}^{[3],0}&=&
- b_0b_1\int_{Q,Q',p} \eta_Q\eta_{-Q}\delta_{Q+Q',0}\sum_{\tau,\tau'=\pm 1}\nn
&&\times
\biggl\lbrack
\c{G}(p_\pe^2,p_\pa)G(p_\pe^2,p_\pa^2)
\c{G}(p_\pe^2,p_\pa+\tau'P) {\mf{V}}( p_\pe^2,(p_\pa+\tau'P)^2,p_\pe^2,p_\pa^2)
\nn
&&\times
 G(p_{\pe}^2,(p_{\pa}+\tau P)^2) 
 {\mf{V}}(p_\pe^2,p_\pa^2,p_{\pe}^2,(p_{\pa}+\tau P)^2)\nn
 &&\times\c{G}( (p-Q)_\pe^2, p_\pa-Q_\pa+\tau P)
G( (p_\pe-Q_\pe)^2, (p_\pa-Q_\pa)^2)\nn
&&\times\mf{V}( (p_\pe-Q_\pe)^2,(p_\pa-Q_\pa+\tau P)^2, (p_\pe-Q_\pe)^2,(p_\pa-Q_\pa)^2)\nn
&&+
\c{G}(p_\pe^2,p_\pa) G(p_\pe^2, p_\pa^2)
G(p_\pe^2,(p_\pa+\tau P)^2) 
{\mf{V}}(p_\pe^2,p_\pa^2,p_\pe^2,(p_\pa+\tau P)^2)\nn
&&\times\c{G} ( p_\pe^2, p_\pa+\tau' P)
\mf{V}(  p_\pe^2, (p_\pa+\tau' P)^2,p_\pe^2, p_\pa^2) \nn
&&\times 
\c{G}( (p_\pe-Q_\pe)^2, p_\pa-Q_\pa)
 G((p_\pe-Q_\pe)^2,( p_\pa-Q_\pa+\tau'P)^2)\nn
&&\times
\mf{V}( (p_\pe-Q_\pe)^2, (p_\pa-Q_\pa)^2,(p_\pe-Q_\pe)^2, (p_\pa-Q_\pa+\tau' P)^2)
\biggr\rbrack {\dot R},\nn
\eea

\bea
T_{2B}^{[3],1}&=&
-b_0^2\sum_{\tau=\pm} \int_{Q,Q',p}\eta_Q\eta_{Q'}\delta_{-Q-Q'+\tau Pe,0}
\nn
&&\times
\biggl\lbrack
\c{G}(p_\pe^2,p_\pa) G(p_\pe^2,p_\pa^2)
\c{G}(p_\pe^2,p_\pa-\tau P)
 {\mf{V}}(p^2_\pe,(p_\pa-\tau P)^2, p_\pe^2,p_\pa^2)\nn
&&\times
 \sum_{\tau'=\pm} 
G(p_\pe^2,(p_\pa+\tau' P)^2)  {\mf{V}}(p_\pe^2,p_\pa^2,p_\pe^2,(p_\pa+\tau' P)^2)
\nn
&&\times
\c{G}((p_\pe-Q_\pe)^2,p_\pa-Q_\pa+\tau' P)G((p_\pe-Q_\pe)^2,(p_\pa-Q_\pa)^2) \nn
&&\times
 {\mf{V}}((p_\pe-Q_\pe)^2,(p_\pa-Q_\pa+\tau' P)^2,(p_\pe-Q_{\pe})^2,(p_\pa-Q_\pa)^2)
\Biggr\rbrack{\dot R}\nn
&&
- b_1^2\sum_{\tau}\int_{Q,Q',p} \eta_Q\eta_{Q'}\delta_{-Q-Q'+\tau Pe,0}\nn
&&\times
\biggl\lbrack
 \c{G}(p_\pe^2,p_\pa)G(p_\pe^2,p_\pa^2)
\c{G}( p^2_\pe, p_\pa-\tau P)
\mf{V}(  p^2_\pe,( p_\pa-\tau P)^2,p_\pe^2,p_\pa^2)
\nn
&&\times
\biggl(\sum_{\tau''=\pm}
 G(p_{\pe}^2,(p_{\pa}-\tau ''P)^2) {\mf{V}}(p_\pe^2,p_\pa^2,p_{\pe}^2,(p_{\pa}-\tau'' P)^2)  \biggr)\nn
 &&\times \c{G}( (p_\pe-Q_\pe)^2,p_\pa-Q_\pa)\nn
&&\times
\biggl(\sum_{\tau'}
G( (p_\pe-Q_\pe)^2, (p_\pa-Q_\pa-\tau'P)^2)\nn
&&\times\mf{V} (  (p_\pe-Q_\pe)^2, (p_\pa-Q_\pa)^2 , (p_\pe-Q_\pe)^2,(p_\pa-Q_\pa-\tau'P)^2 )
\biggr)
\biggr\rbrack{\dot R}.\nn
\eea
 Since no field-dependence of the derivative couplings have been taken into account, the trace $T_{2B}$ has to be taken at vanishing homogeneous background field $\Phi=0$.

In order to identify the terms  containing the integrals of the types \\
 $\int_{Q,Q'} \eta_Q\eta_{Q'} Q_\pe^{2n'}Q_\pa^{2m'} \delta_{Q+Q',0}$ and
 $\int_{Q,Q'} \eta_Q\eta_{Q'} $ $ Q_\pe^{2n'}Q_\pa^{2m'} \delta_{-Q-Q'+Pe,0}$ 
for $(n',m')=(1,0),~ (0,1),~ (2,0),~(1,1), (0,2)$  one has to perform the Taylor expansion of $T_{2B}$ with respect to $Q_{\pe i}$ $(i=1,\ldots, d-1)$ and
$Q_\pa$ and take the second derivatives $\sum_{i=1}^{d-1} \partial^2/\partial Q_{\pe i}^2$, $\partial^2/\partial Q_\pa^2$ and the fourth derivatives\\
$(\sum_{i,j=1}^{d-1} \partial^2/\partial Q_{\pe i}^2 )(\partial^2/\partial Q_{\pe i}^2)$,
 $(\partial^2/\partial Q_\pa^2)(\sum_{i=1}^{d-1} \partial^2/\partial Q_{\pe i}^2)$,
and  $\partial^4/\partial Q_\pa^4$, respectively, at vanishing momenta $Q_\pe=Q_\pa=0$. Then the kernels $\c{T}^{(m)}(Q_\pe^2, Q_\pa^2)$ are identified as the
truncated Taylor-expansions in momenta $Q_\pe$ and $Q_\pa$,
\bea
T_{2B}&=&\sum_{m=0}^1 \sum_{n=1}^3 T_{2B}^{ [n],(m)}= \sum_{m=0}^1 \int_{Q, Q'}\eta_Q\eta_{Q'} \c{T}^{(m)}(Q_\pe^2, Q_\pa^2)\delta_{-Q-Q'+m\tau Pn_1,0}.\nn
\eea

The ordering of the contributions to the trace $T_{2B}$ according to the power
of the explicit occurrence of the induced vertex $\mf{V}$ enables one to 
perform an expansion in powers $l$ $(l=0,1,2,3)$ of $\mf{V}$. Then, however, one has to take into account that $\c{G}^{-1}$ also contains terms of the orders $\mf{V}^0$ and $\mf{V}^2$.

In the zeroth order of the induced vertex $\mf{V}$ the flow equations for 
the momentum-dependent wave function renormalization $\c{Z}(Q_\pe^2, Q_\pa^2)$ and the periodic condensate induced  wave function renormalization $\c{E}(Q_\pe^2, Q_\pa^2)$ are given as
 \bea\label{wfreneqA}
 \int_{Q,Q'}  \gamma_k^{(2)m} (Q_\pe^2, Q_\pa^2)\sum_\tau\delta_{-Q-Q'+m\tau Pn_1,0}\eta_Q\eta_{Q'} &=& \hf T_{2B}^{[0],m}
\eea
with $m=0$ and $1$, respectively.
More explicitly these equations can be written as
\bea\label{wfreneqZ}
&& \int_Q \biggl\lbrack
  \hf {\dot {\c{Z}}}(Q_\pe^2,Q_\pa^2) 
+ \hf {\dot U}'' + \hf \sigma {\dot V}'''+\hf \sigma^2 {\dot U}''''
\biggr\rbrack 
\eta_Q\eta_{-Q}
=
\hf\int_{Q,p}G^2(p_\pe^2,p_\pa^2) \biggl\{
b_0^2G((p_\pe-Q_\pe)^2,(p_\pa-Q_\pa)^2)
\nn
&&
~~~+ b_1^2  \sum_{\tau=\pm} 
G((p_\pe-Q_\pe)^2,(p_\pa-Q_\pa+\tau P)^2) \biggr\}{\dot R}\eta_Q\eta_{-Q} \nn
\eea
and
\bea\label{wfreneqE}
&& \int_{Q,Q'} \biggl\lbrack
\frac{1}{4}  {\dot {\c{E}}}(Q_\pe^2, Q_\pa^2) 
 +\frac{1}{4} {\dot V}''+  \hf \sigma {\dot U}''' 
+ \frac{3}{8} \sigma^2 {\dot V}'''' 
\biggr\rbrack \sum_\tau \delta_{-Q-Q'+\tau Pe,0} \eta_Q\eta_{Q'}
\nn
&&
~~~=
\hf b_0b_1\sum_{\tau=\pm} \int_{Q,Q',p} \eta_Q\eta_{Q'} \delta_{-Q-Q'+ \tau Pe,0} G^2(p_\pe^2,p_\pa^2)
\nn
&&
~~~\times\biggl\{G((p_\pe-Q_\pe)^2,(p_\pa-Q_\pa)^2)
+ G((p_\pe-Q_\pe)^2,(p_\pa-Q_\pa+\tau P)^2) \biggr\}{\dot R}. \nn
\eea
It is worthwhile mentioning that for the EAA preserving $Z_2$ symmetry  and
using Taylor-expansion of the potentials at $\Phi=0$, the induced vertex $\mf{V}
$ vanishes, so that the right-hand sides of Eqs. \eq{wfreneqZ} and \eq{wfreneqE} cannot accommodate contributions depending on the induced vertex $\mf{V}$. Then 
Eq. \eq{wfreneqE} turns to the identity $0=0$ for identically vanishing induced
wave function renormalization $\c{E}$, that is supposed to vanish for keeping
the $Z_2$ symmetry of the EAA.

\section{Extrema of the EAA}\label{minimi}

The positions of the extrema of the rEAA \eq{gamfib} with respect to the parameter $\sigma$, i.e., the amplitude of the periodic condensate are given  by the cubic Eq. \eq{cubiceqsig}. With the help of the new variable
$  s= \sigma + [V'''/(2U'''')]$ Eq. \eq{cubiceqsig} can be recast into the form
\bea\label{cubiceq}
  \c{P}(s)&=& s^3 + 3\mf{p} s+2\mf{q}=0
\eea
with
\bea\label{mfrakp}
\mf{p}&=& 
  \frac{2}{3U''''} \biggl(\c{Z}(0,P^2)+ U'' -\frac{3(V''')^2 }{8U''''} \biggr),\nn
\mf{q} &=& 
\frac{1}{2U''''} \biggl( V' + \frac{(V''')^3}{4\lbrack U''''\rbrack^2} 
-  \frac{ V'''[\c{Z}(0,P^2)+U'']}{U''''} \biggr),
\eea
where we suppressed the notation of the dependences  on the variables $\Phi$ and $k$. As to the roots of Eq. \eq{cubiceq}, let us consider first the case with
  $V\equiv 0$ that implies  $\mf{q}=0$. Then Eq. \eq{cubiceq}
 reduces to
\bea\label{redcubiceq}
  s(s^2+3\mf{p})&=&0.
\eea
This equation has   {\em (i)} the trivial real root $\sigma^{(0)}=s_0^{(0)}=0$ 
and  {\em (ii)} the roots $\sigma^{(\pm)}=s_0^{(\pm)}=\pm\sqrt{ -3\mf{p}}$.  The latter ones are real and non-vanishing if and only if it holds $\mf{p}<0$. 
Then the non-trivial real roots do exist for $g_4(k)>0$, $Z_k<0$ if and only if it holds the inequality
\bea\label{ineq}
\c{Z}(0,P^2)+ g_2&=& g_2-\frac{1}{4Y}<0.
\eea
 It is straightforward to show that the rEAA \eq{gamfib}
takes identical minimum values  for the non-trivial real roots if those exist, because it holds
\bea
 V^{-1}( \Gamma_k[\phi_B]|_{\sigma^{(\pm)}}- \Gamma_k[\phi_B]|_{\sigma^{(0)}})
&=&
 -\frac{1}{g_4}\biggl\lbrack  \c{Z}(0,P^2) + U''\biggr\rbrack^2 <0.
\eea
Since $g_4>0$, $\Gamma_{Bk}$ vs. $\sigma$ has to minima and a maximum in the case, when \eq{cubiceq} has three real roots.

In the general case with non-vanishing induced potential $V$ the number of the real roots of Eq. \eq{cubiceq} can be discussed as follows.
The polynomial $\c{P}(s)$ may have real critical points $s_\pm = \pm\sqrt{ -\mf{p}}$ at which it holds
\bea
  \frac{d\c{P}(s) }{ds} &=&3s^2 + 3\mf{p} =0 .
\eea
Then the following cases can be distinguished:
\begin{enumerate}
\item
For $\mf{p}>0$ there are no critical points, $\c{P}(s)$ is strictly monotonically increasing and Eq. \eq{cubiceq} has a single real root which is the minimum place of  $\Gamma_{Bk}$ vs. $\sigma$.
\item
For $\mf{p}=0$ there is a single critical point, an inflection point of  $\c{P}(s)$, which is still strictly monotonically increasing (except of the critical point) and has a single real root  which is the minimum place of  $\Gamma_{Bk}$ vs. $\sigma$.

\item
For $\mf{p}<0$ there are 2 critical points of $\c{P}(s)$, $s_-=-s_+<s_+=\sqrt{ -\mf{p}}$. Then $\c{P}(s)$ starts at $s\to -\infty$ to increase strictly monotonically, then reaches a maximum at $s_-$ followed by a minimum at $s_+$ and 
 increases again strictly monotonically for $s>s_+$, so that it holds the inequality $\c{P}(s_-)> \c{P}(s_+)$. In this case there are 3 real roots if and only if $\c{P}(s_-)> )$ and $\c{P}(s_+)<0$, otherwise there is only a single real root. Since we have $\c{P}(s_\pm)=
2 \lbrack \mf{q} \mp (-\mf{p})^{3/2} \rbrack$, the necessary and sufficient condition for the existence of 3 real roots $s_i$ $(i=1,2,3)$ is given by the inequality
\bea
&& \mf{p}\sqrt{-\mf{p}} < \mf{q} < -\mf{p} \sqrt{-\mf{p}} .
\eea
The roots  should satisfy the inequalities $s_1<s_-< s_2 <s_+<s_3$,
and then $s_1$ and $s_3$ are the minima of $\Gamma_{Bk}$ vs. $\sigma$. Then one has to decide by computing their values, which one is smaller.

\end{enumerate}

\section{Flow equations in LPA}\label{a:lpaflow}

In the LPA the derivative couplings keep their bare values, $Z_{\pe~k}=Z_{\pa~k}\equiv Z$, $Y_{\pe~k}=Y_{X~k}=Y_{\pa~k}\equiv Y$,
the regulators given by Eqs. \eq{litreg} and \eq{regsh} reduce to
\bea 
&& R_k(p_\pe^2, p_\pa^2) = \biggl\lbrack Z\biggl(k^2 -p_\pe^2 -p_\pa^2\biggr)
~~~+ Y \biggl( k^4 - (p_\pe^2+p_\pa^2)^2\biggr)\biggr\rbrack\Theta(\hf k^2 -p_\pe^2) \Theta(\hf k^2 -p_\pa^2),
\\
&&
R_k^{[\pm]}\bigl(p_\pe^2, (p_\pa\pm P_k)^2\bigr) =     
\Biggl\{ Z \biggl\lbrack \hf k^2 +   \biggl( \frac{k}{\sqrt{2}}+P_k\biggr)^2
- p_\pe^2 -(p_\pa\pm P_k)^2 \biggr\rbrack 
\nn
&&
~~~+ Y \Biggl\lbrack 
\biggl\lbrack \frac{k^2}{2} +  \biggl( \frac{k}{\sqrt{2}}+P_k\biggr)^2
\biggr\rbrack^2
- \biggl( p_\pe^2 + (p_\pa\pm P_k)^2 \biggr)^2 
\Biggr\rbrack  \Biggr\} 
\Theta(\hf k^2-p_\pe^2)\Theta(\hf k^2-p_\pa^2)
\eea   
 the regulated reduced propagators given by Eqs. \eq{greg} and \eq{gregsh}
can be recast into the forms
\bea\label{regredprop}
&&G(p_\pe^2, p_\pa^2,\Phi) =\biggl\lbrack Zk^2 +Yk^4
  + U_k''(\Phi) +\sigma_k(\Phi) V_k'''(\Phi) +\hf \sigma_k^2(\Phi) U_k''''(\Phi)
\biggr\rbrack^{-1},\nn
\eea
\bea\label{regredpropsh}
&&G\bigl(p_\pe^2, (p_\pa \pm P_k)^2,\Phi\bigr) =
\Biggl\lbrack
Z \Biggl(  \hf k^2 +   \biggl( \frac{k}{\sqrt{2}}+P_k\biggr)^2\Biggr)
+ Y \Biggl( \frac{k^2}{2} +  \biggl( \frac{k}{\sqrt{2}}+P_k\biggr)^2\Biggr)^2
\nn
&&
~~~+ U_k''(\Phi) +\sigma_k(\Phi) V_k'''(\Phi) +\hf \sigma_k^2(\Phi) U_k''''(\Phi)
\Biggr\rbrack^{-1},\nn
\eea
which are independent of the momenta,
and the derivative of the regulator given in Eq. \eq{dotr} reduces to
\bea\label{dotreglpa}
  {\dot R}_k (p_\pe^2,p_\pa^2)&=& 2Z k^2+ 4Yk^4.
\eea
Following the strategy of deriving the separate flow equations for the
ordinary potential $U_k(\Phi)$ and the induced potential $V_k(\Phi)$, one finds
from Eq. \eq{potflow} the equations
\bea\label{ulpa1}
\dot{U}_k(\Phi) 
&=&
\hf 
\int_p \biggl\lbrack 
1 -  \frac{1}{4}\lbrack V_k''(\Phi)\rbrack^2 G(p_\pe^2,p_\pa^2,\Phi)
\sum_{\tau=\pm 1} G(p_\pe^2,(p_\pa+\tau P)^2,\Phi)
\biggr\rbrack^{-1}G(p_\pe^2,p_\pa^2,\Phi){\dot R}_k
\nn
\eea
and
\bea\label{potlpa1}
&&\dot{U}_k(\Phi) + \dot{V}'(\Phi)\sigma_k(\Phi) 
+ \dot{U}_k''(\Phi) \sigma_k^2 (\Phi)+ \hf \dot{V}_k'''(\Phi) \sigma_k^3(\Phi) + \frac{1}{4} \dot{U}_k''''(\Phi)\sigma^4(\Phi)
\nn
&& 
~~~=\hf  \int_p \biggl\lbrack 
1 -  \mf{V}^2(\Phi) G(p_\pe^2,p_\pa^2,\Phi)\sum_{\tau=\pm 1} G(p_\pe^2,(p_\pa+\tau P)^2,\Phi)
\biggr\rbrack^{-1} G(p_\pe^2,p_\pa^2,\Phi){\dot R}_k,
\nn
\eea
respectively, where the LPA form of the induced vertex is
\bea\label{vaulpa}
  \mf{V}(\Phi)&=&\hf \biggl\lbrack V_k''(\Phi) + 2\sigma_k(\Phi) U'''_k(\Phi)
 + \frac{3}{2} \sigma_k^2(\Phi) V_k''''(\Phi) \biggr\rbrack  .
\eea

The dimensionless quantities are introduced via the relations $\Phi= k^{1/2} \t{\Phi}$, $\sigma_k=k^{1/2} \t{\sigma}_k$, $U_k=k^3\t{U}_k$, $V_k=k^3\t{V}_k$, and
$Yk^2=\t{Y}$, yielding the expressions for the dimensionless regulated reduced propagators
\bea
\t{G}(p_\pe^2,p_\pa^2,\Phi)
&=&
\biggl( Z + \t{Y} +\t{U}_k''(\t{\Phi}) 
+ \t{\sigma}_k(\t{\Phi})\t{V}_k'''(\t{\Phi}) 
+ \hf  \t{\sigma}_k^2(\t{\Phi})\t{U}_k''''(\t{\Phi}) \biggr)^{-1}
\equiv  \t{g}(\t{\Phi},\t{\sigma}_k(\t{\Phi}) ),
\eea

\bea
&&\t{G}(p_\pe^2,(p_\pa\pm P)^2,\Phi)
=\Biggl\lbrack Z \Biggl( \hf +\biggl( \frac{1}{\sqrt{2}}+ \t{P}_k\biggr)^2\Biggr)
+ \t{Y}\Biggl(   \hf +\biggl( \frac{1}{\sqrt{2}}+ \t{P}_k\biggr)^2\Biggr)^2
\nn
&&
~~~+\t{U}_k''(\t{\Phi}) 
+ \t{\sigma}_k(\t{\Phi})\t{V}_k'''(\t{\Phi}) 
+ \hf  \t{\sigma}_k^2(\t{\Phi})\t{U}_k''''(\t{\Phi})
\Biggr\rbrack^{-1}\equiv\tilde{\mf{g} } (\t{\Phi},\t{\sigma}_k(\t{\Phi}) ),
\eea
and the dimensionless induced vertex
\bea\label{mfVlpa}
 { \tilde {\mf{V}} }(\t{\Phi})&=&\hf \biggl\lbrack 
\t{V}_k''(\t{\Phi}) + 2\t{\sigma}_k(\t{\Phi}) \t{U}'''_k(\t{\Phi})
 + \frac{3}{2} \t{\sigma}_k^2(\t{\Phi}) \t{V}_k''''(\t{\Phi}) \biggr\rbrack.  
\eea
Then we find the dimensionless forms of Eqs. \eq{ulpa1} and \eq{potlpa1},
\bea\label{ulpadiml}
 {\dot{\t{U}}}_k
 &=& -3\t{U}_k 
 +\hf \t{\Phi}\t{U}'_k+
\frac{ Z + 2\t{Y}}{ 8\pi^2 \sqrt{2}}
\biggl( 1 -\hf \lbrack {\tilde {V}}_k'' \rbrack^2 \t{g}(\t{\Phi},0)
\tilde {\mf g}(\t{\Phi},0) \biggr)^{-1}\t{g}(\t{\Phi},0),
\eea
\bea\label{potlpadiml}
&&
{\dot{\t{U}}}_k
+ \t{\sigma_k} {\dot{\t{V}}}_k'
+ \t{\sigma_k}^2  {\dot{\t{U}}}_k'' 
+\hf \t{\sigma_k}^3 {\dot{\t{V}}}_k'''
+\frac{1}{4} \t{\sigma_k}^4  {\dot{\t{U}}}_k'''' 
=
-3\biggl( \t{U}_k +\t{\sigma_k} \t{V}_k'
+ \t{\sigma_k}^2 {\t{U}}_k'' 
+\hf \t{\sigma_k}^3 {\t{V}}_k''' + \frac{1}{4} \t{\sigma_k}^4 \t{U}_k''''\biggr)
\nn
&&
+\hf \t{\Phi}\biggl( \t{U}'_k + \t{\sigma_k} \t{V}_k''
+\t{\sigma_k}^2  {\t{U}}_k''' 
+ \hf \t{\sigma_k}^3 {\t{V}}_k'''' 
\biggr)
+\frac{ Z + 2\t{Y}}{ 8\pi^2 \sqrt{2}}
\biggl( 1 -\hf \lbrack {\tilde { \mf{V}}}(\t{\Phi},\t{\sigma}_k) \rbrack^2 \t{g}(\t{\Phi},\t{\sigma}_k)
\tilde {\mf g}(\t{\Phi},\t{\sigma}_k) \biggr)^{-1}\t{g}(\t{\Phi},\t{\sigma}_k)
\eea
 Here we suppressed in the notation the dependence of the potentials and $\t{\sigma}_k$ on $\t{\Phi}$.

\section{Flow equations in NLO of GE}\label{a:nloflow}

 In NLO of the GE the couplings
$Z_{\pe ~k} $ and $Z_{\pa~k}$ are running, but the quartic derivative couplings keep their bare values $Y_{\pe~k}=Y_{X~k}=Y_{\pa~k}\equiv Y$.
The regulators given by Eqs. \eq{litreg} and \eq{regsh} reduce to
\bea\label{litregnlo}
&&  R_k(p_\pe^2,p_\pa^2) =
\biggl\lbrack Z_{\pe ~k}\biggl(\hf k^2-p_\pe^2\biggr)+ Z_{\pa~k}\biggl(\hf k^2-p_\pa^2\biggr)
+ Y\biggl( k^4-(p_\pe^2+p_\pa^2)^2\biggr)
\biggr\rbrack\Theta(\hf k^2-p_\pe^2)\Theta(\hf k^2-p_\pa^2),
\eea
\bea\label{regshnlo} 
&& R_k^{[\pm  ]}\bigl(p_\pe^2,(p_\pa\pm P_k)^2\bigr) =
\biggl\lbrack Z_{\pe~ k}\biggl(\hf k^2-p_\pe^2\biggr)
+ Z_{\pa~k}\biggl\lbrack \biggl( \frac{k}{\sqrt{2}}+P_k\biggr)^2-(p_\pa\pm P_k)^2\biggr\rbrack 
\nn
&&
~~~+ Y \Biggl\lbrack 
\biggl\lbrack \frac{k^2}{2} +  \biggl( \frac{k}{\sqrt{2}}+P_k\biggr)^2
\biggr\rbrack^2
- \biggl( p_\pe^2 + (p_\pa\pm P_k)^2 \biggr)^2 
\Biggr\rbrack \Biggr\}
\Theta(\hf k^2-p_\pe^2)\Theta(\hf k^2-p_\pa^2).
\eea   
The regulated reduced propagators given by Eqs. \eq{greg} and \eq{gregsh}
can be recast into the forms
\bea\label{gregnlo}
 G(p_\pe^2,p_\pa^2) &=&\biggl\lbrack (Z_{\pe~k} + Z_{\pa~k}) \hf k^2
  +Yk^4 + U_k''(\Phi) +\sigma_k(\Phi) V_k'''(\Phi)+\hf \sigma_k^2(\Phi) U_k''''(\Phi)\biggr\rbrack^{-1},\\
\label{gregshnlo}
 G\bigl(p_\pe^2, (p_\pa\pm P_k)^2\bigr)&=& \Biggl\lbrack
Z_{\pe~k} \hf k^2 + Z_{\pa~k}\biggl(  \frac{k}{\sqrt{2}}+P_k\biggr)^2
+ Y \Biggl( \frac{k^2}{2} +  \biggl( \frac{k}{\sqrt{2}}+P_k\biggr)^2\Biggr)^2
\nn
&&
+ U_k''(\Phi) +\sigma_k(\Phi) V_k'''(\Phi) +\hf \sigma_k^2(\Phi) U_k''''(\Phi)\Biggr\rbrack^{-1}.
\eea
The scale-derivative of the regulator function is given as
\bea\label{dotregnlo}
{\dot R}_k(p_\pe^2, p_\pa^2)&=& 
 (Z_{\pe~k}+Z_{\pa ~k})k^2 + 4Y k^4 + \rho_k(p_\pe^2,p_\pa^2)
\eea
where
\bea\label{rhonlo}
  \rho_k(p_\pe^2,p_\pa^2)&=& {\dot Z}_{\pe~k}\biggl( \hf k^2- p_\pe^2\biggr)
 + {\dot Z}_{\pa~k}\biggl( \hf k^2- p_\pa^2\biggr).
\eea
The flow Eq. \eq{ulpa1} for the ordinary potential $U_k(\Phi)$ holds even now, and the flow equation for the induced potential takes the form
\bea\label{potflnlo} 
&& {\dot U}_k(\Phi)+ \sigma_k^2(\Phi) \biggl({\dot{\c{Z}}}_k(0,P_k^2)+{\dot U}_k''(\Phi)\biggr)
+\sigma_k(\Phi){\dot V}_k'(\Phi)+ \sigma_k^2(\Phi) {\dot U}_k''(\Phi)
+ \hf \sigma_k^3(\Phi) {\dot V}_k'''(\Phi)
+ \frac{1}{4}\sigma_k^4(\Phi) {\dot U}_k'''' (\Phi)
\nn
~~~&=&
\hf \int_p \c{G}(p_\pe^2, p_\pa,\Phi){\dot R}_k(p_\pe^2,p_\pa^2),
\eea
As compared to the corresponding Eq. \eq{potlpa1} in LPA the only modification is that the term $\sigma_k^2(\Phi){\dot{\c{Z}}}_k(0,P_k^2)$ appears in NLO on the right-hand side of Eq. \eq{potflnlo}. In Appendix \ref{trquadeta} we have derived  the flow equation \eq{wfreneqZ} for the momentum-dependent wave function renormalization in zeroth order of  the induced vertex $\mf{V}$. For the latter Eq. \eq{mfVlpa} holds even in the NLO approximation. In the case of the usage of the $Z_2$-symmetric ansatz  \eq{eaaon} for the EAA with $Z_2$ symmetric ordinary potential, the Taylor-expansion at $\Phi=0$, and neglecting the induced wave function renormalization $\c{E}_k(Q_\pe^2,Q_\pa^2)$, one finds $\mf{V}(0)=0$, so that the terms depending explicitly on the induced vertex $\mf{V}$ fall off in the general flow equation for the momentum-dependent wave function renormalization and one arrives at the flow Eq.  \eq{wfreneqZ}, where also the factor $b_0^2=0$ for  $\Phi=0$, Therefore Eq.
 \eq{wfreneqZ} reduces to the equation 
\bea\label{wfrflnlo}
&& \int_Q \biggl\lbrack
\hf  ( {\dot Z}_{\pe~k}Q_\pe^2+  {\dot Z}_{\pa~k}Q_\pa^2)
+ \hf {\dot U}_k''(0) + \hf \sigma_k(0) {\dot V}_k'''(0) +\hf \sigma_k^2(0) {\dot U}_k''''(0)
\biggr\rbrack 
\eta_Q\eta_{-Q}
\nn
~~~&=&
b_1^2 \hf\int_{Q,p}G^2(p_\pe^2,p_\pa^2,0) \biggl\{\sum_{\tau=\pm} 
G((p-Q)_\pe^2,(p_\pa-Q_\pa+\tau P)^2,0) \biggr\}{\dot R}_k(p_\pe^2,p_\pa^2)\eta_Q\eta_{-Q} 
\eea
in NLO of the GE,
where both sides are taken at $\Phi=0$.
The corresponding flow equations for the wave function renormalizations $Z_{A~k}$
 $(A=\pa,~\pe)$ are read off by comparing the terms linear in  $Q_\pe^2 $ and $Q_\pa^2$ on both sides of Eq. \eq{wfrflnlo},
\bea\label{wfrAflnlo}
{\dot Z}_{A~k}&=&b_1^2 \hf  \biggl\lbrack \partial_{Q_A}^2 \int_{p}G^2(p_\pe^2,p_\pa^2,0)
\sum_{\tau=\pm} G((p-Q)_\pe^2,(p_\pa-Q_\pa+\tau P)^2,0){\dot R}_k(p_\pe^2,p_\pa^2) \biggr\rbrack_{Q =0 },
\eea
where the momentum-derivatives $ \partial_{Q_A}^2$ are to be understood as \\ $\sum_{\mu=2}^3\partial_{Q_{\pe,\mu}}\partial_{Q_{\pe,\mu}}$ for $A=\pe$ and $\partial_{Q_\pa}^2$ for $A=\pa$.
The change to dimensionless quantities is performed similarly as in the LPA. Explicit forms of the beta-functions for the various couplings have been obtained by computer algebraic routine.

\bibliographystyle{elsarticle-num}

\end{document}